\documentclass{aastex631}

\usepackage{graphicx}
\usepackage{comment}
\usepackage{amsmath}

\newcommand{\beq}{\begin{equation}}
\newcommand{\eeq}{\end{equation}}

\accepted{for publication in the Astrophysical Journal on 25 June 2025}

\begin{document}

\title{Deterministic and Stochastic Study of the X-ray Emission from the TeV Blazar Mrk~421}

\author[0000-0001-7614-7672]{Radim P{\'a}nis}
\affiliation{Research Centre for Theoretical Physics and Astrophysics, Institute of Physics in Opava, Silesian University in Opava,\\
Bezru{\v c}ovo n{\'a}m. 13, CZ-74601 Opava, Czech Republic}
\email{radim.panis@physics.slu.cz}

\author[0000-0002-0705-6619]{Gopal Bhatta}
\affiliation{Janusz Gil Institute of Astronomy, University of Zielona Góra, ul. Szafrana 2, 65-516 Zielona Góra, Poland}

\author[0000-0003-4586-0744]{Tek P. Adhikari}
\affiliation{CAS Key Laboratory for Research in Galaxies and Cosmology, Department of Astronomy, University of Science and Technology of China, Hefei, Anhui 230026, China}
\affiliation{School of Astronomy and Space Science, University of Science and Technology of China, Hefei, Anhui 230026, China}
\email{tek@ustc.edu.cn}

\author{Maksym Mohorian}
\affiliation{School of Mathematical and Physical Sciences, Macquarie University, Sydney, NSW 2109, Australia}
\affiliation{
Astronomy, Astrophysics and Astrophotonics Research Centre, Macquarie University, Sydney, NSW, 2109, Australia}

\author[0000-0002-9126-1817]{Suvas Chandra Chaudhary}
\affiliation{Inter-University Center for Astronomy and Astrophysics (IUCAA), Pune, India}
\affiliation{Department of Physics, University of the Free State, 205 Nelson Mandela Dr., Bloemfontein, 9300, South Africa}
\author{Adithiya Dinesh}
\affiliation{University of St. Andrews, Scotland}

\author{Rajesh K. Bachchan}
\affiliation{Department of Physics, Patan Multiple Campus, Tribhuvan University, Nepal}

\author{Niraj Dhital}
\affiliation{Central Department of Physics, Tribhuvan University, Kirtipur 44613, Nepal}

\author{Zden\v{e}k Stuchl{\'i}k}
\affiliation{Research Centre for Theoretical Physics and Astrophysics, Institute of Physics in Opava, Silesian University in Opava,\\
Bezru{\v c}ovo n{\'a}m. 13, CZ-74601 Opava, Czech Republic}

\begin{abstract}
We present a comprehensive timing analysis of X-ray data from the {\it XMM-Newton} satellite, examining 50 light curves covering 17 years of observations of the blazar Mrk~421.  This work uses classical deterministic and  stochastic methods in a novel way, enabling the distinction of  temporal scales and offering essential insights through correlations 
among parameters. Deterministic behaviors are primarily explored through recurrence quantification analysis (RQA), used innovatively by varying the threshold input parameter to 
examine variability at multiple temporal scales.  To investigate behavior across various scales from a stochastic perspective, we apply both autoregressive moving average (ARMA) and autoregressive integrated moving average (ARIMA) models, with results from ARIMA more tightly related to short scales. Our findings reveal that Mrk~421's X-ray emission is a multifaceted process, driven by both deterministic and  stochastic patterns, indicating a complex interplay of physical phenomena. Our study demonstrates that deterministic patterns are more pronounced at small temporal scales, which are disconnected from large scales.  On the other hand, stochastic processes with memory propagate from large to small time scales, while noise affects both scales, as indicated by the correlation analysis. These results underscore the importance of advanced methodologies for interpreting astrophysical data, contributing to ongoing discussions in blazar physics by exploring connections between our calculated parameters and established models. The same approach can potentially be applied to other sources, enhancing our general understanding of variability and emission mechanisms in blazars.
\end{abstract}



\section{Introduction} \label{sec:intro}
Active Galactic Nuclei (AGNs) are compact regions at the center of galaxies with intense radiation emission caused by accretion of matter onto supermassive black  holes. Among AGNs, blazars are characterized by prominent jets pointing within a few degrees of our line of sight, with apparent emission amplified by an order of magnitude - or more - due to relativistic beaming.  Blazars are distinguished in BL Lacertae objects (BL Lacs) and flat spectrum radio quasars (FSRQs) due to the evidence of distinct features  in their optical spectra, with BL Lacs exhibiting weak or absent emission lines compared to FSRQs \citep{1995PASP..107..803U,2015ApJ...810...14A,1999NuPhS..69..389M,2015mgm..conf.1015G,2017AIPC.1792b0007T,1999ASPC..159..311G}.

      X-ray observations are crucial for understanding the high-energy dynamics around supermassive black holes in AGNs and blazars. They provide a unique window into the high-energy particles present in the relativistic jet of blazars, offering insights into their spectral energy distribution (SED).   Their SED shows two distinct energy  peaks, indicating specific emission processes potentially linked to the acceleration mechanisms and interactions of relativistic particles within the jets.
Furthermore, X-ray observations grant us access to crucial information about the physical conditions within the jet itself, including the strength of the magnetic field, the temperature of the emitting plasma, and the density of relativistic particles \citep{2001ASPC..234..437M}.

Here we present our analysis of the X-ray variability of one of the most observed blazar,  Mrk~421, using all the relevant X-ray observations from the {\it XMM-Newton} satellite. In this analysis, we prioritize methods with minimal parameter selection, including both deterministic and stochastic modeling. Specifically, we employ autoregressive moving average (ARMA) and autoregressive integrated moving average (ARIMA) models to capture stochastic processes, while recurrence quantification analysis (RQA) is used to explore deterministic structures in the light curves. These approaches allow us to distinguish different temporal scales and provide a detailed examination of Mrk~421 variability \citep{2019PASP..131f3001M,2020ApJ...905..160B,2005A&A...435..773V,PhysRevA.61.052306}. These methods are applied to a large dataset of 50 light curves, minimizing human bias and maximizing the generalizability of our results.

This study also emphasizes the importance of considering different temporal scales when analyzing light curves, as these scales are crucial to understanding observed astrophysical phenomena \citep{2019MNRAS.482..743R,2007ApJ...662..900T,2020AdSpR..65..720K}.  We apply RQA in a unique manner to identify deterministic patterns at different temporal scales, enhancing our capacity to analyze the complex variability of Mrk~421. Unlike traditional approaches, which often focus on a single scale, we leverage the flexibility of RQA by varying the threshold input parameter, which was chosen to reflect a specific amount of recurrence points. These amounts are later averaged to represent behavior corresponding to recurrence levels of $5\%$ and $50\%$, thereby capturing dynamics across different temporal scales.  In addition, we adopt a novel approach by using both ARMA and ARIMA models together, whereas traditionally only one method is employed. This dual approach enables a more comprehensive understanding of the stochastic nature and memory effects across both short and long temporal scales, with ARIMA providing insights into short scales due to its  detrending process, while ARMA retains information from long-term  trends, allowing it to capture behavior associated with large scales. Furthermore, the correlations calculated between all the parameters\footnote{In this context, we use "parameters" broadly to refer to the different numerical outputs analyzed in this study, including model parameters from autoregressive time series modeling, RQA outcome measures, and statistical test statistics. Where appropriate, we also use the term "quantities" to collectively refer to these results when emphasizing their calculated nature rather than their modeling origin.}
, specifically RQA measures, autoregressive parameters, additional statistical tests, and the length and mean of specific observations, provide valuable insights into the complex interplay between deterministic and stochastic processes.

The structure of the paper is as follows: in Section~\ref{s2} we introduce the blazar Mrk~421 and provide an overview of its key properties. In Sec.~\ref{s3} we present details of X-ray observations and data processing. In Sec.~\ref{s4} we describe the deterministic and stochastic methods used for variability analysis, along with additional statistical tests. In Sec.\ref{s5} we present the results of applying these methods to the X-ray light curves of Mrk~421.  In Sec.~\ref{s6} we summarize the correlation analysis among the computed parameters and discuss our findings in the context of possible emission mechanisms and their physical implications.

\section{Blazar Mrk~421}  \label{s2}
The BL Lac object Mrk~421 has been intensively monitored in the last decades, using  multi wavelength observatories operating from both space and ground, due to its relevant variability from the radio to the TeV energy bands  \citep{2022ApJS..262....4N,2022MNRAS.513.1662M,2021A&A...647A..88A, 2018ApJ...854...66K, 2015A&A...576A.126A, 2002BASI...30..301J} and also to investigate the blazar neutrino connection \citep{2016APh....80..115P, 2011ApJ...736..131A, 1999APh....11...49M}, as highlighted in more recent epochs.

The electromagnetic (EM) emission mechanisms in Mrk~421 and other blazars involve the acceleration of high-energy particles within relativistic jets pointed directly to Earth, leading to variable emission across the EM spectrum. Relativistic electrons in these jets are energized through processes like shock acceleration, magnetic reconnection, and turbulence, emitting radiation via synchrotron emission and inverse Compton scattering \citep{Bhatta_2021, 2011ApJ...729..104K}.

In the domain of very high-energy $\gamma$-rays, Mrk~421 shows rapid flux changes on time scales shorter than an hour  \citep{2020ApJ...890...97A, 2021A&A...655A..89M}. A decade-long study of the source with Fermi/LAT $\gamma$-ray and ground-based optical observations reveals a lognormal flux distribution, long-term memory in the power spectral density, and year-long quasi-periodic oscillations \citep{Bhatta_2021, bhatta2020}.

Study of time resolved X-ray emission for different models can reveal possible correlations between various fitting parameters. \citet{2021MNRAS.508.5921H} have studied the time resolved X-ray emission for different models. Stochastic and time domain analysis of the blazar Mrk~421 has been studied in \citep{2020ApJ...897...25B, 2018ApJ...863..175G, 2014ApJ...786..143S}. Both deterministic and nondeterministic study using different timing analysis techniques discussed in \citet{2007PhDT........70E} are widely used to investigate periodicity and stationarity in blazars.

\begin{table*}
  \centering
  \small
  \begin{tabular}{|c|c|c|c|c|c|c|}
    \hline
    Obs ID & EPIC Instrument & Start time [MJD] & End time [MJD] & Mode & Exposure [Ks] & Expo. ID \\
    (1) & (2) & (3) & (4) & (5) & (6) & (7) \\
    \hline
    0099280101 & PN & 51689.1624 & 51689.4251 & T & 16.4 & S008 \\
    0099280101 & PN & 51689.4414 & 51689.8176 & I & 12.3 & S010 \\
    0099280201 & PN & 51850.0060 & 51850.4342 & I & 24.2 & S010 \\
    0099280301 & PN & 51861.9324 & 51862.4729 & I & 25.6 & S010 \\
    0136540101 & PN & 52037.3989 & 52037.8352 & I & 25.7 & S008 \\
    0136540301 & M1 & 52582.0308 & 52582.3028 & T & 22.8 & S003 \\
    0136540401 & M1 & 52582.3202 & 52582.5922 & T & 22.9 & S003 \\
    0136540801 & PN & 52592.8741 & 52592.9840 & I & 5.5 & S008 \\
    0136541001 & PN & 52609.9727 & 52610.7817 & T & 56.8 & S008 \\
    0136541101 & PN & 52610.8351 & 52610.9451 & I & 7.2 & S008 \\
    0136541201 & PN & 52611.0031 & 52611.1130 & I & 7.1 & S008 \\
    0150498701 & PN & 52957.6897 & 52958.2417 & T & 19.0 & S003 \\
    0153950601 & M1 & 52398.6795 & 52399.1297 & T & 38.4 & S003 \\
    0153950701 & PN & 52399.1911 & 52399.3890 & I & 15.9 & S005 \\
    0153951201 & PN & 53681.8447 & 53681.9465 & T & 3.8 & S005 \\
    0153951301 & PN & 53681.7058 & 53681.8041 & T & 8.3 & S005 \\
    0158970101 & M1 & 52791.5570 & 52792.0269 & T & 39.9 & U002 \\
    0158970201 & PN & 52792.0603 & 52792.2721 & I & 14.6 & S009 \\
    0158970701 & M1 & 52797.8970 & 52798.4607 & T & 48.1 & S010 \\
    0158971201 & PN & 53131.1251 & 53131.8762 & T & 12.8 & S003 \\
    0158971301 & PN & 53683.7759 & 53684.4553 & T & 30.8 & S003 \\
    0162960101 & PN & 52983.8975 & 52984.2459 & I & 13.4 & S007 \\
    0302180101 & M2 & 53854.8676 & 53855.3479 & T & 39.8 & S002 \\
    0411080301 & PN & 53883.0932 & 53883.8849 & I & 29.6 & S003 \\
    0411080701 & PN & 54074.5064 & 54074.7113 & T & 17.4 & S003 \\
    0411081301 & PN & 54230.1689 & 54230.3668 & I & 9.5 & S003 \\
    0411081401 & PN & 54230.4143 & 54230.4964 & I & 4.8 & S003 \\
    0411081501 & PN & 54230.5439 & 54230.6261 & I & 5.8 & S003 \\
    0411081601 & PN & 54230.6735 & 54230.7557 & I & 2.9 & S003 \\
    0411081901 & M1 & 54423.5489 & 54423.7653 & I & 18.3 & S001 \\
    0411082701 & PN & 54617.1091 & 54617.2098 & I & 6.3 & U002 \\
    0411083201 & PN & 55151.7552 & 55151.8501 & I & 6.5 & S600 \\
    0502030101 & PN & 54593.0812 & 54593.5673 & T & 27.7 & S003 \\
    0510610101 & PN & 54228.6283 & 54228.9061 & T & 11.0 & S003 \\
    0510610201 & PN & 54228.3529 & 54228.6017 & T & 16.7 & S003 \\
    0560980101 & PN & 54792.6095 & 54792.7160 & I & 8.5 & S600 \\
    0560983301 & PN & 54976.1719 & 54976.2783 & I & 8.5 & S600 \\
    0656380101 & PN & 55319.3278 & 55319.4112 & I & 6.4 & S600 \\
    0656380801 & PN & 55512.8889 & 55512.9850 & I & 7.6 & S600 \\
    0656381301 & PN & 55514.8844 & 55514.9805 & I & 7.6 & S600 \\
    0670920301 & PN & 56776.1859 & 56776.3363 & T & 8.6 & S003 \\
    0670920401 & PN & 56778.1597 & 56778.3310 & T & 13.5 & S003 \\
    0670920501 & PN & 56780.1518 & 56780.3231 & T & 11.3 & S003 \\
    0791780101 & PN & 57695.5677 & 57695.7529 & I & 11.2 & S001 \\
    0791780601 & PN & 57877.1860 & 57877.3134 & I & 7.7 & S001 \\
    \hline
  \end{tabular}
  \caption{The \textit{XMM-Newton} observational information for Mrk~421. (1) The observation ID, (2) the observation EPIC instrument: EPN (PN), EMOS1 (M1) and EMOS2 (M2), (3) the start time of the observation, (4) the stop time of the observation, (5) the observation mode: T (Timing) and I (Imaging), (6) the total exposure time, and (7) the exposure ID.}
  \label{tab:obs}
\end{table*}

\section{XMM Observations and Data Processing}\label{data}  \label{s3}

X-ray Multi-Mirror Mission ({\it XMM-Newton}) is a space telescope with three EPIC (European Photon Imaging Cameras) cameras along with a spectrometer named as Reflection Grating Spectrometer (RGS) and an Optical Monitor (OM) with a Ritchey-Chretien design which was launched by the European Space Agency. The EPIC cameras have an outstanding combination of energy range (0.1-12~keV) and effective area (1500~cm$^2$). 

Our analysis is based on a selection of 50 observations found in the HEASARC Data Archive \footnote{\url{http://nxsa.esac.esa.int/nxsa-web}, provided with EPIC exposures and Science files \footnote{\url{https://heasarc.gsfc.nasa.gov/db-perl/W3Browse/w3query.pl}};  basic information about these observations are given in Table~\ref{tab:obs}.} We reduced the raw data files using Science analysis system (SAS) 16.1.0 to get concatenated and calibrated event lists using the usual SAS procedures \footnote{\url{https://www.cosmos.esa.int/web/xmm-newton/sas-threads}}. {\it EPPROC} and {\it EMPROC} task from SAS are used to produce clean event files. Most of the OBSIDs are affected by pile-up, thus we used {\it epatplot} command  to minimize pile-up effects  by excising PSF core upto certain radii until pile-up effects become minor. We selected rectangular source regions for Timing mode and circular source region for Imaging mode using DS9 \citep{2000ascl.soft03002S}. The high background flaring is assessed by following standard procedures. By selecting background rate thresholds for EPIC-PN (RATE $\leq$ 0.4) and EPIC-MOS (RATE $\leq$ 0.35), the corresponding good time interval (GTI) files are created to obtain clean event files. Additionally, source light curves are extracted from pileup-corrected annular regions (rectangular for Timing mode), while background light curves are obtained from source-free regions (circular for Imaging mode and rectangular for Timing mode) in the same images. After getting the background-corrected light curves, the SAS task {\it epiclccorr} is utilized to remove known contaminant effects such as quantum efficiency, vignetting, and bad pixels, which can affect detection efficiency \citep[see also][]{2023ApJ...955..121D,2022MNRAS.510.5280M}.
    
    In our research, we processed a set of 50 generated light curves with the aim of pre-processing the data effectively. A key part of this pre-processing involved the removal of outliers, which were present in small quantities within the dataset. These outlying data points in the light curves were excluded using 5-$\sigma$ filtering for each individual OBSID to ensure the robustness of our analysis.

Additionally, since many analytical techniques require uniform data spacing for their applicability, we also used linear interpolation to fill in missing values \citep{2018FrP.....6...80F,bhattacharyya2020blazar, 2023ApJ...951..106B, 2023MNRAS.518.4372P, 2002A&A...382L...1P}. These interpolated values are visually distinguished by their representation in red color within the plots, see  Figure~\ref{LCS0} -- \ref{LCS5}.

    It is important to note that the proportion of interpolated points in our data remains relatively low, not exceeding $ 8.2\%$ of the dataset. This percentage is considered small and unlikely to significantly impact the collective results, particularly given that our primary focus is on patterns and correlations derived from the 50 observations as a whole. However, results of individual observations, particularly those with extreme or missing values, may be more influenced by the interpolated data points.

\begin{figure*}[h]  
		\centering 
		\includegraphics[width=1.\linewidth]{ 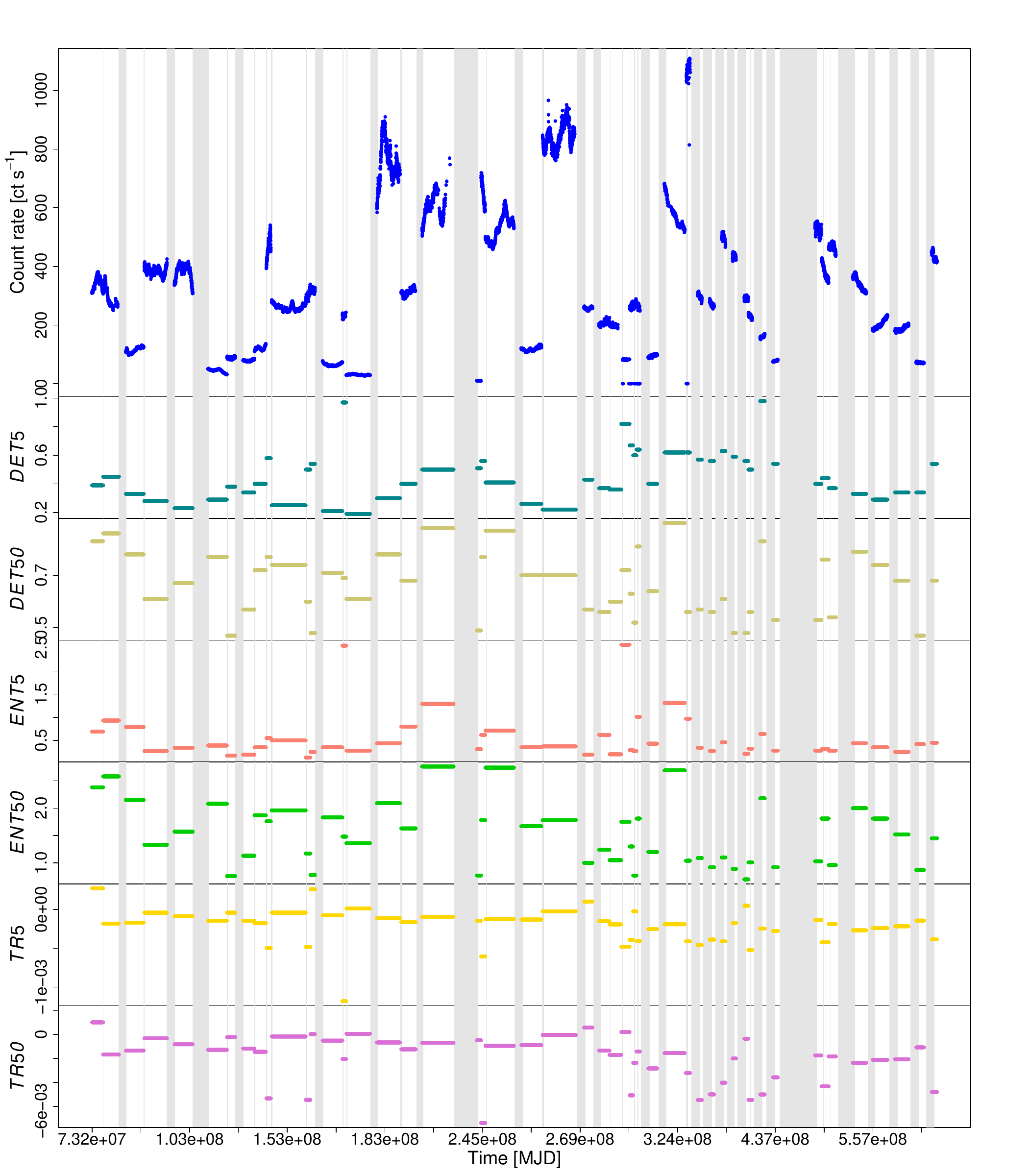} 
		\caption{Light curves of Mrk~421 (top), and calculated RQA measures of the light curves (gradually down). This figure shows all the analyzed light curves of Mrk~421 depicted in blue. Gray bands correspond to gaps due to missing data between observations. For visual clarity, these gaps have been proportionally shortened, resulting in an uneven time distribution. The calculated scalar quantities of deterministic nature from Table~\ref{tab:meas}, namely $DET5$, $DET50$, $ENT5$, $ENT50$, $TR5$, and $TR50$, are depicted below the light curves and correspond to the epochs of the corresponding observations.
  \label{BP1}}
\end{figure*}

\section{Methodology} \label{s4}

In this section, we introduce the deterministic RQA method alongside the stochastic ARMA and ARIMA models to investigate the multi-timescale variability of the TeV blazar Mrk~421. These methods are carefully chosen to allow distinguishing  the scales on which phenomena at play in the jets operate, providing a  detailed examination of the blazar's behavior. Complementary to these techniques, we employ the Augmented Dickey-Fuller and Tsay’s tests reinforcing the robustness of our analysis by assessing nonstationarity and nonlinearity in the data.

As a representative of stochastic modeling , ARMA is applied to analyze the variability as a whole, including the information from large scales, because it is not integrated as  ARIMA, which, on the other hand, focuses more on the variability which removes trends, and thus more on smaller scales \citep{2018FrP.....6...80F}. The deterministic approach is represented by RQA \citep{2007PhR...438..237M,1998PhLA..237..131Z}, which is associated with the investigation of phase space properties of the underlying physical system.

 The ARMA and ARIMA models are standard tools which require in input just the maximal values of the parameters considered for the model fitting. On the other hand, RQA requires a threshold parameter, the choice of which can noticeably affect the results. To address this issue, we implement RQA in a new way that eliminates the need to select a single fixed threshold. Instead of relying on a fixed threshold, we apply an averaging approach where calculations are performed over multiple threshold values. This method mitigates the sensitivity to a particular threshold selection, leading to a more robust and reliable estimation of the calculated measures \citep{2020ApJ...905..160B,2023EPJST.232...47P}. 
In this approach  the choice of the averaging parameter is related with the scales of potential physical processes, which in this study we distinguish between short and long time scales, based on the maximal threshold, when averaged.

\subsection{Recurrence quantification analysis \label{RQA}}

Since RQA is based on the principles of nonlinear  time series analysis (NLTSA), this section introduces its essential concepts, including embedding parameters, which are a fundamental aspect of this approach and can provide valuable insights into physical properties. NLTSA is introduced here  briefly, since it is less commonly used in data analysis compared to stochastic methods.

NLTSA offers a unique perspective on the deterministic aspects of physical systems, distinguishing itself from common stochastic approaches that deal with randomness and probabilistic behavior  \citep[see][]{2015Chaos..25i7610B,Iwanski1998RecurrencePO}. Its advantage lies in its direct link to physical properties. Even when examining a one-dimensional input that appears random, nonlinear methods can uncover patterns indicative of deterministic chaos – a concept in nonlinear dynamics where systems exhibit a behavior that appears random and unpredictable, but is actually deterministic, arising from the system's inherent nonlinear nature. This means that the system's future behavior is fully determined by its initial conditions, even though it appears random.  

An essential aspect of this approach is the topological relationship between the original and reconstructed phase spaces. By embedding one-dimensional time series into a higher-dimensional space, a technique known as delay-coordinate embedding, and applying advanced methods of NLTSA, deeper insights can be revealed, leading to a more accurate interpretation of the data. This embedding process requires setting up a time lag and determining an appropriate embedding dimension to effectively capture the system's dynamics.

Mutual information,  a measure of the information shared between two random variables, is a  common tool for the estimation of time lags in analyzed time series. A modification of mutual information, the so called Average Mutual
Information (AMI), is considered in this work along with L. Cao algorithm   
\citep[see][for calculation of embedding dimensions]{2020ApJ...905..160B}. L. Cao's practical method for determining the minimum embedding dimension of a scalar time series has been used in order to estimate the embedding parameter. The big advantage of this method is the requirement of just a single input parameter, the time delay parameter $\tau$ \citep{1997PhyD..110...43C}. Also, the implementation of the method requires a lower computational time, compared to other methods (e.g., invariant methods).

While embedding, which involves using a time lag and embedding dimension to map a one-dimensional time series into a higher-dimensional space, is commonly used in nonlinear time series analysis, it is not a required preprocessing step for RQA.
In this study, we chose not to embed the time series before calculating RQA because some light curves are too short for meaningful embedding, and applying it in such cases could reduce the robustness of the analysis rather than improve it.
However, the calculated time lag and embedding dimension still provide valuable insights into the system’s temporal dependencies and complexity, even without being explicitly used as a preprocessing step for RQA. Furthermore, it is worth noting that the definition of appropriate conditions for the application of time series embedding remains a subject of  debates within the scientific community \citep{ Iwanski1998RecurrencePO}. The selection of embedding parameters is often contingent upon the specific characteristics of the time series under investigation, and there is no universally accepted methodology for making these determinations.

The RQA is a widely-used tool of nonlinear analysis, introduced by \citet{1992PhLA..171..199Z} and later improved by \citet{2008EPJST.164....3M}, which
evaluates the nonlinear property of the Recurrence Plot (RP), a graphical tool introduced by \citet{1987EL......4..973E} for analysing the state space trajectories.

In RQA, the  foundation  for calculating RP is established by the binary matrix defined as follows:
\beq   
R_{i,j}= H(\epsilon - \| x_i - x_j\| ) \quad  i, j = 1, ...,N,   \label{Eq:1}
\eeq
where $H(·)$ is the Heaviside function, $N$ is the number of measured points $x_i$, $ \| \cdot \|$ is a norm and $\epsilon$ is a threshold distance which is a crucial value having a strong effect on the result.  

\begin{figure*}[h]
		\centering 
		\includegraphics[width=1.23\linewidth]{ 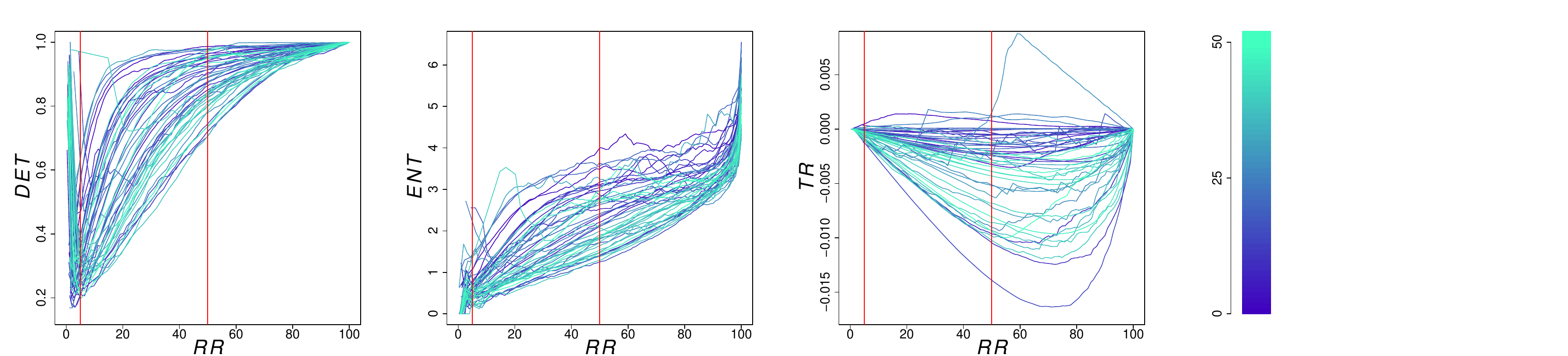}  
		\caption{This figure shows the evolution of the RQA measures, which were calculated for all the  $RR \in [1-99]\%$ for all the 50 light curves. The color bar reflects the chronology of the observations, with lighter blue indicating more recen  observations and darker blue indicating older ones (the oldest is denoted by 0 on the bar). The red vertical lines denote the $5\%$ and the $50\%$ value of $RR$, which were considered the highest limits in the averaging process for RQA measures.\label{RQAplot}}
\end{figure*}

The RP is generated as a graphical representation of this square matrix. The RQA measures we use in this work are:
\begin{enumerate}
 \item Recurrence rate ($RR$), which   characterizes the amount of recurrent points in the phase space trajectory of a system. Mathematically, $RR$ is defined as the percentage of recurrence points in the recurrence plot: \footnote{ In both the text and  Figure~\ref{RQAplot}, we refer to $RR$ in percentages rather than decimals as for other measures. }
 \beq
RR = \frac{1}{N^2} \sum_{i,j=1}^N R_{i,j} =  \frac{1}{\frac{N(N-1)}{2}} \sum_{i=1}^{N} \sum_{j=i+1}^{N} R_{ij} =  \frac{1}{\frac{N(N-1)}{2}} \sum_{i=2}^{N} \sum_{j=1}^{i-1} R_{ij} \label{RR},
 \eeq
 which provides a measure for the density of the recurrence points in the RP. The $RR$ can be expressed using only lower or upper diagonal points.

 \item Determinism ($DET$), which tells how deterministic or well behaved a system is,    is derived from the analysis of diagonal lines in a recurrence plot and reflects the degree to which the dynamics of a system is predictable and repeatable over time. It is defined as:
\beq
DET  = \frac{\sum_{l=l_{min}}^N l P(l)}{ \sum_{i,j=1}^N R_{i,j}}, \label{DET}
\eeq
where $P(l)$ denotes the frequency distribution of the lengths $l$ of the diagonal lines, and represents a weighted sum of these lengths in the recurrence plot. This sum is computed by multiplying each diagonal line length by its frequency of occurrence and then summing these products over all possible line lengths. The minimal line length considered as a line   is usually set up to $l_{min} = 2$, as it often allows for the highest differentiation of determinism values between different dynamic states of the system \citep[see, e.g.,][]{Babaei2014Selection}. The resulting value is then normalized by the total number of recurrent points in the recurrence matrix.
A high $DET$ value indicates strong determinism, suggesting that the system exhibits consistent and predictable patterns in its dynamics.
More specifically, a high $DET$ value would indicate that specific  patterns recur predictably over time, such as daily or seasonal variations.

\item  Entropy ($ENT$), that is the diagonal line length distribution in the recurrence plot  normalized by the total number of lines and then used to estimate the Shannon entropy, which measures the level of uncertainty or randomness within the distribution. High values of RQA entropy indicate more complex and less predictable recurrence structures. RQA entropy is one of several measures in RQA that are used to analyze the properties of a time series. It provides insight into the complexity  of the recurrence structure, making it a valuable tool for understanding the underlying dynamics of the time series. It is defined as:

\beq
   ENT =  - \sum_{l=l_{min}}^N  p(l) \ln p(l),
\eeq
where $p(l)$ is the probability that a diagonal line in the RP is exactly of the length $l$ and it can be estimated from the frequency distribution $P(l)$ with:
\beq
 p(l) = \frac{P(l)}{\sum_{l=l_{min}}^N  P(l)}.
\eeq
\item  Trend Measure ($TR$), which (in RQA) evaluates the decreasing density in the recurrence plot offering insights into the system's trends and stationarity. It is the regression coefficient of a linear relationship between the density of recurrence points in a line parallel to the Line of Interest (LOI) and its distance to the LOI. The recurrence rate in a diagonal line parallel to LOI of distance $k$ (diagonal-wise recurrence rate or $\tau$-recurrence rate) is calculated and the trend is defined as the sum of the deviations of the diagonal-wise recurrence rates from their mean, divided by the sum of the squared deviations of the distance values from their mean: 
\beq
TR = \frac{\sum_{i=1}^{\tilde{A}} (i - {N}/2 )( RR_i - \langle RR_i\rangle )}{\sum_{i=1}^{\tilde{A}} (i - {N}/2 )^2},
\eeq
where  both $\langle .\rangle$ and ${\tilde {A}}$  are lower than $N( )$; \cite[see][for details, including the definition of further RQA measures]{2007PhR...438..237M, webber2014recurrence}. High $TR$ values typically indicate nonstationary systems.
\end{enumerate}

In this work, the implementation of RQA method follows the same approach as in \citet{2020ApJ...905..160B,2023EPJST.232...47P}, where the RQA measures are averaged over a range of thresholds defined as:
 \beq RQA(  \epsilon ) = RQA(   RR( \epsilon, lmin )  ).
  \eeq

The thresholds for RQA are calculated for a wide range of $RR \in [1-99]\%$, as shown in Figure~\ref{RQAplot}. However, for further analysis and discussion, the calculated RQA measures are averaged at two specific $RR$ values:  $RR = 5\%$  (denoted as $DET5$, $ENT5$, $TR5$) and   $RR = 50\%$  (denoted as $DET50$, $ENT50$, $TR50$). 

In our analysis, small scales are represented by $RR$ values up to 5\%, while large scales are represented by $RR$ values up to 50\%. By selecting $RR$ values of 5\% and 50\% to average the RQA measures, we aimed to explore to what extent different dynamic processes rely on different time scales. As shown in Figure~\ref{RQAplot}, RQA parameters, and particularly $DET$,  show different behaviors for  $RR$ higher than $ 5\% $, while the $50\%$ threshold was adopted since, as shown by \citet{2023EPJST.232...47P}, deterministic patterns are well recognized when averaged close to this level.

RQA, like many other methods, produces the best results under certain conditions such as sufficient data length, low noise levels, and uniform sampling. Short data series may not capture the system's full dynamics, potentially leading to less accurate interpretations.
Nonstationarity also influences RQA measures, particularly the behavior of time series on large scales, and is likely to have a more pronounced impact on RQA measures with high threshold values. It generally impacts the RP, causing progressively faded regions at the corners, which in turn affects the RQA. Added dynamics through interpolation may also affect the results. In astronomical time series, missing data points are often present due to observational gaps. In this study, we applied interpolation to address such gaps and enable consistent analysis of the light curves, following similar practices used in other works \citep[see, e.g.,][]{2023MNRAS.518.4372P}.
Despite these limitations, we focus on obtaining relevant information by analyzing many light curves, assuming that consistent results across multiple datasets will enhance the overall relevance and reliability of our findings. Additionally, our averaging procedure helps in mitigating the impact of both intrinsic and observational noise, further supporting the robustness of our results \citep{2023EPJST.232...47P, 2020ApJ...905..160B}.

\begin{table*}
 \centering
 \begin{flushleft}

\hspace*{-5cm}\scalebox{0.90}{%

\begin{tabular}{|c|c|c|c|c|c|c|c|c|c|c|c|c|c|c|c|}
\hline
 Obs ID &  length & mean [$\mathrm{ct\,s^{-1}}$] & $p$ & $q$ & $p_i$ & $q_i$ & $DET5$ & $ENT5$ & $TR5$ & $DET50$ & $ENT50$ & $TR50$ & ADF & Tsay's \\ 
 (1)& (2)& (3)& (4)& (5)& (6)& (7)& (8)& (9)& (10)& (11)& (12)& (13)& (14)& (15)\\
  \hline
 0099280101 & 326 & 301.28 & 6 & 4 & 3 & 4 & 0.45 & 0.93 & -1.80e-04 & 0.86 & 2.58 & -1.67e-03 & -1.5678 & 0.5647 \\ 
  0099280101 & 228 & 343.97 & 7 & 5 & 3 & 5 & 0.39 & 0.69 & 2.70e-04 & 0.83 & 2.38 & 1.02e-03 & -0.6489 & 1.9099 \\ 
  0099280201 & 370 & 115.03 & 8 & 9 & 4 & 7 & 0.33 & 0.79 & -1.70e-04 & 0.78 & 2.15 & -1.35e-03 & -2.0508 & 0.6302 \\ 
   0099280301 & 467 & 383.56 & 4 & 4 & 0 & 1 & 0.28 & 0.27 & -4.00e-05 & 0.61 & 1.33 & -3.10e-04 & -0.9764 & 0.8483 \\ 
   0136540101 & 377 & 383.71 & 6 & 6 & 3 & 8 & 0.23 & 0.34 & -9.00e-05 & 0.67 & 1.57 & -8.20e-04 & 0.5194 & 0.7967 \\ 
   0136540301 & 235 & 78.87 & 3 & 2 & 3 & 3 & 0.34 & 0.19 & -1.40e-04 & 0.57 & 1.13 & -1.17e-03 & -1.0904 & 1.5936 \\ 
  0136540401 & 235 & 120.34 & 8 & 2 & 4 & 3 & 0.40 & 0.35 & -1.80e-04 & 0.72 & 1.87 & -1.44e-03 & -0.3438 & 0.9746 \\ 
   0136540801 & 95 & 469.76 & 4 & 1 & 0 & 3 & 0.58 & 0.55 & -5.00e-04 & 0.77 & 1.76 & -5.33e-03 & 0.0973 & 0.6163 \\ 
   0136541001 & 699 & 260.41 & 4 & 4 & 3 & 4 & 0.25 & 0.50 & -4.00e-05 & 0.74 & 1.96 & -1.70e-04 & -2.0409 & 1.1024 \\ 
   0136541101 & 95 & 304.01 & 1 & 1 & 0 & 2 & 0.50 & 0.13 & -4.80e-04 & 0.60 & 1.17 & -5.46e-03 & -1.9291 & 1.1162 \\ 
 0136541201 & 95 & 317.41 & 3 & 2 & 3 & 2 & 0.54 & 0.25 & 2.60e-04 & 0.48 & 0.78 & 3.00e-05 & -2.8876 & 0.1035 \\ 
   0150498701 & 477 & 751.42 & 9 & 3 & 4 & 4 & 0.30 & 0.44 & -1.10e-04 & 0.78 & 2.09 & -6.60e-04 & -2.7246 & 0.4650 \\ 
   0153950601 & 389 & 44.27 & 10 & 8 & 4 & 3 & 0.29 & 0.39 & -1.40e-04 & 0.77 & 2.08 & -1.28e-03 & 0.0118 & 0.2531 \\ 
   0153950701 & 171 & 88.23 & 9 & 2 & 2 & 3 & 0.38 & 0.17 & -4.00e-05 & 0.47 & 0.76 & -2.20e-04 & -3.7611 & 0.5808 \\ 
   0153951201 & 88 & 650.45 & 1 & 10 & 2 & 6 & 0.56 & 0.62 & -6.10e-04 & 0.77 & 1.78 & -7.39e-03 & -2.3948 & 0.0968 \\ 
  0153951301 & 85 & 10.04 & 0 & 1 & 0 & 1 & 0.51 & 0.31 & -1.50e-04 & 0.49 & 0.77 & -4.80e-04 & -3.1683 & 0.4017 \\ 
   0158970101 & 406 & 64.71 & 6 & 4 & 0 & 2 & 0.21 & 0.35 & -7.00e-05 & 0.71 & 1.83 & -5.10e-04 & -2.4260 & 0.5513 \\ 
   0158970201 & 77 & 231.96 & 3 & 9 & 3 & 10 & 0.97 & 2.55 & -1.18e-03 & 0.69 & 1.48 & -2.04e-03 & -2.2955 & 3.5105 \\ 
   0158970701 & 487 & 30.46 & 8 & 5 & 4 & 3 & 0.19 & 0.28 & 1.00e-05 & 0.61 & 1.36 & 6.00e-05 & -2.4773 & 1.0083 \\ 
   0158971201 & 649 & 593.47 & 10 & 8 & 3 & 4 & 0.50 & 1.29 & -9.00e-05 & 0.88 & 2.76 & -6.90e-04 & -1.0593 & 1.6712 \\ 
   0158971301 & 587 & 528.88 & 6 & 9 & 3 & 4 & 0.41 & 0.71 & -1.20e-04 & 0.87 & 2.74 & -9.50e-04 & -0.7655 & 0.9520 \\ 
   0162960101 & 301 & 311.45 & 3 & 3 & 3 & 3 & 0.40 & 0.80 & -1.60e-04 & 0.68 & 1.63 & -1.24e-03 & -2.8524 & 0.7484 \\ 
   0302180101 & 415 & 117.16 & 7 & 6 & 1 & 3 & 0.26 & 0.35 & -1.30e-04 & 0.70 & 1.67 & -8.80e-04 & -2.0643 & 2.1759 \\ 
  0411080301 & 684 & 841.76 & 10 & 14 & 7 & 13 & 0.22 & 0.37 & -2.00e-05 & 0.70 & 1.78 & -4.00e-05 & -1.5307 & 1.2072 \\ 
  0411080701 & 177 & 257.02 & 4 & 4 & 3 & 4 & 0.43 & 0.19 & 1.00e-04 & 0.57 & 1.00 & 5.80e-04 & -1.8252 & 1.3794 \\ 
  0411081301 & 171 & 82.20 & 3 & 10 & 2 & 4 & 0.82 & 2.57 & -4.80e-04 & 0.72 & 1.75 & 2.10e-04 & -1.9867 & 5.3225 \\ 
   0411081401 & 71 & 264.82 & 1 & 1 & 0 & 1 & 0.67 & 0.29 & -3.90e-04 & 0.63 & 1.30 & -5.08e-03 & -3.2250 & 3.4777 \\ 
 0411081501 & 71 & 270.22 & 1 & 0 & 0 & 1 & 0.60 & 0.27 & -2.00e-05 & 0.52 & 0.77 & -2.36e-03 & -4.8399 & 0.1487 \\ 
   0411081601 & 71 & 237.74 & 2 & 2 & 0 & 1 & 0.64 & 1.01 & -4.10e-04 & 0.81 & 1.81 & -1.42e-03 & -2.5633 & 2.2713 \\ 
  0411081901 & 187 & 94.25 & 2 & 3 & 2 & 2 & 0.40 & 0.43 & -2.50e-04 & 0.64 & 1.20 & -2.83e-03 & -4.4446 & 0.7132 \\ 
   0411082701 & 87 & 1074.39 & 1 & 1 & 1 & 2 & 0.62 & 0.97 & -4.10e-04 & 0.56 & 1.04 & -3.21e-03 & -2.8446 & 1.3046 \\ 
   0411083201 & 82 & 497.55 & 1 & 1 & 0 & 1 & 0.63 & 0.46 & -4.10e-04 & 0.61 & 1.10 & -4.03e-03 & -1.9607 & 1.3283 \\ 
   0502030101 & 420 & 573.52 & 9 & 8 & 3 & 7 & 0.62 & 1.31 & -1.90e-04 & 0.90 & 2.69 & -1.55e-03 & -1.9970 & 0.6224 \\ 
   0510610101 & 240 & 197.31 & 9 & 12 & 9 & 13 & 0.36 & 0.20 & -1.90e-04 & 0.60 & 1.05 & -1.70e-03 & -3.3997 & 0.3828 \\ 
  0510610201 & 215 & 206.74 & 3 & 1 & 2 & 1 & 0.37 & 0.62 & -1.50e-04 & 0.56 & 1.24 & -1.35e-03 & -4.5271 & 1.2928 \\ 
   0560980101 & 92 & 297.49 & 1 & 1 & 0 & 2 & 0.57 & 0.34 & -4.60e-04 & 0.57 & 1.09 & -5.47e-03 & -3.5062 & 1.3718 \\ 
   0560983301 & 92 & 270.23 & 1 & 1 & 1 & 2 & 0.56 & 0.27 & -3.90e-04 & 0.56 & 0.92 & -5.00e-03 & -3.2432 & 1.2060 \\ 
   0656380101 & 72 & 433.45 & 0 & 0 & 4 & 3 & 0.59 & 0.00 & -1.80e-04 & 0.48 & 0.89 & -2.00e-03 & -3.7637 & 0.0000 \\ 
   0656380801 & 83 & 291.18 & 5 & 3 & 4 & 3 & 0.56 & 0.21 & 5.00e-05 & 0.48 & 0.70 & -3.50e-04 & -2.4891 & 1.0849 \\ 
   0656381301 & 83 & 230.22 & 1 & 1 & 0 & 3 & 0.50 & 0.32 & -5.20e-04 & 0.56 & 1.01 & -5.46e-03 & -3.6116 & 2.6803 \\ 
   0658800101 & 97 & 160.94 & 3 & 14 & 2 & 15 & 0.98 & 0.64 & -2.50e-04 & 0.83 & 2.18 & -5.01e-03 & -2.3942 & 1.3346 \\ 
  0658800801 & 99 & 77.72 & 1 & 1 & 0 & 1 & 0.54 & 0.28 & -2.80e-04 & 0.53 & 0.92 & -3.58e-03 & -3.9518 & 0.7683 \\ 
   0658801301 & 275 & 340.84 & 8 & 4 & 1 & 6 & 0.33 & 0.44 & -2.70e-04 & 0.79 & 2.00 & -2.36e-03 & -2.3661 & 0.4578 \\ 
   0658801801 & 303 & 202.67 & 10 & 10 & 4 & 4 & 0.29 & 0.35 & -2.40e-04 & 0.74 & 1.81 & -2.12e-03 & -1.8536 & 0.4889 \\ 
   0658802301 & 279 & 187.09 & 5 & 10 & 3 & 10 & 0.34 & 0.25 & -2.20e-04 & 0.68 & 1.52 & -2.06e-03 & -2.8612 & 1.0038 \\ 
   0670920301 & 130 & 524.56 & 1 & 1 & 0 & 1 & 0.40 & 0.28 & -1.30e-04 & 0.53 & 1.03 & -1.74e-03 & -2.5696 & 2.6344 \\ 
   0670920401 & 148 & 382.04 & 5 & 6 & 1 & 2 & 0.44 & 0.31 & -4.20e-04 & 0.76 & 1.81 & -4.33e-03 & -1.6750 & 1.4097 \\ 
   0670920501 & 148 & 466.39 & 1 & 1 & 0 & 1 & 0.37 & 0.28 & -1.90e-04 & 0.54 & 0.96 & -1.83e-03 & -1.7001 & 1.0409 \\ 
  0791780101 & 160 & 71.63 & 1 & 2 & 0 & 2 & 0.34 & 0.42 & -1.40e-04 & 0.47 & 0.87 & -1.07e-03 & -5.2862 & 1.1046 \\ 
   0791780601 & 110 & 431.95 & 4 & 2 & 1 & 0 & 0.54 & 0.45 & -3.90e-04 & 0.68 & 1.45 & -4.81e-03 & -2.1653 & 1.1069 \\ 
   \hline
\end{tabular}
}

\caption{
Calculated quantities of the $\textit{XMM-Newton}$ observations of Mrk~421 from 2002-2019. This table shows the (1) Observation ID, (2) length - number of data points in the interpolated light curve, (3) mean count rate, (4-5) ARMA $p$ and $q$ measures, (6-7) ARIMA $p_i$ and $q_i$ measures, (8-10) RQA measures $DET$, $ENT$, $TR$ averaged to $5\%$ of $RR$, (11-13) RQA measures $DET$, $ENT$, $TR$  averaged to $50\%$ of $RR$, (14) ADF test statistic value, (15) Tsay’s test statistic value. \label{tab:meas}}
\end{flushleft}
\end{table*}

\subsection{ARMA and ARIMA}  \label{ARIMA}

ARMA-related models became popular in the 70's with the publication of  \citet{1976tsaf.conf.....B}. The main idea is that the current value of the series, $x_t$, can be explained as a function of number $p$ its past values, $x_{t-1}, x_{t-2}, \dots , x_{t-p}$, what describes the auto-regressive (AR) part. The moving average (MA) process coefficients quantify the dependence of current values on recent past random shocks to the system $\epsilon_{t-1}, \epsilon_{t-2}, \dots , \epsilon_{t-q}$, where $\epsilon_t$ is the error term for the $t$-th time point.

\begin{enumerate}

\item  Auto-regressive process of the order  $p$, AR(p), is described as:
\beq
y_t = a_0 + \sum_{j = 1}^p  a_j y_{t-j}  + \epsilon_t.
\eeq

\item Moving average process of the order $q$, MA($q$), is described as:
\beq
y_t = \sum_{j = 0}^q  \beta_j \epsilon_{t-j}.
\eeq

\item Auto-regressive moving average process of the orders $p$ and $q$, ARMA($p,q$), is described as:
\beq
y_t = a_0 + \sum_{j = 1}^p  a_j y_{t-j} + \sum_{j = 0}^q \beta_j \epsilon_{t-j}.
\eeq

\end{enumerate}

In time series modeling with autoregressive fitting, achieving a good fit requires ensuring stationarity, characterized by a constant mean and variance. ARIMA models are a type of  autoregressive models, where "I" stands for integrated \footnote{In the Figures~\ref{BP1}, \ref{cor}  and the Table~\ref{tab:meas}, the $p$ and $q$ quantities are related to ARMA, while the quantities with lower $i$ index, $p_i$ and $q_i$ are related to ARIMA.}, meaning that the series has been differenced a number of times (of order "$n$") to make it stationary. The ARIMA($p,n,q$) model is estimated by fitting an ARMA($p,q$) model after differencing the time series $n$-times.

The differencing is used to make a time series stationary, leading to a better fit. Let $x_i$ represent the time series, and $y_i$  its first difference, defined as:
\beq
y_i = \nabla x_i = x_i - x_{i-1} = (1 - L)x_i,
\eeq
where L is the lag operator $L^{n}x_{t}=x_{t-n}$ for which $n$ denotes the order of the lag.

This method of analysis is often used in time series analysis, where it also serves for making forecasts, as well as across various fields of science, such as in the study of blazars, as seen in \citet{2020ApJ...897...25B}. The stochastic element in astronomical data should also be noted \citep{2005A&A...435..773V}. 
In our analysis, the ARMA and ARIMA model parameters were calculated using the R Stats Package. We conducted a detailed parameter optimization process for both ARMA and ARIMA models. For the ARIMA model, we systematically varied $p$, $q$, and the differencing order $n$ (with $n = 0$ for ARMA) from 0 to 16 in a three-loop cycle (none of these parameter values exceeded 15).  The objective of this iterative search was to identify the combination of $p$, $q,$ and $n$ that minimizes the Akaike information criterion ($AIC$) which is broadly recognized as a standard metric for model comparison and is defined in the literature \citep{1974ITAC...19..716A, 10.5555/1088844}, providing a quantitative basis to determine the most appropriate model configuration.  It is given as:
 \beq
AIC = \log  \widehat{\sigma}^2 _k + \frac{m + 2k}{m},
\eeq
 where 
  \beq
   \widehat{\sigma}^2 _k  = \frac{SSE_k}{m}
 \eeq
 and $SSE_k$ denotes the residual sum of squares under the model with $k$ regression coefficients and the sample size of length $m$.
$AIC$ is based on the maximum-likelihood estimate of $k$, which serves as an estimator of prediction error and evaluates the quality of statistical models for a given dataset. $AIC$ estimates the relative amount of information lost by a given model -- the less information a model loses, the higher the quality of that model.

Nonstationarity refers to a characteristic where the statistical properties of the process generating the time series change over time.  It can manifest in various ways, such as the presence of a trend in the data, changes in the mean or variance, or alterations in the data's overall structure.

When comparing the results of ARMA  and ARIMA models fitted to the same data, any differences in the calculated quantities are likely attributable to the nonstationarity of the data. For instance, if the data exhibit nonstationarity due to the presence of a trend, the ARIMA model, which involves differencing to remove the trend, is likely to produce more accurate forecasts than the ARMA model, which does not incorporate differencing. However, it is important to note that the nonstationarity in blazar light curves may not have a clearly defined origin, whether deterministic or stochastic.

In the context of analyzing irregularly spaced data, there is also the option to use Continuous-time Autoregressive Moving Average (CARMA) models  instead of the traditional ARMA and ARIMA models. CARMA models, which are based on stochastic differential equations, represent a more sophisticated method for handling such data \citep{2018FrP.....6...80F}, where light curve variability features and power spectral density are examined in great detail. However, the implementation and detailed discussion of CARMA models are beyond the scope of our intentions in this article. Nevertheless, using both ARMA and ARIMA models provides insights into the underlying scales within the data, which are discussed in later parts of this article.

\subsection{Augmented Dickey–Fuller test} \label{ADF}
The Augmented Dickey-Fuller (ADF) test  checks for the presence of a unit root in a time series sample \citep{Dickey1979DistributionOT}. 
A unit root indicates that the time series follows a random walk, meaning its statistical properties, such as mean and variance, can change over time, leading to nonstationarity. 
The null hypothesis asserts that a unit root is present, while the alternative hypothesis contends that the series is stationary. The ADF test statistic is a negative number: the lower it is, the stronger the evidence against the null hypothesis, reinforcing the rejection of the unit root hypothesis at a certain confidence level.

The ADF test is a very popular test for stationarity in time series data and it is a popular choice for both researchers and practitioners. It is relatively easy to implement and interpret, and it is quite powerful against a variety of nonstationarities. However, in the context of the ADF test, it is important to consider the length of the time series. Short time series are more likely to appear stationary, as there might not be sufficient duration for nonstationary characteristics, such as long-term trends or cyclical patterns, to develop fully. Consequently, short series may not reveal underlying trends or structural breaks due to their limited span. This implies that when a time series is observed over a short period, the ADF test may have limited ability to detect long-term nonstationarity, potentially affecting the reliability of the test results.

\begin{figure*}[h]   
		\centering 
		\includegraphics[width=1.\linewidth]{ 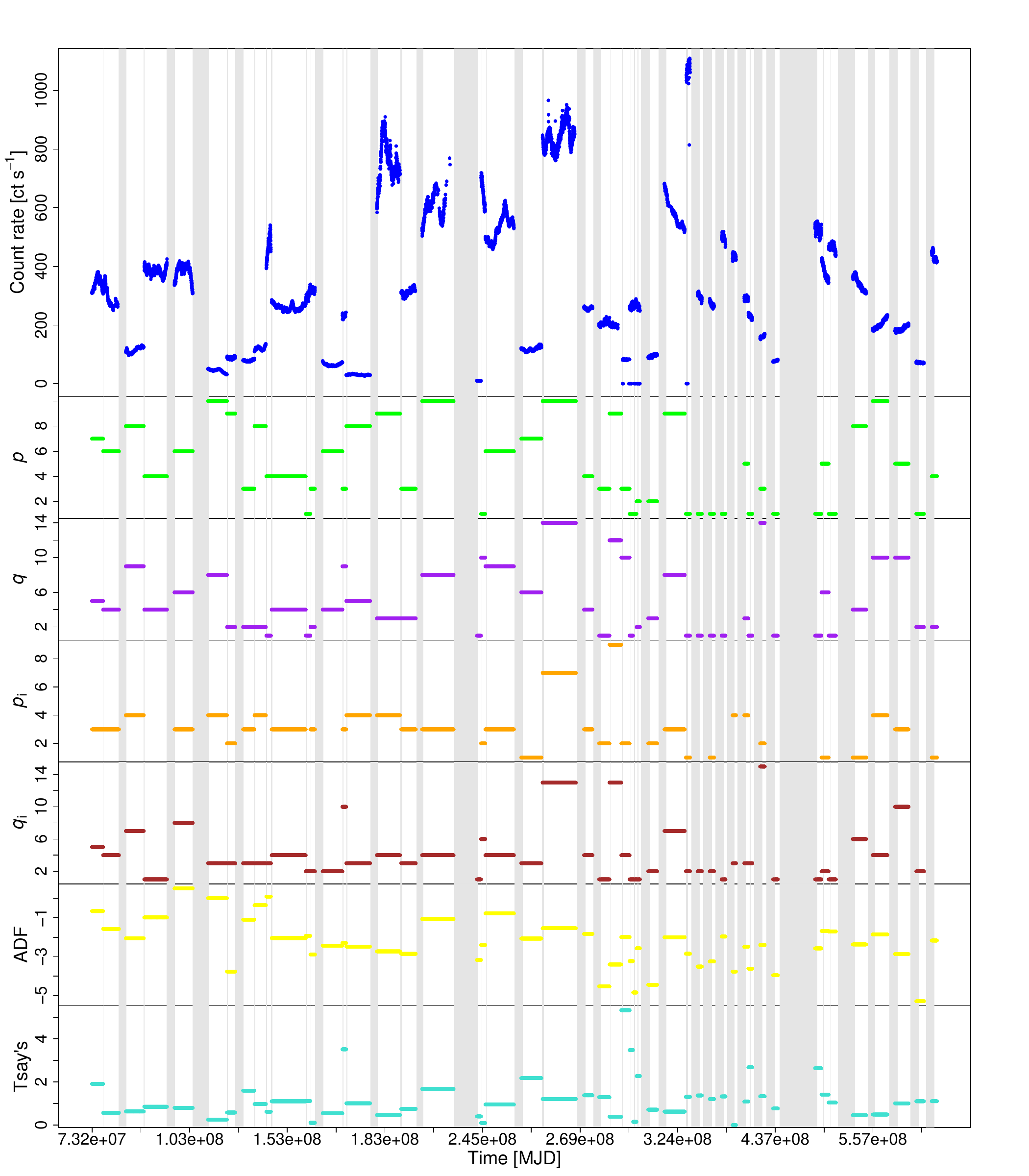} 
\caption{This is the continuation of Figure~\ref{BP1}. Below, the calculated scalar quantities from Table~\ref{tab:meas}, namely $p$, $q$, $p_i$, $q_i$, ADF test statistic and Tsay’s test statistic, are shown}.
      \label{BP2}
\end{figure*}

\subsection{Tsay’s nonlinearity test} \label{Tsay}

The Tsay’s test is a statistical method used to detect nonlinearity in time series data. It evaluates the null hypothesis that a linear model (e.g., ARMA) adequately describes the time series. If the resulting F-statistic exceeds the critical threshold (typically at a 5\% significance level), the null hypothesis is rejected, indicating that the time series exhibits nonlinear behavior \citep{10.1093/biomet/73.2.461}.  The test is nonparametric, meaning it does not assume that the time series follows any specific distribution. This allows for greater flexibility in application, as it can be applied to a wide range of time series without requiring a particular underlying distribution.

However, the Tsay’s test may have limitations in detecting certain types of nonlinearity, such as threshold effects or component interactions, and its power depends on the characteristics of the tested series \citep{Psaradakis2002Power}. For instance, short time series may not provide sufficient information to detect nonlinear structure, while missing values and linear interpolation (used to achieve regular sampling) can introduce artificial dynamics that may influence the test statistic. In our analysis, we examine 50 light curves of varying lengths, some of which are quite short (fewer than 100 points) and include interpolated segments. These characteristics may influence the resulting statistics, and thus the test results should be interpreted with caution. The Tsay’s test statistic follows an F-distribution, and its critical value depends on the chosen significance level and degrees of freedom. If the statistic does not exceed the threshold, there is insufficient evidence to reject the null hypothesis, and a linear model is considered adequate.

The Tsay’s test is implemented in various statistical software packages, including R, where it can be accessed through functions designed for time series analysis. These implementations facilitate the application of the Tsay’s test in practical settings, allowing researchers and analysts to rigorously test for nonlinearity in time series datasets. The test is available through the R package \texttt{nonlinearTseries} \citep{garcia2nonlineartseries,10.1093/biomet/72.1.39}.

\section{Results}   \label{s5}

The complete list of results obtained in this study is presented in  Table~\ref{tab:meas}.  To better visualize the calculated quantities and facilitate comparison, Figures~\ref{BP1} and \ref{BP2} have been created, showing all the analyzed light curves along with their respective calculated measures. 
The 50 observations in this study cover a period of approximately 17~years, but they are unevenly distributed including some missing values in observations. 

Figure~\ref{BP1} and \ref{BP2} share the same quantity (time) in the X-axis, and this is not evenly spaced due to these gaps, which are condensed to fit within the graphics, with the shrink applied only to time intervals with no data using a consistent scale factor.  The deterministic RQA measures of $DET$, $ENT$, and $TR$ in both small and large scale versions are displayed in Figure~\ref{BP1}, while the stochastic parameters $p, q$, and ARIMA $p_i, q_i$ along with ADF and Tsay’s test statistics can be found in Figure~\ref{BP2}. The light curves shown in Figure~\ref{BP1} and \ref{BP2} illustrate the overall variability behavior. Their mean count rate and the length of the time series (i.e., the number of data points after interpolation) are considered quantitatively in the correlation analysis. As discussed later, the length of the time series plays a significant role in shaping some of the extracted measures, whereas the mean count rate shows only weak correlations and does not substantially affect the outcomes. The gaps between the observations are more frequent in the latter half of the time period, and the highest levels of count rate are concentrated around the midpoint.

\begin{figure*}[h]
		\centering 
		\includegraphics[width=1.\linewidth]{ 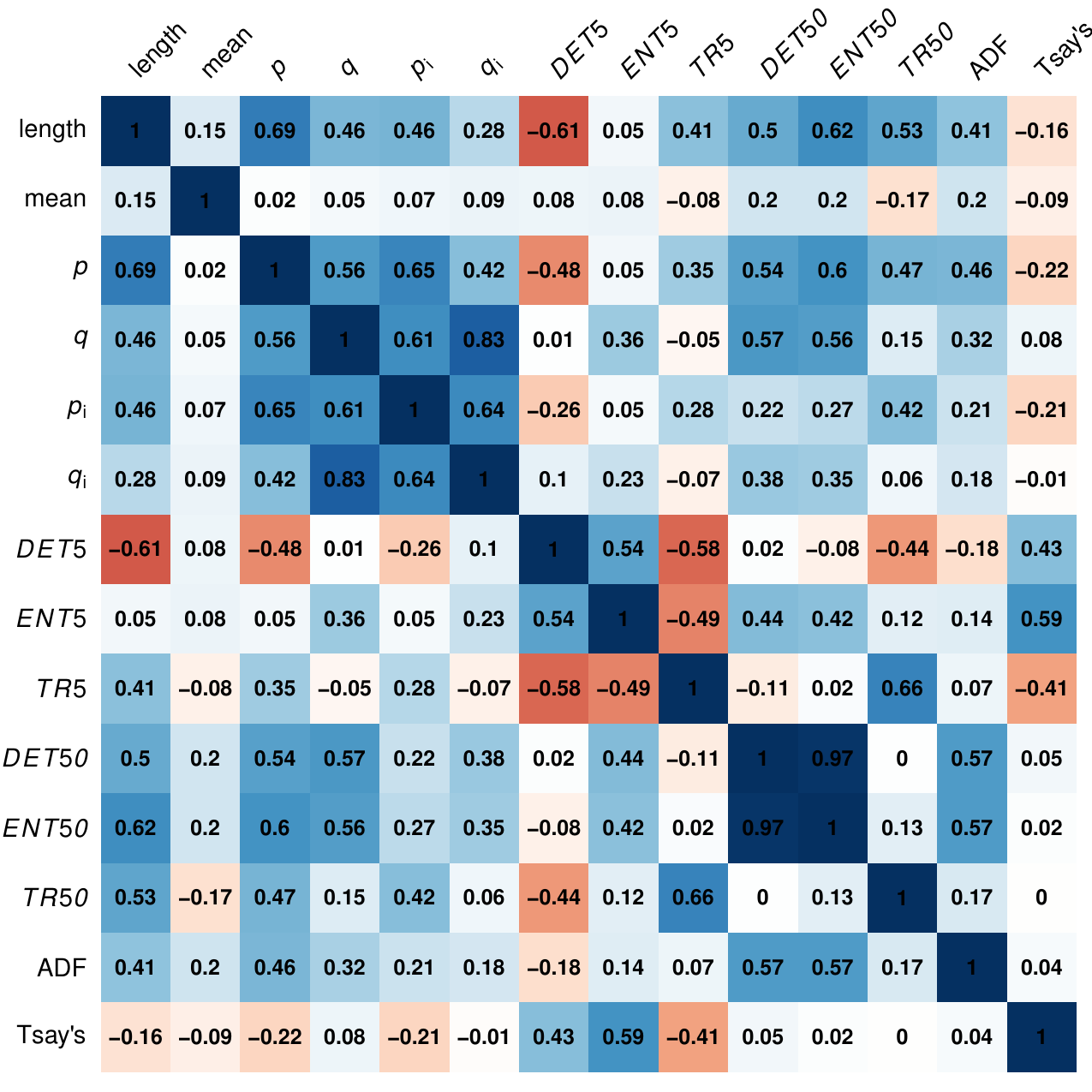}
		\caption{This figure illustrates the correlations between the computed measures from Table\,\ref{tab:meas} for all 50 analyzed light curves of Mrk\,421, where shades of red signify negative correlations and shades of blue indicate positive ones. The intensity of the color corresponds to the correlation strength.\label{cor}}
\end{figure*}

The following subsections present, method by method, the results of the applied analysis, focusing on numerical outcomes such as parameter distributions, averages, variances, and ranges, and highlighting significant relationships from a technical perspective. A correlation analysis among all calculated quantities, as depicted in Figure~\ref{cor}, is carried out as a central component of our study, allowing us to extract more meaningful insights from the data. A correlation matrix is utilized as an efficient method for evaluating and discussing the results \citep{2019arXiv190208704H, 2020A&A...642A.156V, 2023arXiv230713063M}. Its purpose is to provide a clear and concise summary of the relationships between the extracted quantities across the large dataset. To maintain clarity and avoid redundancy, we restrict the correlation discussion in this Section to relationships within each method. Broader correlations involving comparisons across calculated parameters from different methods, including RQA, ARMA, ARIMA, ADF, and Tsay’s test results, are discussed separately in Section~\ref{s6}, where a deeper interpretation of the variability processes is developed.

\subsection{Recurrence Quantification Analysis}

This section presents the results of a comprehensive RQA applied to 50 light curves, providing insights into their determinism, complexity, and trends. As mentioned in Section~\ref{RQA}, we do not apply embedding as a preprocessing step; in other words, we do not transform the one-dimensional light curves into a higher-dimensional phase space before calculating the RQA measures. Although embedding is not applied in our analysis, we report the average estimated embedding dimension across all light curves as 7.3 with variance 2.79, which serves as an estimate of the phase space dimensionality and reflects the system's degrees of freedom. The average time lag is 4.24 with variance 19.66.

The calculated RQA measures are displayed in Figure~\ref{RQAplot}, where the color bar reflects the  observation chronology—lighter blue marks more recent  observations, while darker blue denotes older ones. The $DET$, $ENT$, and $TR$ measures display variations over time, and the clustering of lighter blue lines suggests a possible link between stationarity and observation length, as tested by the ADF test. A notable difference emerges around $RR$ = 5\%, which defines the lower boundary for averaging RQA measures at small scales, while $RR$ = 50\% serves as the upper boundary for examining large-scale structures.

The $DET5$ and $DET50$ measures provide an estimate of the deterministic content within the light curves. The $DET5$ values have a mean of $0.46$ with variance $0.031$, ranging from $0.19$ to $0.98$, suggesting that some light curves exhibit highly predictable behavior. Meanwhile, $DET50$ has a mean of $0.66$ with variance $0.02$, ranging from $0.47$ to $0.90$, highlighting long-term patterns in the data. The variance of $DET5$ at $0.031$ is higher than the variance of $DET50$ at $0.02$. A correlation analysis shows that $DET5$ and $DET50$ have a correlation of $0.02$, indicating that deterministic properties at small and large scales are not strongly related.

The entropy measures $ENT5$ and $ENT50$ quantify the complexity of the light curves. The $ENT5$ values have a mean of $0.54$ with variance $0.25$, ranging from $0$ to $2.57$, while $ENT50$ has a mean of $1.51$ with variance $0.32$, spanning from $0.70$ to $2.76$. A correlation of $0.42$ is found between $ENT5$ and $ENT50$, suggesting a measurable relationship between entropy at different scales.

The TR5 and TR50 measures assess the presence of trends within the light curves. TR5, capturing small-scale trends, ranges from  -1.18e-03 to 2.70e-04 with a mean of -2.19e-04 and variance 5.49e-08. TR50, reflecting large-scale temporal trends, ranges from -7.39e-03 to 1.02e-03 with a mean of -2.06e-03 and variance 3.82e-06.

Large TR50 values indicate greater nonstationarity, with higher variability across light curves compared to TR5. The variance of TR5 at 5.49e-08 is two orders of magnitude lower than the variance of TR50 at 3.82e-06, suggesting greater uniformity in small-scale trends and higher variability in large-scale trends. There is a significant positive correlation of 0.66 between TR5 and TR50, suggesting that trend behavior at small and large temporal scales is systematically related.

\subsection{Autoregressive modeling} 
 
We present the results of the ARMA and ARIMA modeling of the light curves, focusing on the estimated parameters and their statistical properties.

The estimated autoregressive parameter \( p \) for ARMA models spans a wide range, with a mean value of 4.38 and variance 10.24, ranging from 0 to 10. In contrast, the ARIMA \( p_i \) values are lower, with a mean of 2.10 and variance 3.77, ranging from 0 to 9. The majority of observations have ARIMA \( p_i \) values between 0 and 4, while ARMA \( p \) values show a broader distribution, with a substantial fraction exceeding 4. This confirms that ARMA models tend to rely more on past observations than ARIMA models.

The moving average parameters $q$ and $q_i$ show similar distributions, with mean values of $4.44$ and variance $14.62$ for ARMA $q$ compared to mean $3.86$ and variance $11.22$ for ARIMA $q_i$. Both parameters cover wide ranges with ARMA $q$ values spanning from $0$ to $14$ while ARIMA $q_i$ values extend from $0$ to $15$.

We present here the $AIC$ model selection criterion to assess the relative performance of ARMA and ARIMA. The ARMA model exhibits an average $AIC$ of 1243.71, while the ARIMA model achieves 1233.39, indicating that ARIMA only slightly outperforms ARMA in terms of goodness of fit.

A correlation analysis reveals that the moving average parameters \( q \) and \( q_i \) are strongly correlated, with \(0.83 \), indicating that short-term dependencies remain stable between ARMA and ARIMA. Similarly, the autoregressive parameters \( p \) and \( p_i \) show a strong correlation of \(0.69 \), reinforcing the idea that ARIMA captures similar autoregressive structures, albeit at lower values due to differencing.

\subsection{ ADF test }

The ADF test indicates that observations in the second half of the time series tend to be slightly more stationary, with a mean test statistic of $-2.35$ and variance $1.55$, ranging from $-5.29$ to $0.52$. Additionally, these same time series are somewhat shorter and this, as previously noted, can limit the ADF test's ability to detect underlying trends or structural breaks, as previously noted.

\subsection{ Tsay’s test }
Analysis of Tsay's test statistics from the X-ray emission light curves of Mrk~421 presents a nuanced picture of its variability, with values ranging from 0 to 5.32  with a mean of 1.20 and a variance of 0.96. While the exact values vary, there are instances where the test statistics exceed high values of 3, which could suggest evidence of nonlinearity. However, such indications should be interpreted with caution, as they do not conclusively prove nonlinearity across the board. The indications of potential nonlinearity align with expectations, considering the diverse and complex nature of blazar emission. These expectations arise from various proposed deterministic scenarios and models, suggesting likely nonlinear behavior. The inherent variability in the light curves may be indicative of dynamic and intricate processes actively influencing the observed phenomena. Even if the light curves display clear deterministic processes, they are fundamentally measurements that are influenced by other stochastic processes, which obscures the ability to clearly determine the deterministic content.

\section{Discussion and Conclusions}   \label{s6}
Our study portrays the X-ray emission from Mrk~421 as a complex process, where both stochastic and deterministic processes are at play.  The distinction  in modeling scales and the correlation between models, including RQA, ARMA and ARIMA modeling, along with the results from ADF  and Tsay's test statistics, highlight the presence of both short- and long-term dependencies within the data. These methods and techniques are attempting to provide a thorough understanding of the primary physical processes that govern the fluctuating emission of this blazar. Furthermore, the investigated correlations between calculated quantities not only deepen our insight into the significance of these metrics but also, in a broader sense, enable us to extract more nuanced information. This enhanced understanding underscores the complexity and interconnectivity of the phenomena under study.

Nonlinear modeling:  the L. Cao method estimates the degrees of freedom for most of the light curves to be approximately 7 with a small variance, suggesting some uniformity in the dynamical complexity across these observations. In contrast, the average time lag calculated using the AMI method is around 4, but with considerable variability. This variability in the time lags could point to complex and possibly fluctuating astrophysical processes within the blazar, which might involve the dynamics of its jet, interactions within its magnetic field, the mechanisms of accretion, or interactions with the external medium. Such factors might contribute to the observed differences in the calculated time lags across the light curves and the associated X-ray emission.

The averaged RQA measures of determinism, entropy, and trend at different scales (averaged to 5 and 50 \% of $RR$) provide insight into the deterministic properties of the Mrk~421 light curves. Among these measures, the smallest correlation is observed between the small- and large-scale variants of the determinism measure, suggesting that the deterministic processes responsible for the light curves might work independently at various scales. This indicates that the dynamics driving short-term behaviors may be distinct from those influencing long-term behaviors.

High determinism at small scales $DET5$ can be associated with more predictable processes, possibly periodic, quasi-periodic, or even chaotic \citep{2019EPJC...79..479P}. These processes could include orbiting hot spots in the accretion disk, as indicated by \citet{2020ApJ...900..100R} and \citet{1997MNRAS.290..139B}. In contrast, large-scale determinism $DET50$ might reflect global, structured behaviors in the jet or disk \citep[see, e.g.,][]{2017A&A...601A..52B}.

The high entropy measures $ENT5$ and $ENT50$ indicate the complexity and possible randomness of the system, with higher values suggesting  complex dynamics, possibly due to turbulence or magnetic field interactions. The observation that small-scale variability tends to show more ordered and structured behavior is a relevant point, that could imply that at these scales the system exhibits more deterministic dynamics, possibly governed by specific physical processes or interactions.

On the other hand, this finding supports the hypothesis that large-scale variability could be attributed to random shocks driven by stochastic processes. Since there is a noticeable correlation between $ENT5$ and $ENT50$, as well as between ARMA \( q \) and ARIMA \( q_i \), it can be inferred that noise processes are affecting both scales. This suggests that the relationship between complexity at different scales could be influenced by the presence of noise and stochasticity rather than purely deterministic mechanisms.

We find a notably high correlation between $ENT50$ and $DET50$, which could be indicative of a relationship between complexity and determinism at larger scales. A plausible interpretation of this correlation is that with the increase in the scale of analysis, the deterministic patterns exhibit greater complexity. This case could have implications for our understanding of the physical processes driving light curve variability, suggesting that the large-scale variability has a complex deterministic nature with multiple degrees of freedom. Another view on this is that the correlation is due to the fact that both measures are sensitive to large scales in the data. As mentioned above, averaging over more thresholds can increase the magnitude of $ENT50$ and $DET50$ simultaneously, which could make them more correlated.  The coexistence of deterministic and complex patterns in Mrk 421 light curves is further complicated by noise processes \citep{2016ApJ...833..208M}, underscoring the need for advanced modeling to account for their intricate variability.

The trend measure, at both small $TR5$ and large $TR50$ scales, evaluates the presence of trends in these light curves, with higher values indicating increased nonstationarity. $TR5$ quantifies temporal trends on a smaller scale, showing varied trends and transitions across the light curves. $TR50$ extends this analysis to larger-scale temporal dynamics, with higher values signifying more prominent nonstationarity associated with these larger-scale trends.

The mean value of $TR5$ is approximately one order of magnitude higher than that of $TR50$, while the variance of $TR5$ is two orders of magnitude lower compared to $TR50$, indicating that smaller-scale trends are more uniform, whereas larger-scale trends exhibit greater variability.

The significant positive correlation between $TR5$ and $TR50$ suggests that nonstationary processes may propagate across both small and large scales, pointing to a dynamic relationship between these trends. This implies that processes driving nonstationarity at one scale may influence trends at other scales, or that similar mechanisms may operate across scales. These results highlight the interconnected nature of trends and the importance of considering multiple scales when analyzing temporal dynamics.

The positive correlation of both $TR5$ and $TR50$ with parameters $p$ and $p_i$ indicates that these trends are likely driven by stochastic processes with memory, implying that nonstationarity is influenced by past data points. $TR50$, in particular, shows a stronger correlation with $p$ and $p_i$, highlighting that larger-scale trends have a more substantial impact on smaller ones.

Similarly, the observed negative correlation between both trend measures and the small-scale determinism measure $DET5$ reveals an inverse relationship, where higher determinism on smaller scales corresponds to reduced nonstationary behavior. Furthermore, the significant negative correlation between $TR5$ and Tsay’s nonlinearity measure, as well as $DET5$, contrasts with the zero correlation observed between Tsay’s nonlinearity and $TR50$. This suggests that as trends increase at smaller scales, deterministic nonlinearity decreases, and vice versa.

The correlation of $TR5$ and $TR50$ with the length of observations further emphasizes the importance of the observation period in understanding the dynamics of these light curves. Although the correlations between nonlinear measures $TR5$ and $TR50$ with ADF test statistics are only slightly significant on larger scales, this finding highlights the need for more advanced analysis techniques to fully capture the complexity of these phenomena.

There are several physical implications suggested by the observed correlations. 
The varying degrees of stationarity and trends
 identified by the ADF and RQA $TR$ measures could be indicative of the blazar's jet dynamics, where periods of stability are interspersed with episodes of enhanced activity of multifaceted nature, possibly affected by shock waves or magnetic reconnection events \citep{2018ApJ...864..164Y}. The variability across different scales, as revealed by the RQA, suggests that multiple physical processes are possibly at play, ranging from the microphysics of particle interactions to the macroscopic behavior of the jet and the accretion disk \citep{2018Galax...6....2B}.

Autoregressive modeling:  The use of ARMA and ARIMA models in our analysis helps in understanding the stochastic memory and randomness of the Mrk~421 light curves. The $p$ and $q$ parameters from these models indicate the extent to which past values and noise (residuals) in the time series influence future values. In the context of blazars, a high $p$ value may suggest that the emission process retains a memory of past states, potentially linked to the persistence of physical conditions such as magnetic field structures or stochastic particle acceleration mechanisms. The memory effects identified by these models are also relevant to the characteristic timescales of particle acceleration and cooling \citep{2022ApJS..262....4N}.

An interesting finding of our analysis is the negative correlation between the RQA $DET5$ measure and the ARMA and ARIMA $p$ and  $p_i$ parameters. This relationship indicates that higher levels of determinism at smaller scales, as denoted by high $DET5$ values, correspond to a diminished temporal dependence or memory within the data. In astrophysical terms, this suggests that certain short-term patterns at small scales in Mrk~421's X-ray emission are more deterministic and less influenced by their past states. Such patterns could reflect deterministic physical processes, even chaotic ones, perhaps related to specific emission mechanisms within the blazar's jet or relatively stable physical conditions in its accretion disk \citep{2019EPJC...79..479P}. There, interactions between ionized plasma, magnetic fields, and relativistic effects are shown to produce periodic or chaotic motions, which could modulate the variability of the emitted radiation and influence the observed light curves. 

In the case of ARIMA, the autoregressive order $p_i$ is relatively low for most observations, which might reflect that the physical conditions within the blazar, such as magnetic field structures or particle acceleration mechanisms, are not persistent over long time scales. In contrast, ARMA $p$ values exhibit a broader distribution, with a substantial number exceeding 4, suggesting stronger memory effects in the undifferenced data. Instead, these conditions might be changing relatively quickly, leading to a less stable and more variable emission process. In general, the moving average $q$ parameter, which tends to be stable, could suggest that short-term random fluctuations are consistent. These fluctuations might arise from smaller, localized events such as turbulence or minor reconnection events in the blazar's jet \citep{2003NewAR..47..513L,2021ApJ...912..109K,2014ApJ...780...87M}. However, our findings suggest that there is significant variation in the $q$ parameter across both ARMA and ARIMA models, which indicates a high degree of variability in the noise characteristics of the blazar's light curves. This suggests that the nature and intensity of transient, localized events contributing to the noise in the observations vary significantly. Despite this variability, the correlation pattern in the noise structure observed in both ARIMA and ARMA models suggests that the integration in ARIMA does not significantly alter the short-term noise characteristics captured by the $q$ parameter. This similarity might point to a consistent underlying mechanism influencing these fluctuations, underscoring the complexity and dynamism of the processes occurring in the blazar's jet. It indicates that the transient and localized events contributing to the noise are inherently similar in nature, irrespective of long-term trends or shifts in the time series.

The moving average parameters $q$ and $q_i$ are highly correlated, indicating that the short-term noise characteristics captured by these parameters are consistent across the two models. Furthermore, the modest difference in the $AIC$ between ARMA and ARIMA models suggests that the data required a low degree of differencing, indicating that short-term variations dominate, but long-term influences still contribute to the overall behavior of the light curves of blazar jets \citep{2017ApJ...843...81O}.

ADF and Tsay's Test: The ADF test results, revealing varying degrees of stationarity  among the light curves, offer insights into the dynamics. Stationary periods, as indicated by the ADF test, could correspond to stable phases in blazars, potentially linked to consistent accretion rates or stable magnetic field structures within the blazar system. These stable periods might be influenced by long-term processes such as jet precession \citep{2021MNRAS.503.3145B} or the gradual accumulation and release of energy in the accretion disk.  Nonstationarity, particularly notable in the first half of the observation period, might be also associated with transient events like jet instabilities, shocks, or a change in the underlying processes. This could signify the onset of new dynamics within the blazar system, reflecting a shift in the accretion dynamics or magnetic field structures that govern the jet emission \citep{2018SSRv..214...81P}.

According to the Tsay's test results, we observe an absence of a  consistent pattern across the examined dataset of 50 time series,  instead revealing varying degrees of nonlinearity with a minority exhibiting  pronounced extensions. This could mean that the emission processes are not simply additive or proportional but may involve complex interactions, such as feedback mechanisms between the jet and the accretion disk or nonlinear particle acceleration processes \citep{2005MNRAS.359..345U, 2019EPJC...79..479P,2016EPJC...76...32S}. 
In the case of Mrk~421's emission, nonlinear processes often dominate within a complex mixture of phenomena, particularly at small scales. Their presence is closely tied to increased complexity, as indicated by the significant correlation between the Tsay's test statistic and the $DET5$ and $ENT5$ measures. This suggests that deterministic processes in the time series are rarely simple and are likely to manifest as complex deterministic dynamics or in combination with stochastic elements.

In conclusion, the intricate nature of the X-ray light curves from Mrk~421 presents a complex challenge. Our analysis indicates that the X-ray emission from blazars, such as Mrk~421, are not merely random occurrences but also encompass deterministic patterns, with evidence of these patterns being more pronounced at short time scales. 
 
It appears that deterministic trends on both high and low scales are independent, but it also seems that stochastic trends might be common to both, with indications that short-memory stochastic processes may propagate from high to low scales.

We must also acknowledge that the inherent complexity of these processes in Mrk~421 involves multiple degrees of freedom and a multitude of stochastic elements. The collected data present a range of potential features and scenarios, as previously discussed, which complicates the task of supporting a singular physical explanation. Instead, they suggest that the observed variability likely results from a combination of multiple scenarios, each contributing to the complex interplay of deterministic and stochastic elements observed in the emission, with interesting relationships identified between scales as revealed by our analysis of 50 observations. This complexity underscores the challenges in astrophysical data interpretation and the need for continued advancements in observational technologies and analytical methodologies. Future developments in technical instruments promise to yield higher quality and quantity of data. Coupled with sophisticated methods of data analysis, these advancements could provide deeper insights and a clearer understanding of these phenomena.
Moreover, the application of our methodology to other sources holds potential in advancing our understanding of blazar's intricate nature, contributing to the broader knowledge of these phenomena.

\section*{Acknowledgements}
RP would like to express his acknowledgement to the institutional support provided by the Research Centre for Theoretical Physics and the Institute of Physics of Silesian University in Opava, SGS/24/2024 ‘Astrophysical processes in strong gravitational and electromagnetic fields of compact object’, and for the support from the project of the Czech Science Foundation GAČR 23-07043S. TPA acknowledges the support of the National Natural Science Foundation of China ( grant nos. 12222304, 12192220, and 12192221).
This work was partially supported by a program of the Polish Ministry of Science under the title ‘Regional Excellence Initiative’, project no. RID/SP/0050/2024/1. The authors would like to express their gratitude to the referee for the thoughtful comments and valuable suggestions.

\begin{appendix}  \label{App}

\begin{figure*}[!htb] 
		\centering 
	\begin{minipage}{.38\textwidth} 
		\centering 
		\includegraphics[width=0.891\linewidth]{ 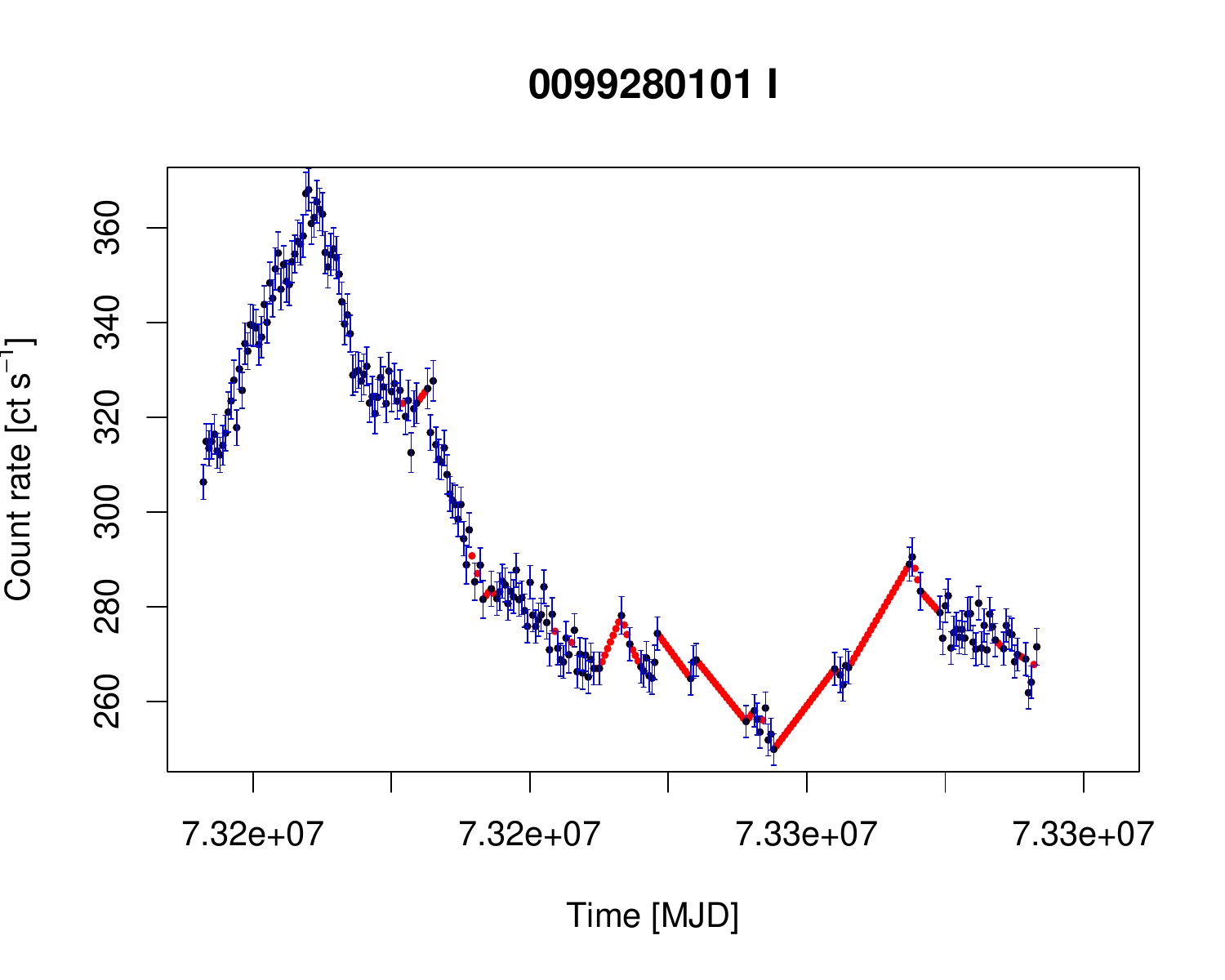}
	\end{minipage}
	\begin{minipage}{.38\textwidth} 
		\centering 
		\includegraphics[width=0.891\linewidth]{ 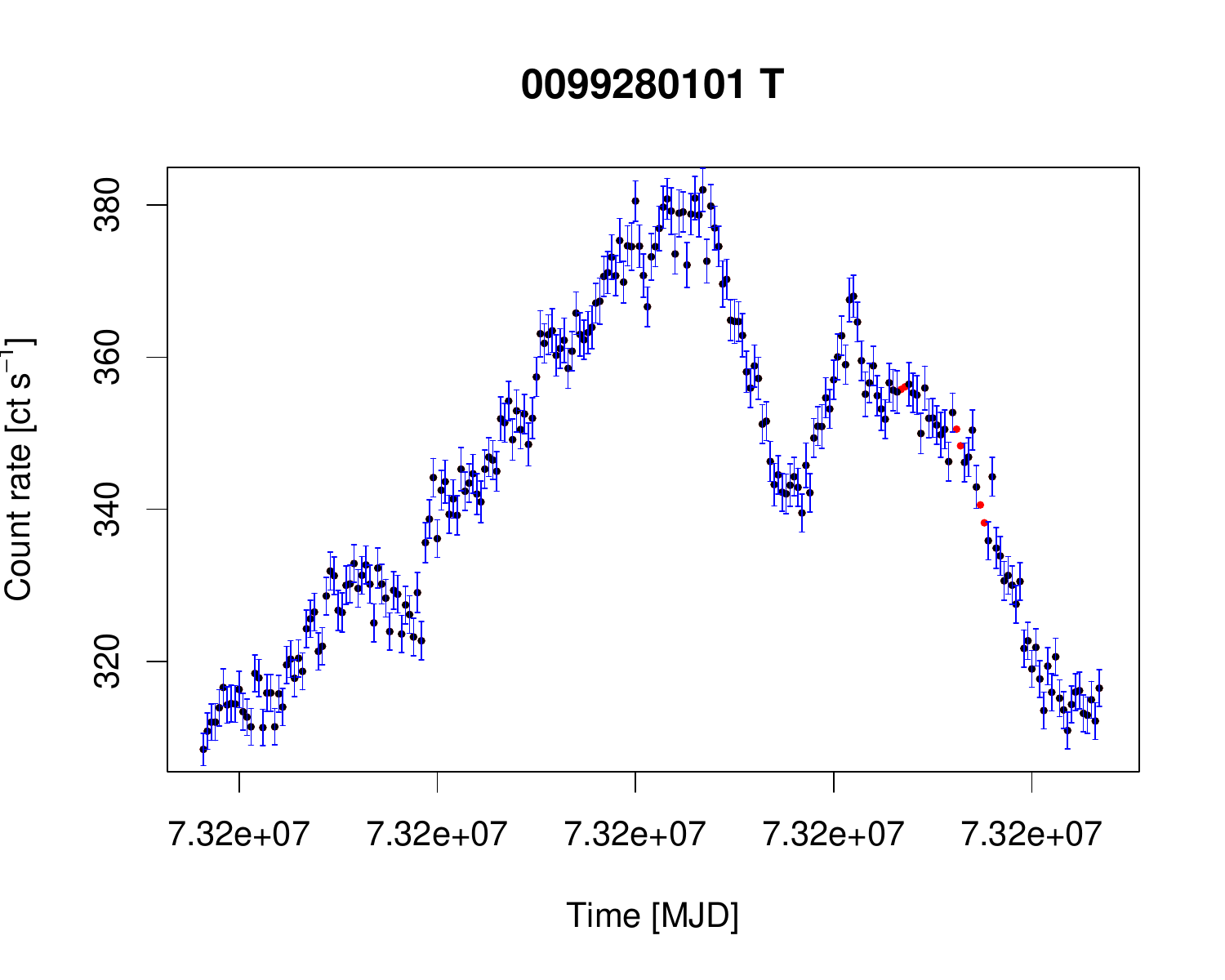}
	\end{minipage}
	\begin{minipage}{.38\textwidth} 
		\centering 
		\includegraphics[width=0.891\linewidth]{ 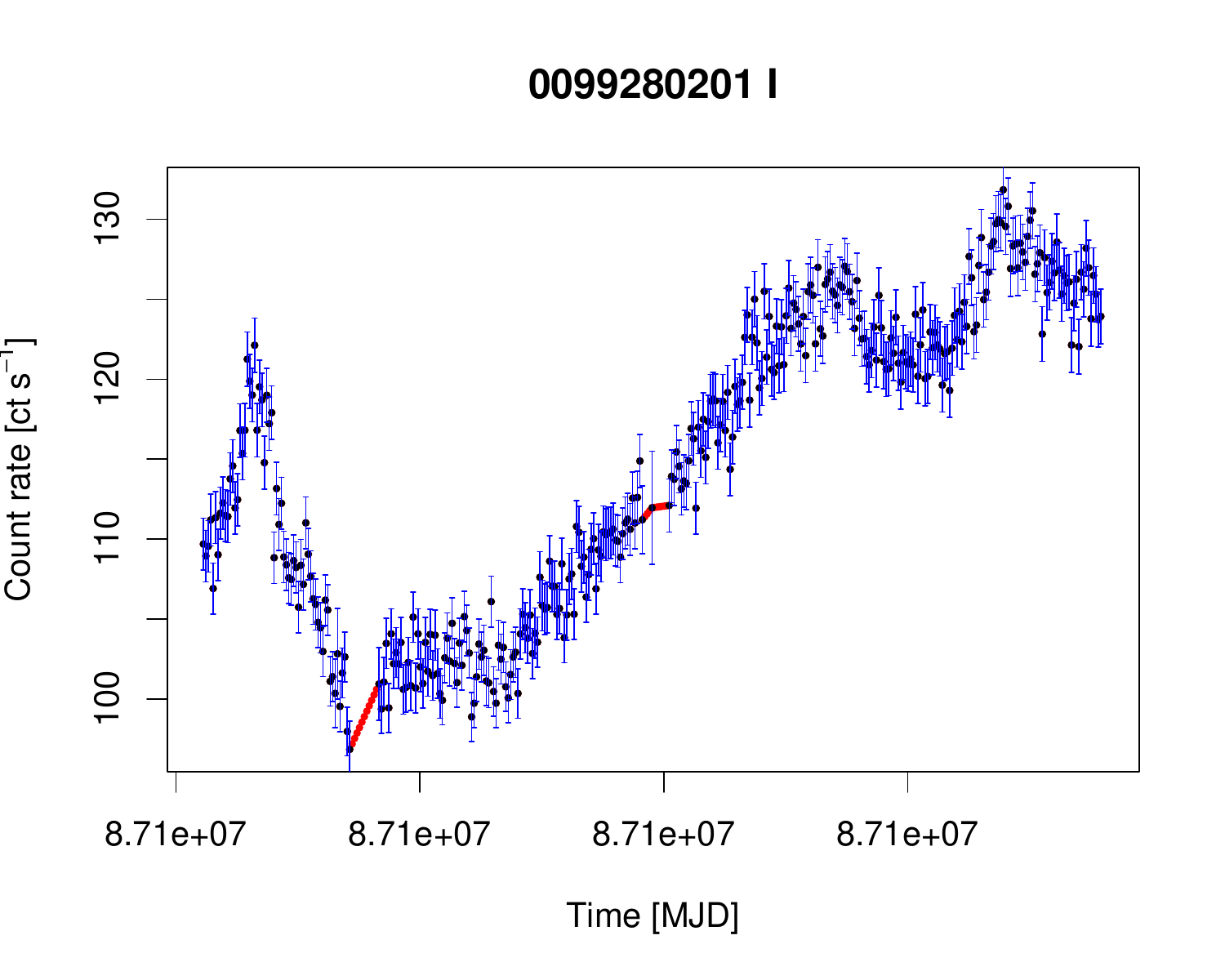}
	\end{minipage}
	\begin{minipage}{.38\textwidth} 
		\centering 
		\includegraphics[width=0.891\linewidth]{ 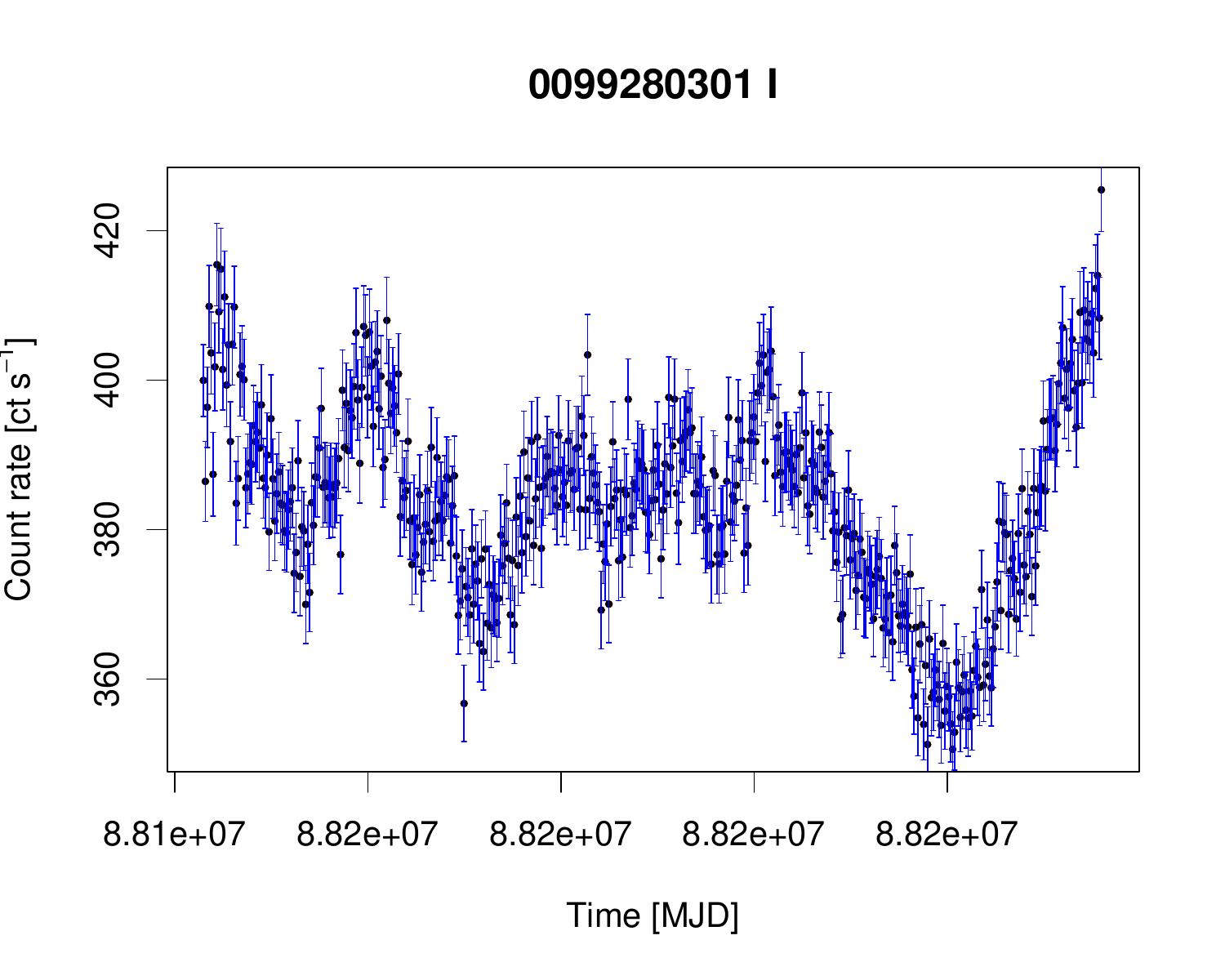}
	\end{minipage}
		\begin{minipage}{.38\textwidth} 
		\centering 
		\includegraphics[width=0.891\linewidth]{ 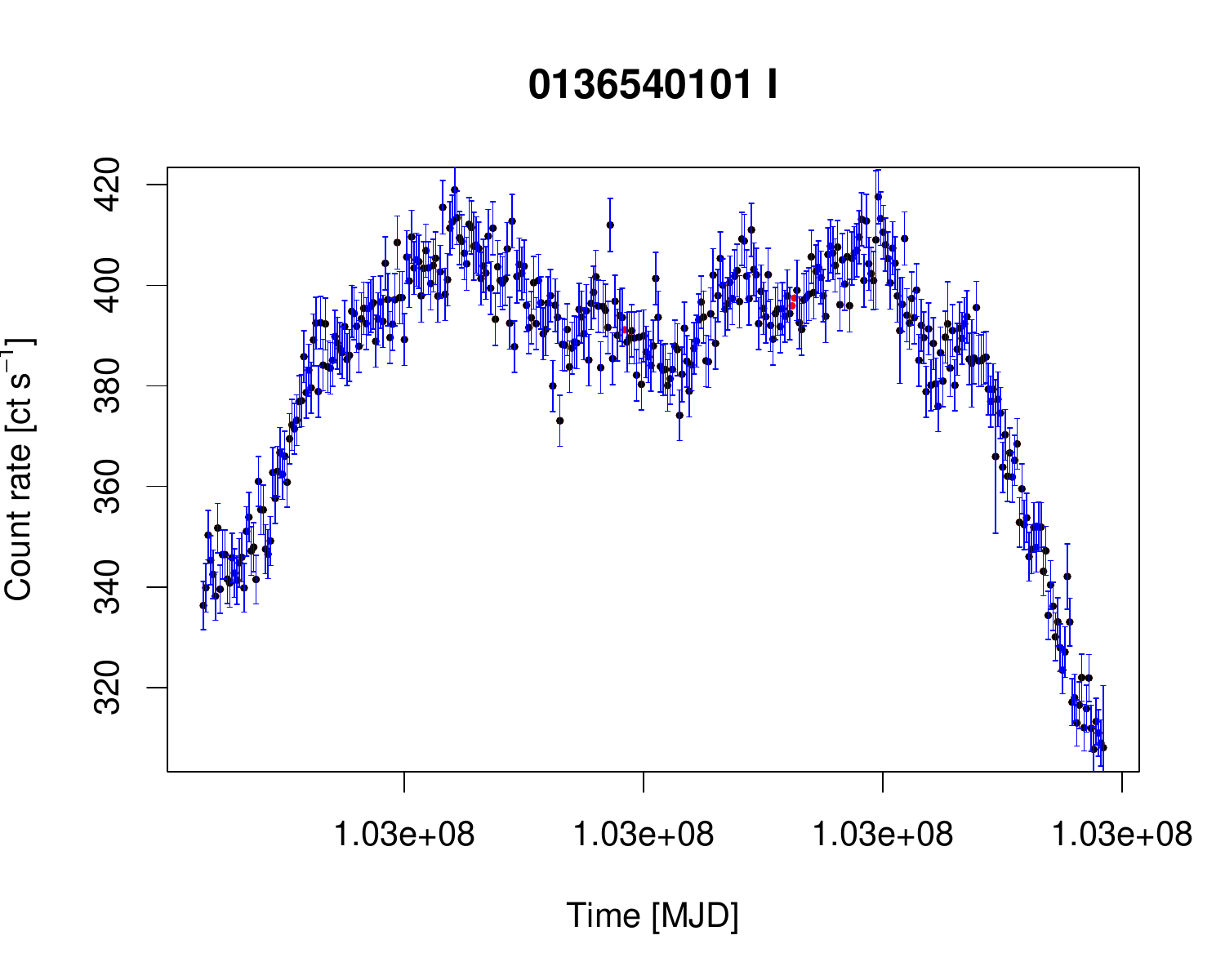}
	\end{minipage}
		\begin{minipage}{.38\textwidth} 
		\centering 
		\includegraphics[width=0.891\linewidth]{ 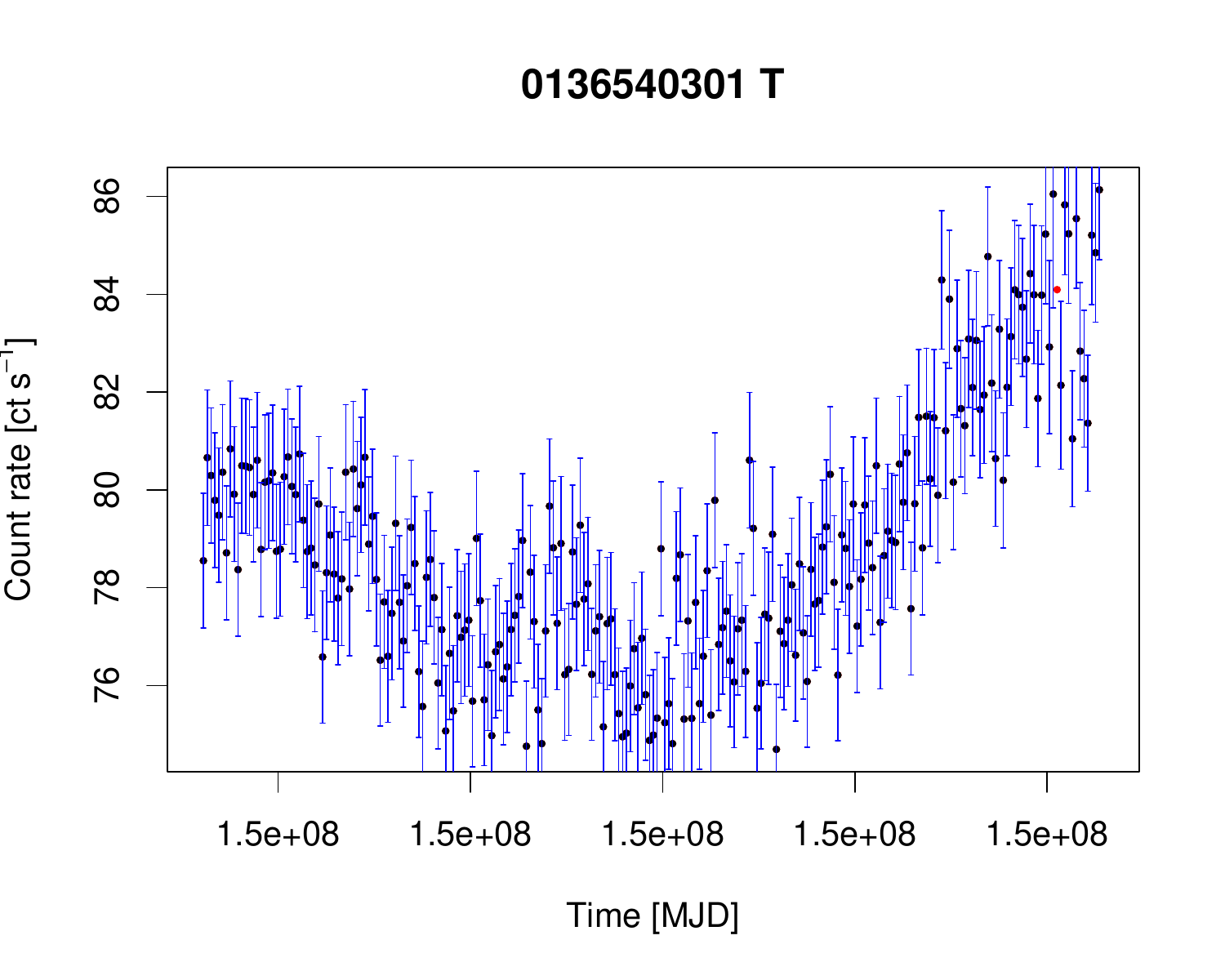}
	\end{minipage}
	\begin{minipage}{.38\textwidth} 
		\centering 
		\includegraphics[width=0.891\linewidth]{ 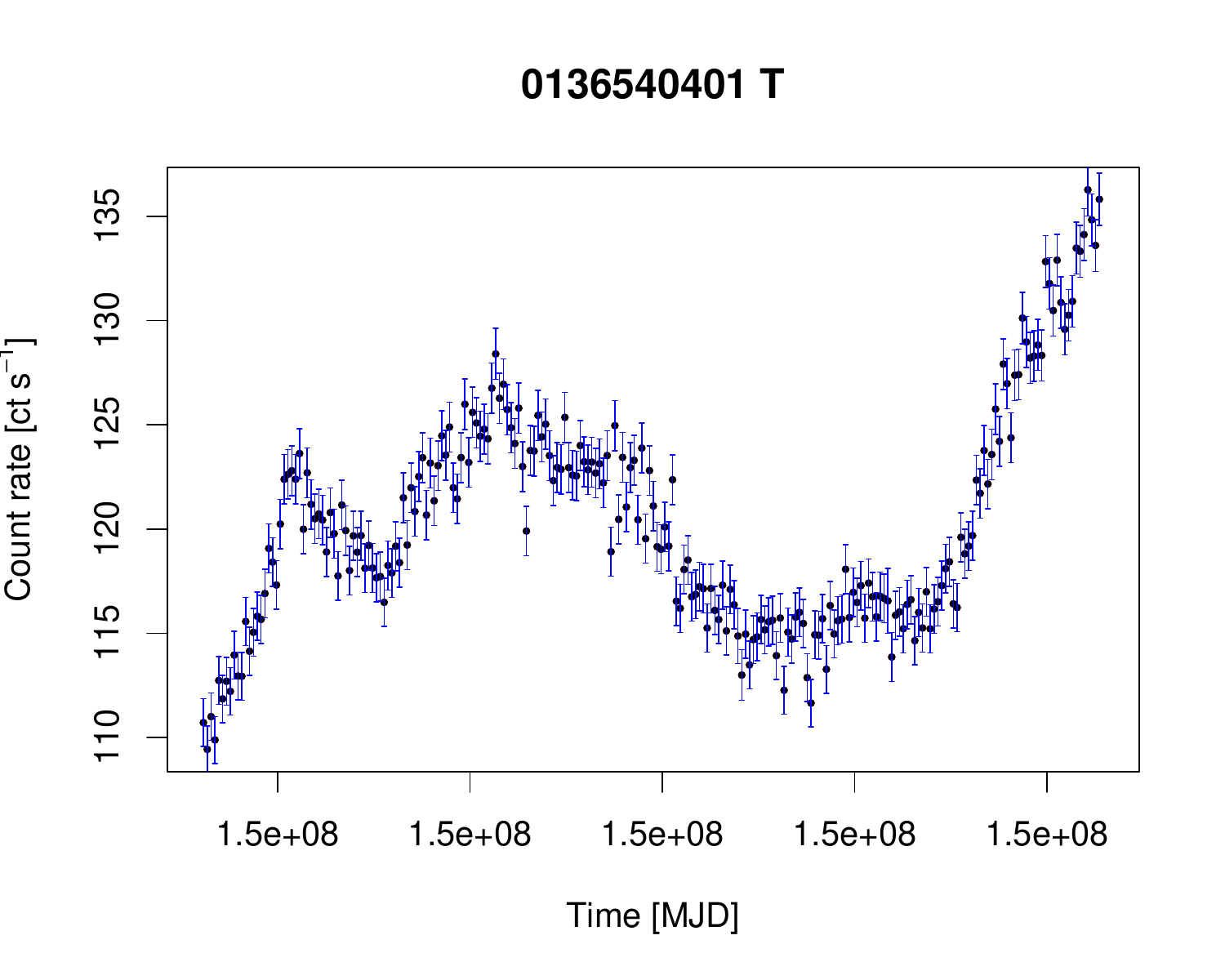}
	\end{minipage}
	\begin{minipage}{.38\textwidth} 
		\centering 
		\includegraphics[width=0.891\linewidth]{ 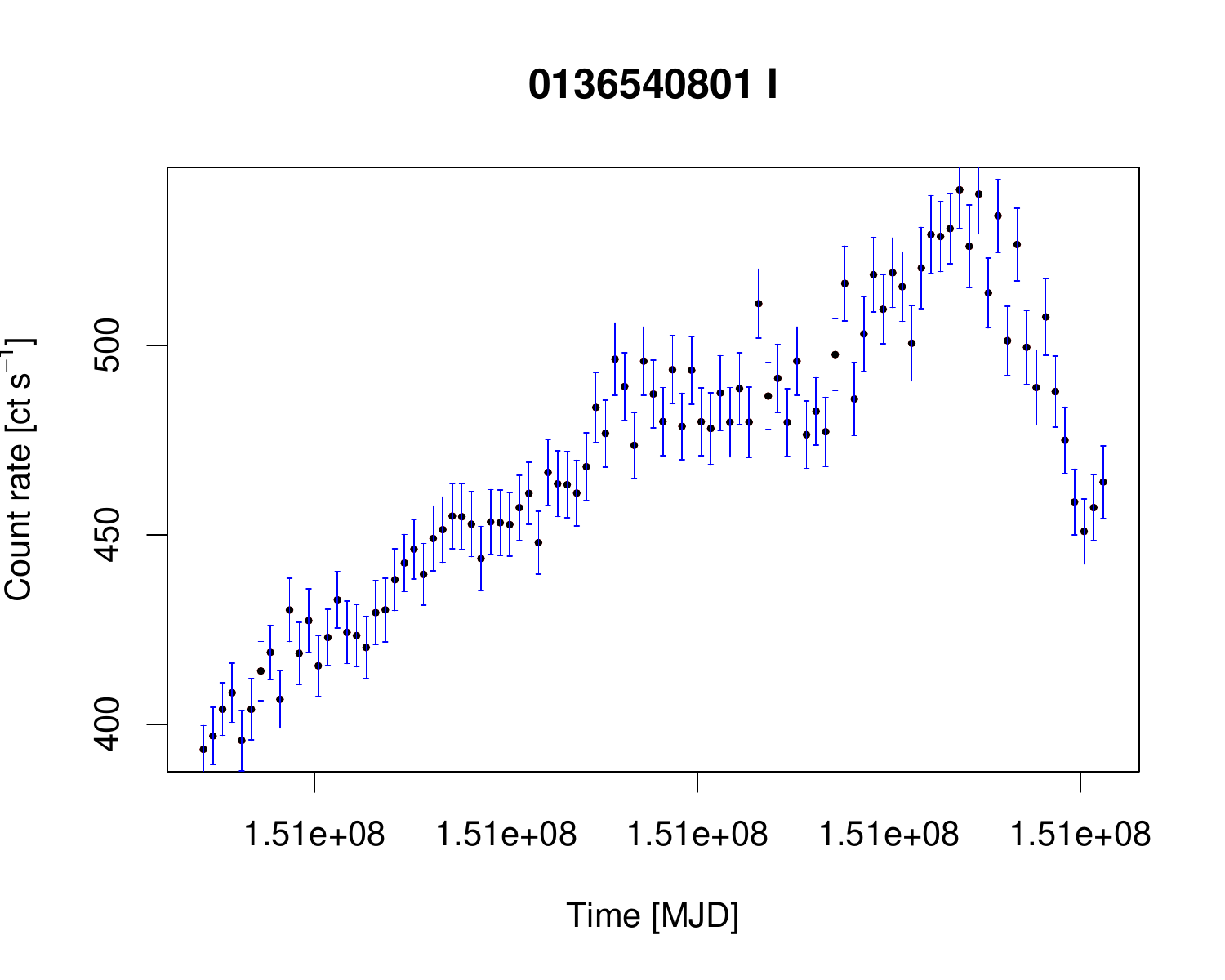}
	\end{minipage}
  \caption{Analyzed light curves 1-8 of Mrk~421, the red color denotes linearly interpolated data.  \label{LCS0}}
\end{figure*}

\begin{figure*}[!htb]  
		\centering 
	\begin{minipage}{.38\textwidth} 
		\centering 
		\includegraphics[width=0.891\linewidth]{ 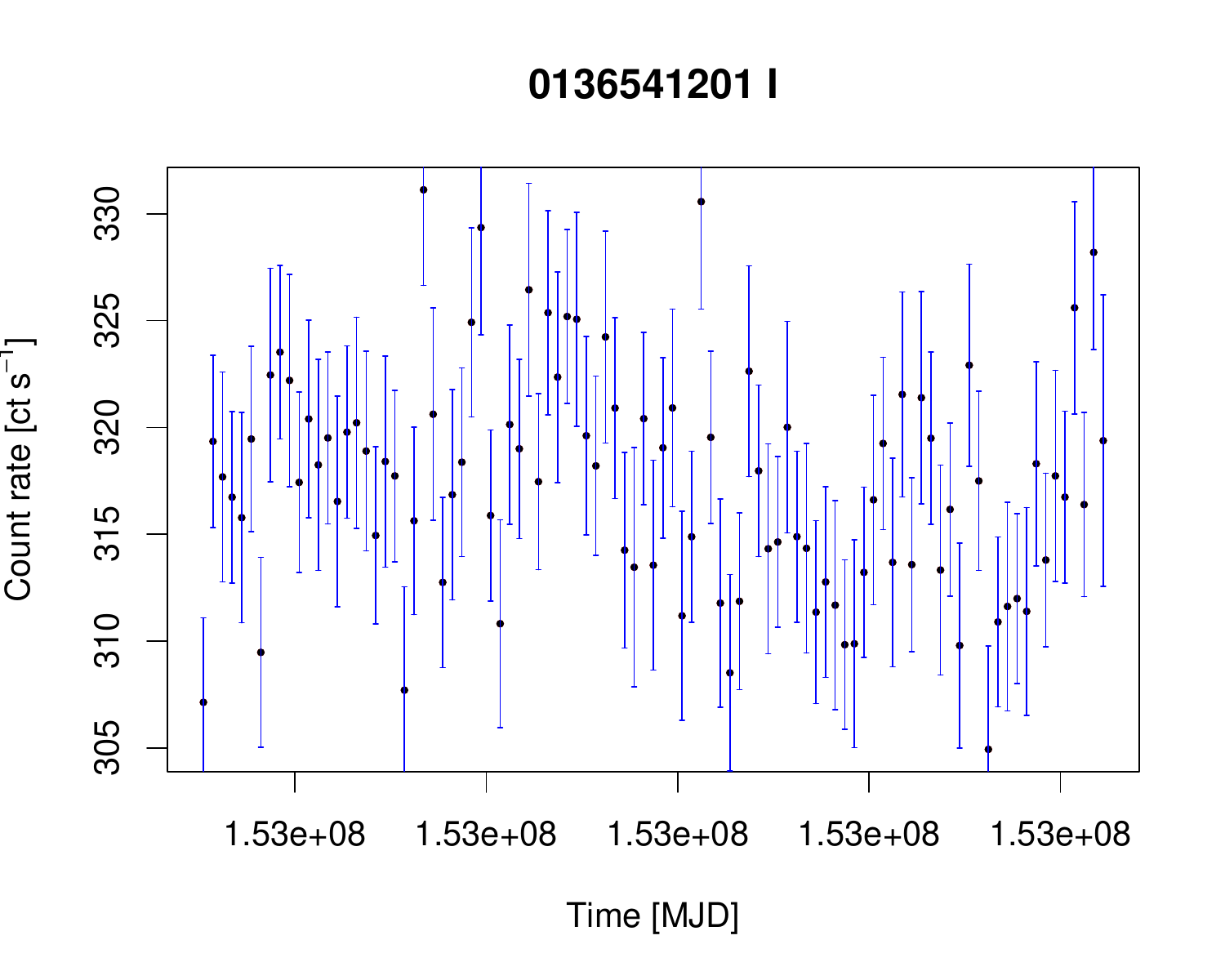}
	\end{minipage}
	\begin{minipage}{.38\textwidth} 
		\centering 
		\includegraphics[width=0.891\linewidth]{ 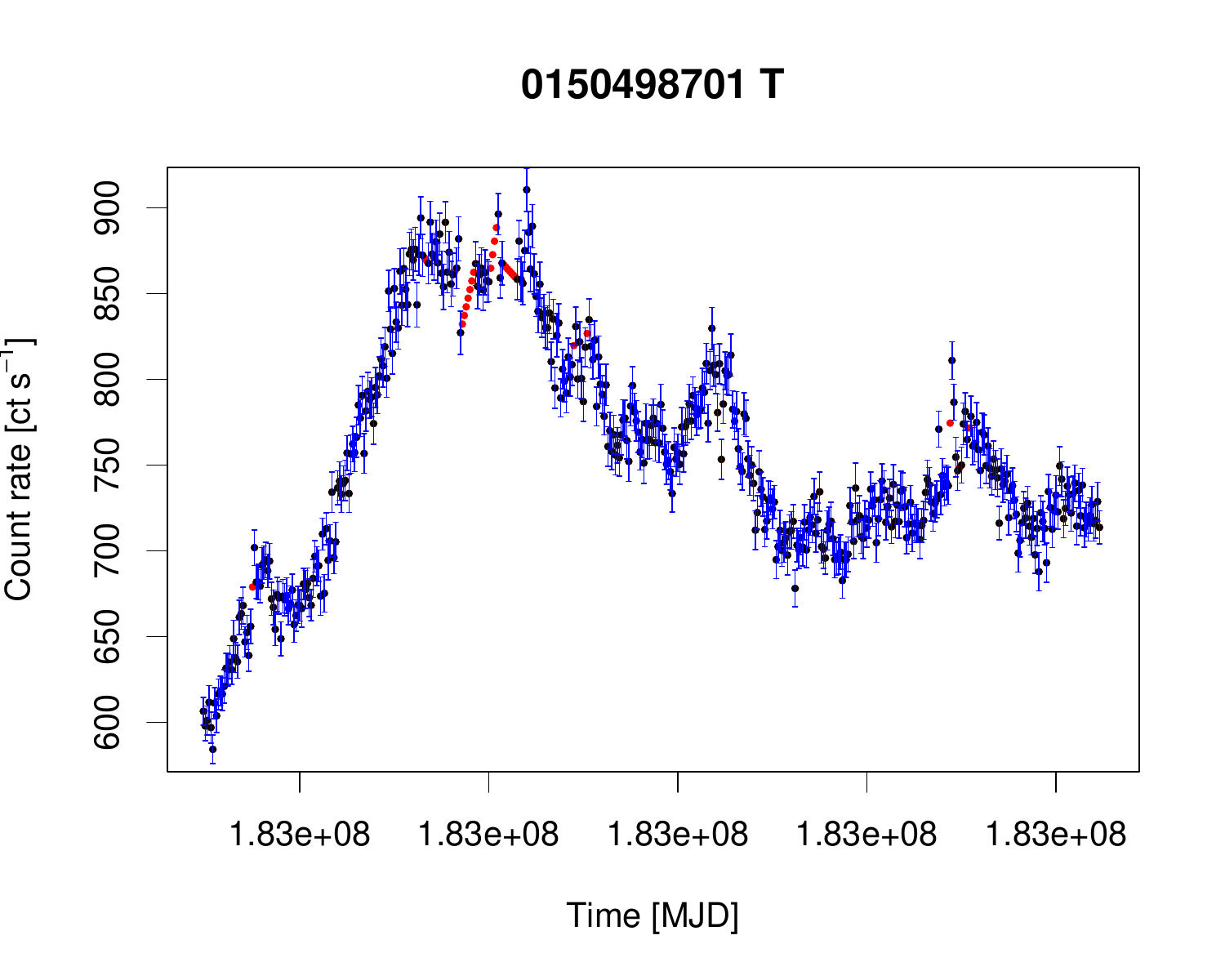}
	\end{minipage}
	\begin{minipage}{.38\textwidth} 
		\centering 
		\includegraphics[width=0.891\linewidth]{ 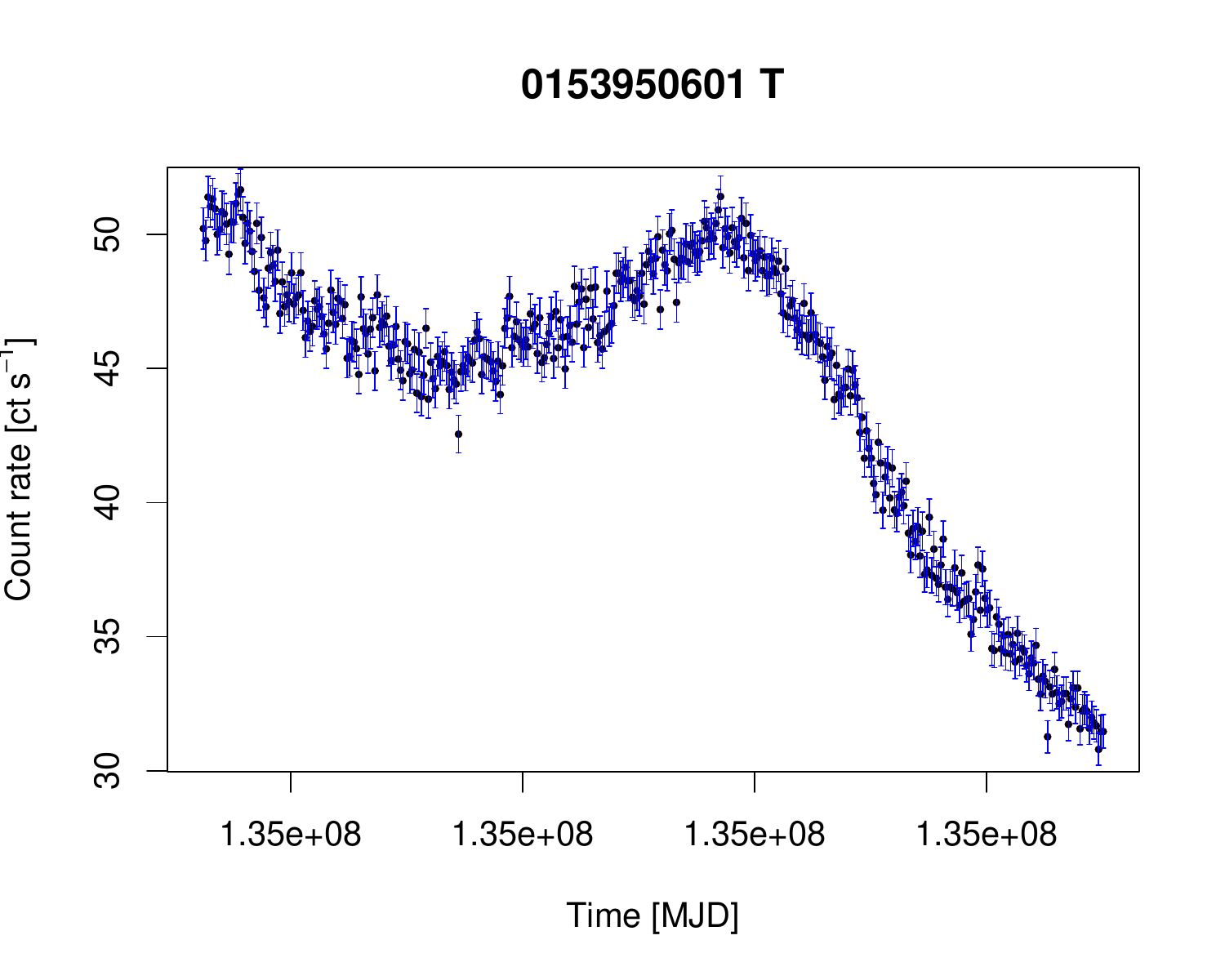}
	\end{minipage}
	\begin{minipage}{.38\textwidth} 
		\centering 
		\includegraphics[width=0.891\linewidth]{ 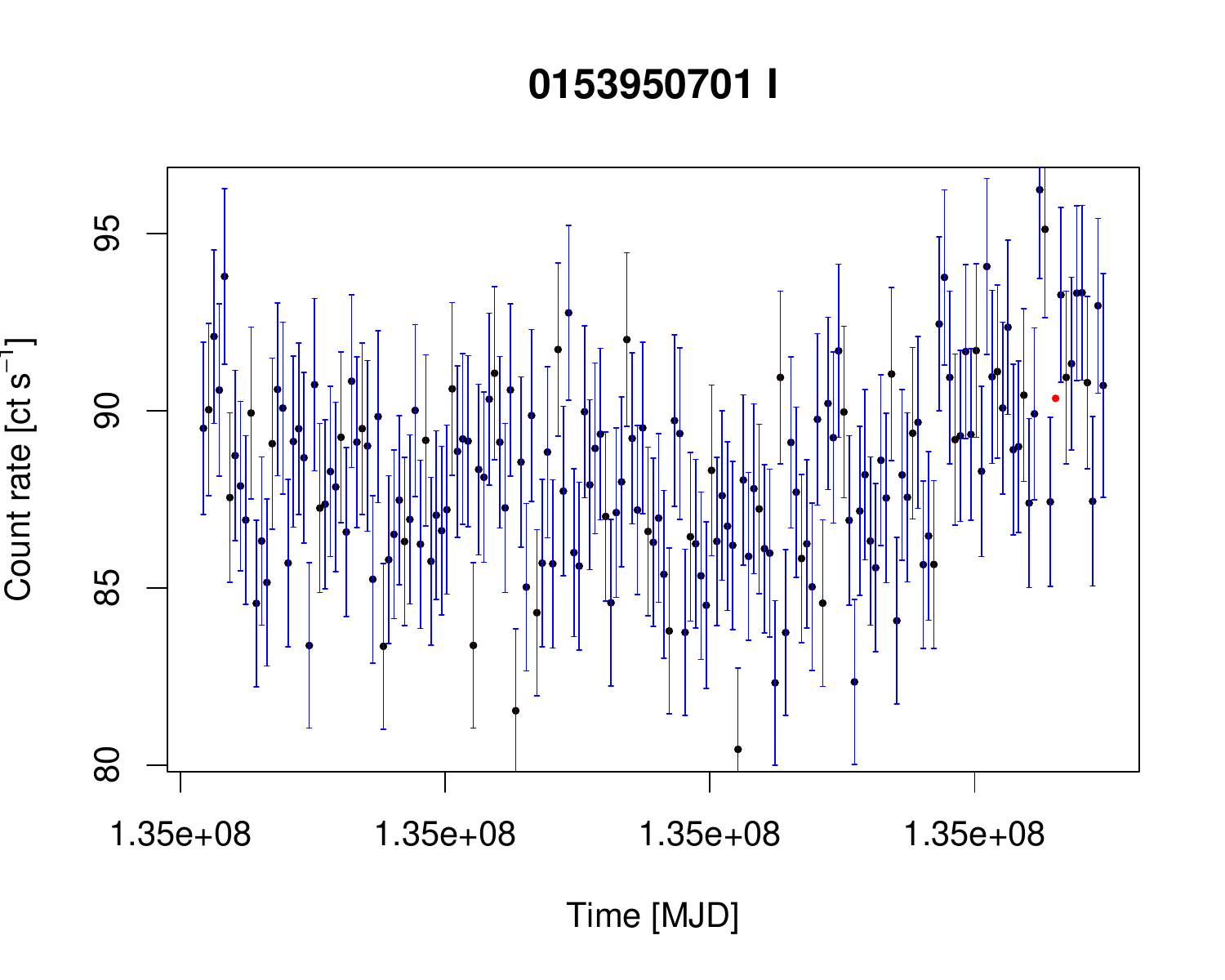}
	\end{minipage}
		\begin{minipage}{.38\textwidth} 
		\centering 
		\includegraphics[width=0.891\linewidth]{ 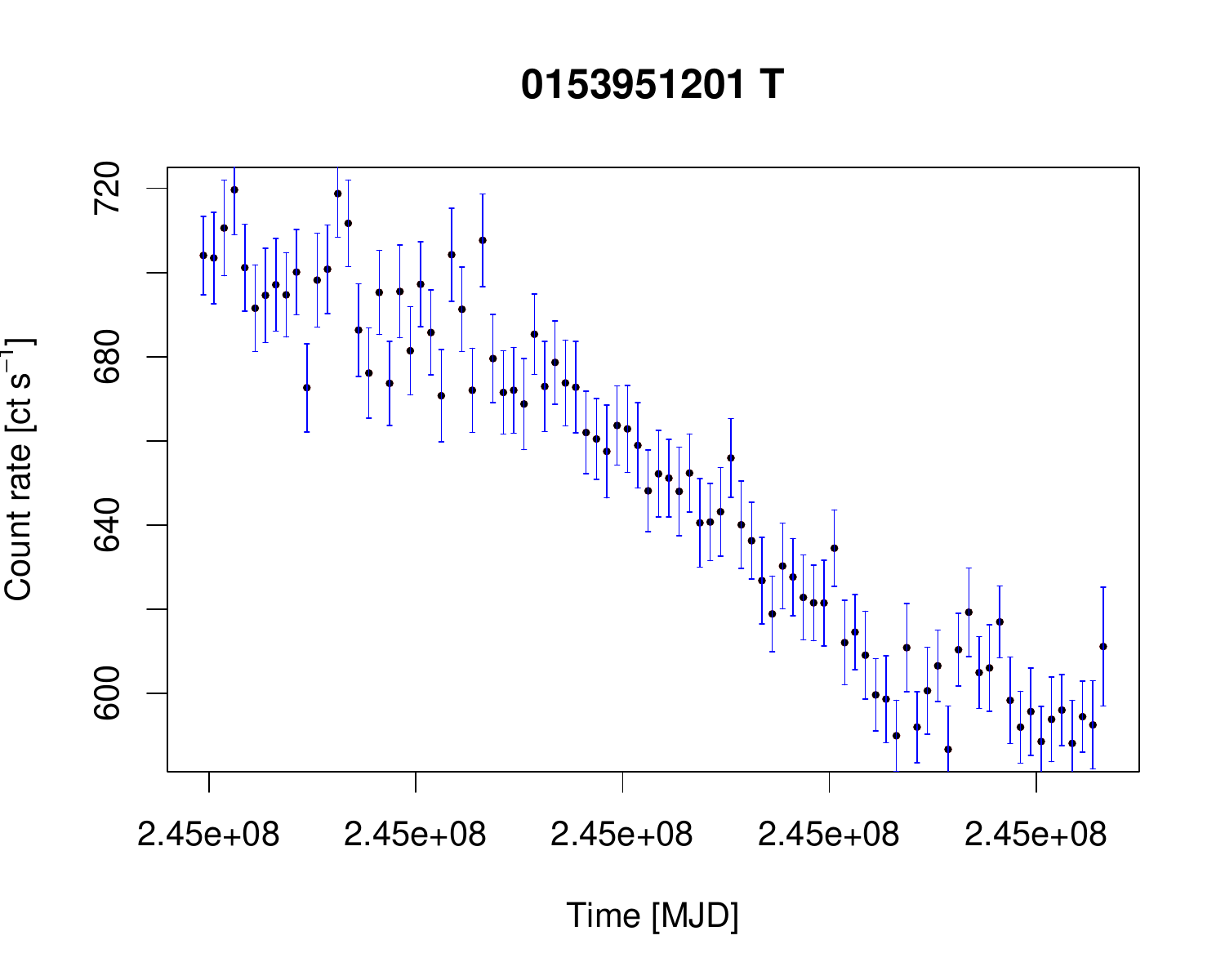}
	\end{minipage}
		\begin{minipage}{.38\textwidth} 
		\centering 
		\includegraphics[width=0.891\linewidth]{ 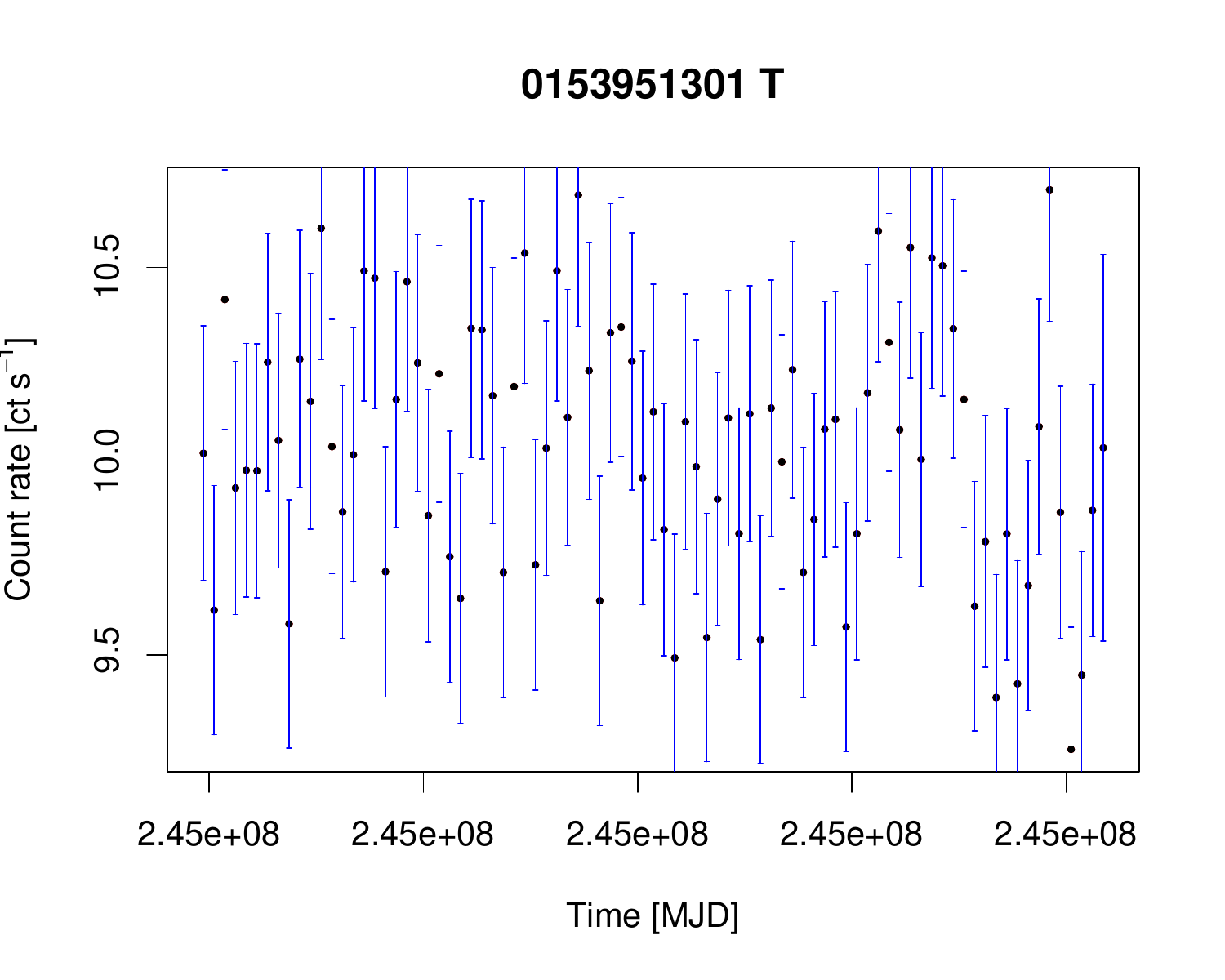}
	\end{minipage}
	\begin{minipage}{.38\textwidth} 
		\centering 
		\includegraphics[width=0.891\linewidth]{ 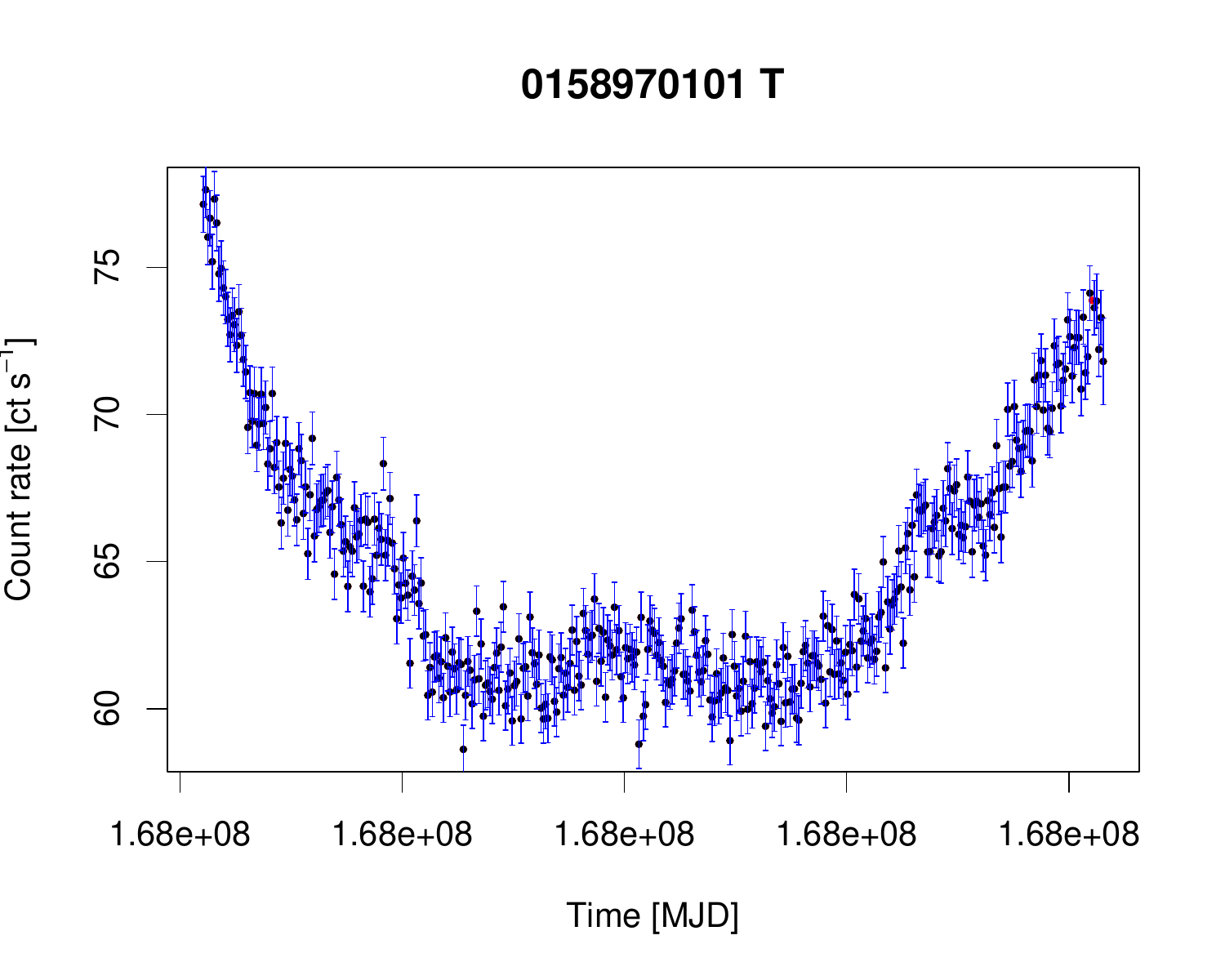}
	\end{minipage}
	\begin{minipage}{.38\textwidth} 
		\centering 
		\includegraphics[width=0.891\linewidth]{ 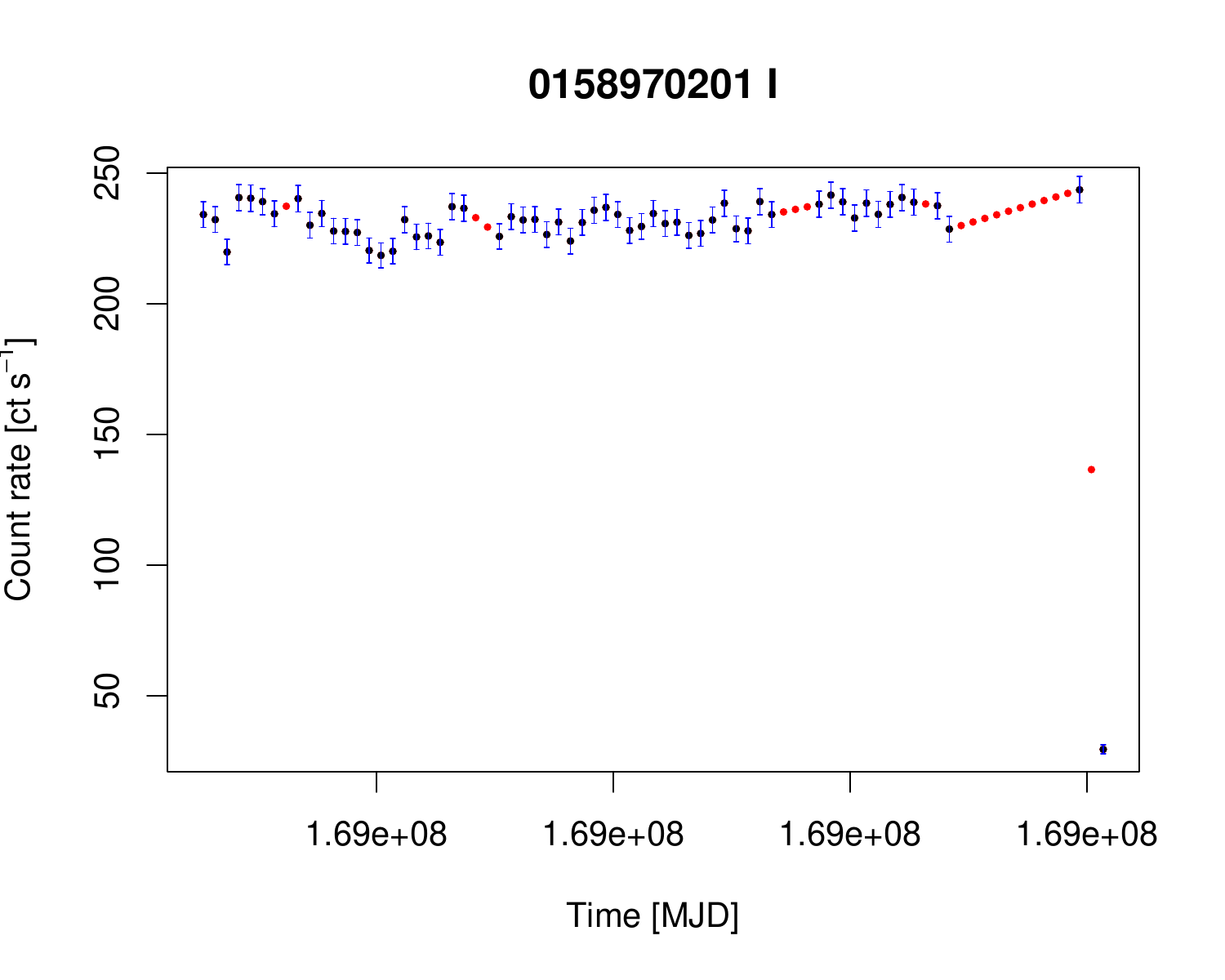}
	\end{minipage}
	\begin{minipage}{.38\textwidth} 
		\centering 
		\includegraphics[width=0.891\linewidth]{ 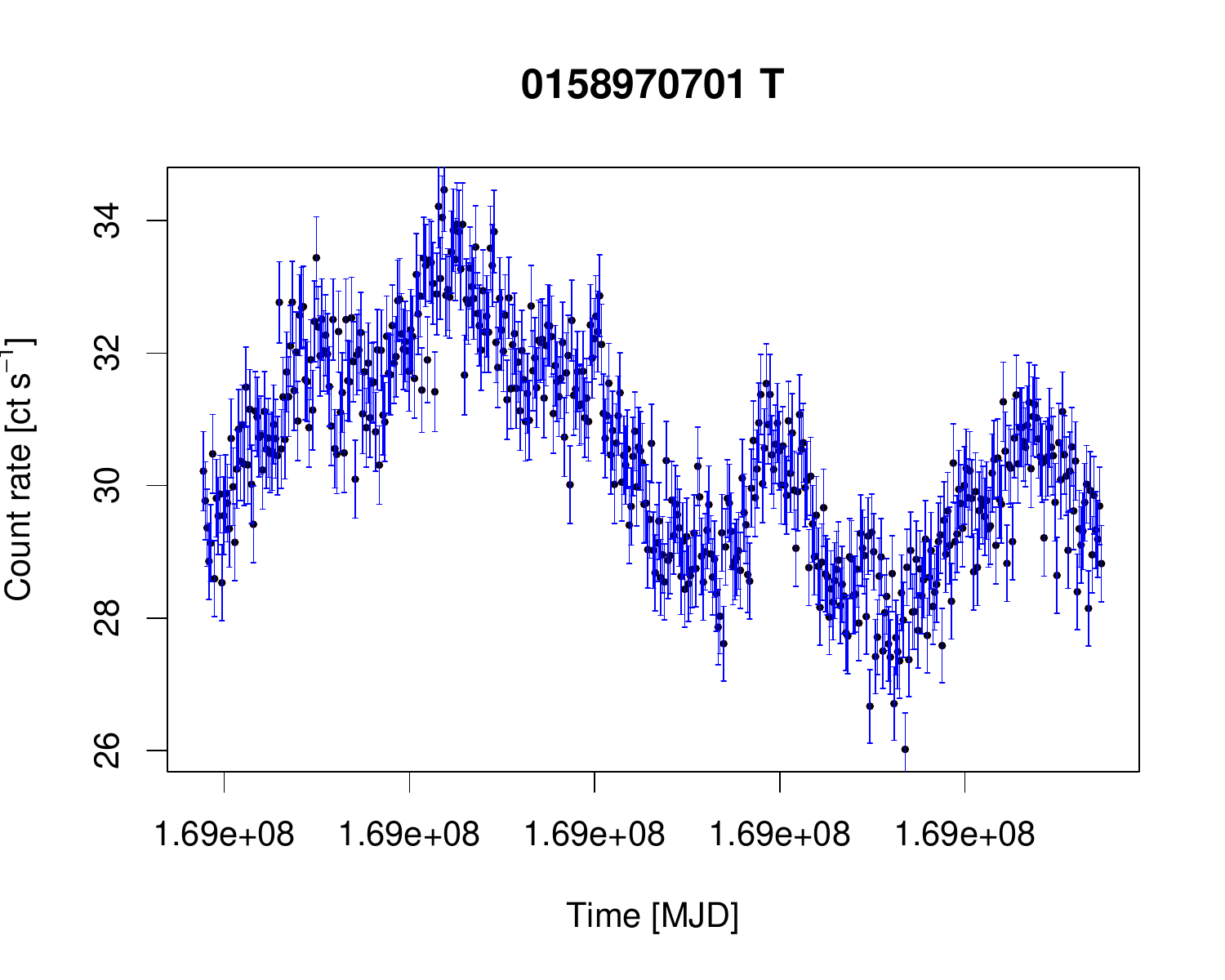}
	\end{minipage}
	\begin{minipage}{.38\textwidth} 
		\centering 
		\includegraphics[width=0.891\linewidth]{ 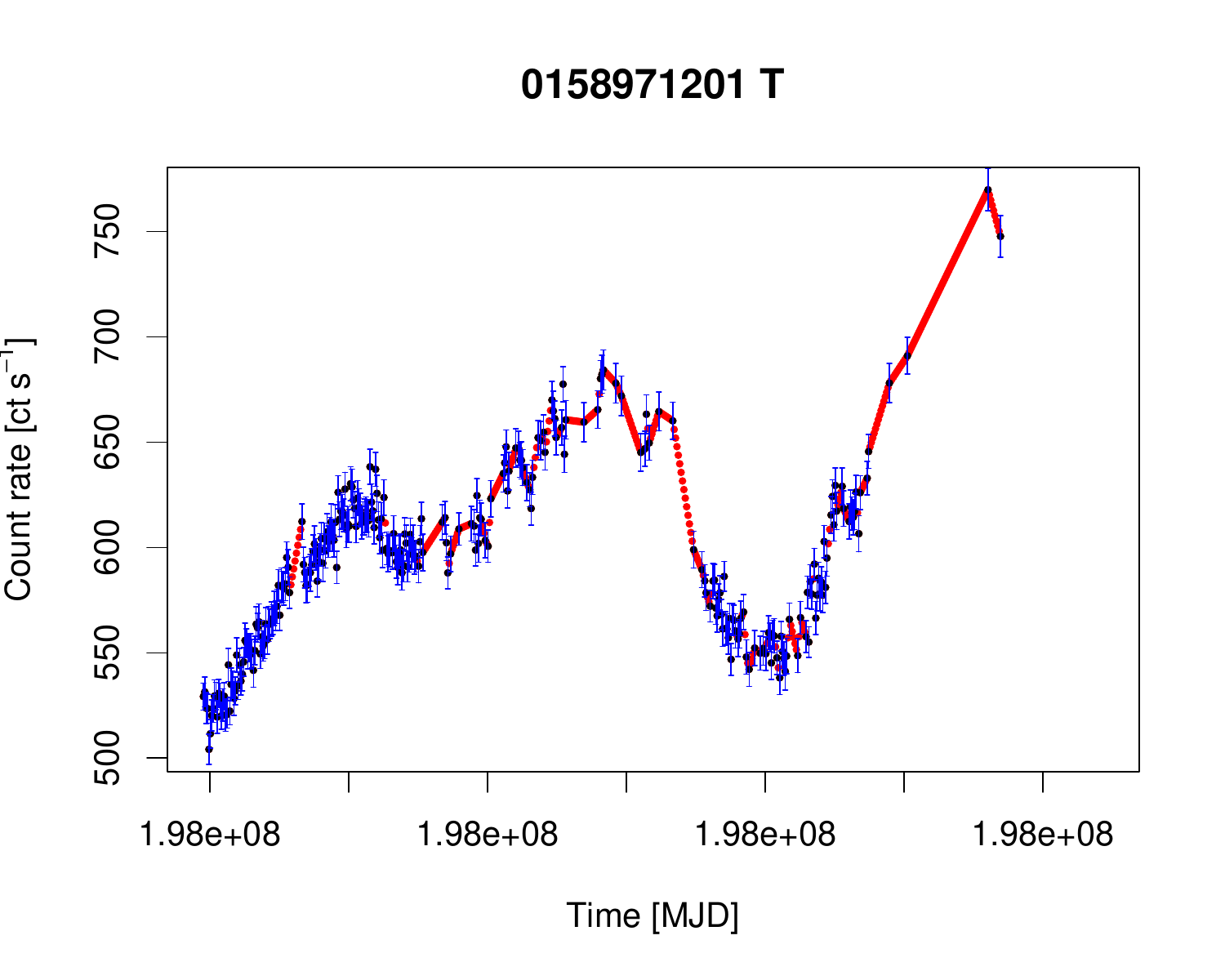}
	\end{minipage}
  \caption{Analyzed light curves 9-18 of Mrk~421, the red color denotes linearly interpolated data.  \label{LCS1}}
\end{figure*}

\begin{figure*}[!htb]  
		\centering 
	\begin{minipage}{.38\textwidth} 
		\centering 
		\includegraphics[width=0.891\linewidth]{ 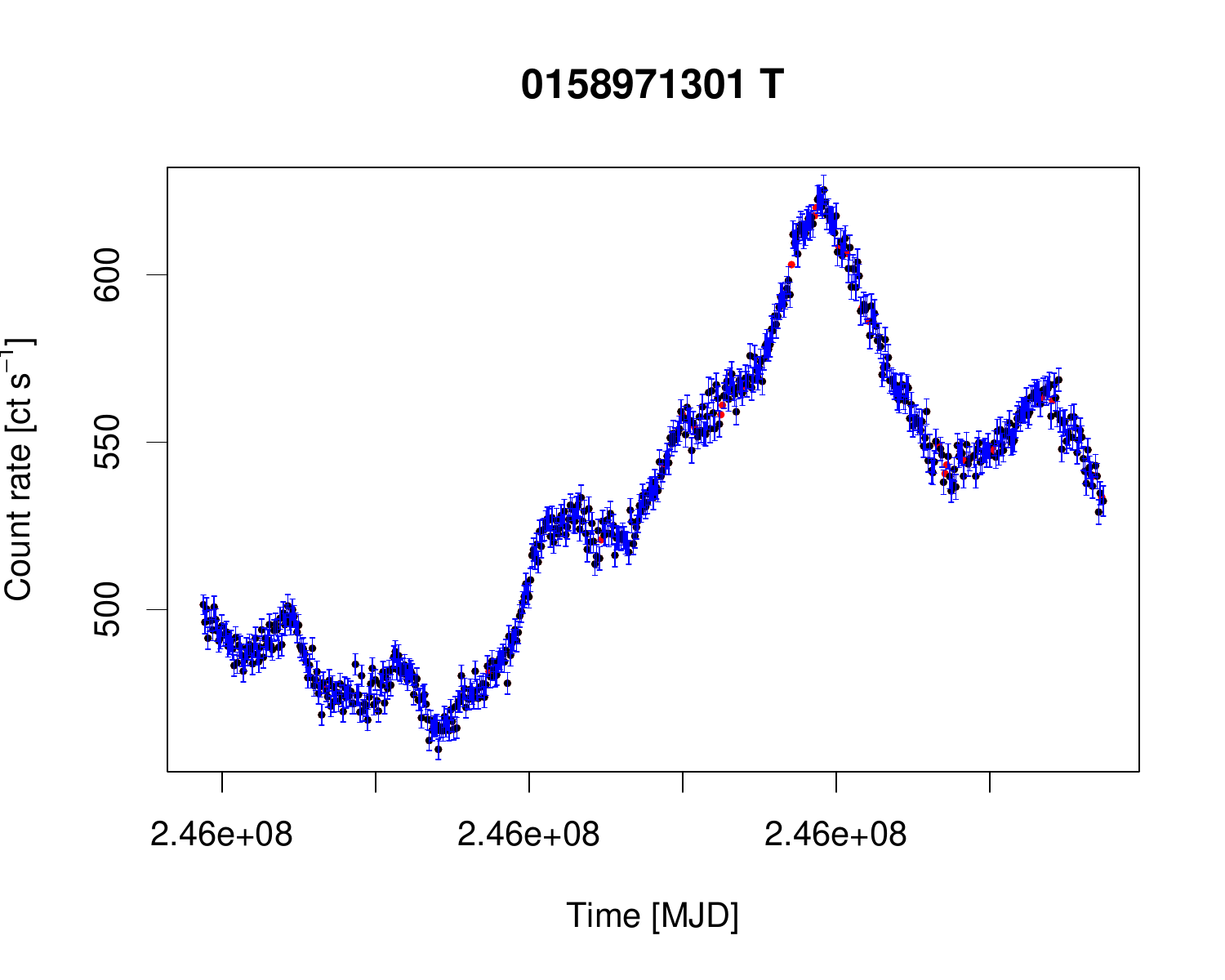}
	\end{minipage}
	\begin{minipage}{.38\textwidth} 
		\centering 
		\includegraphics[width=0.891\linewidth]{ 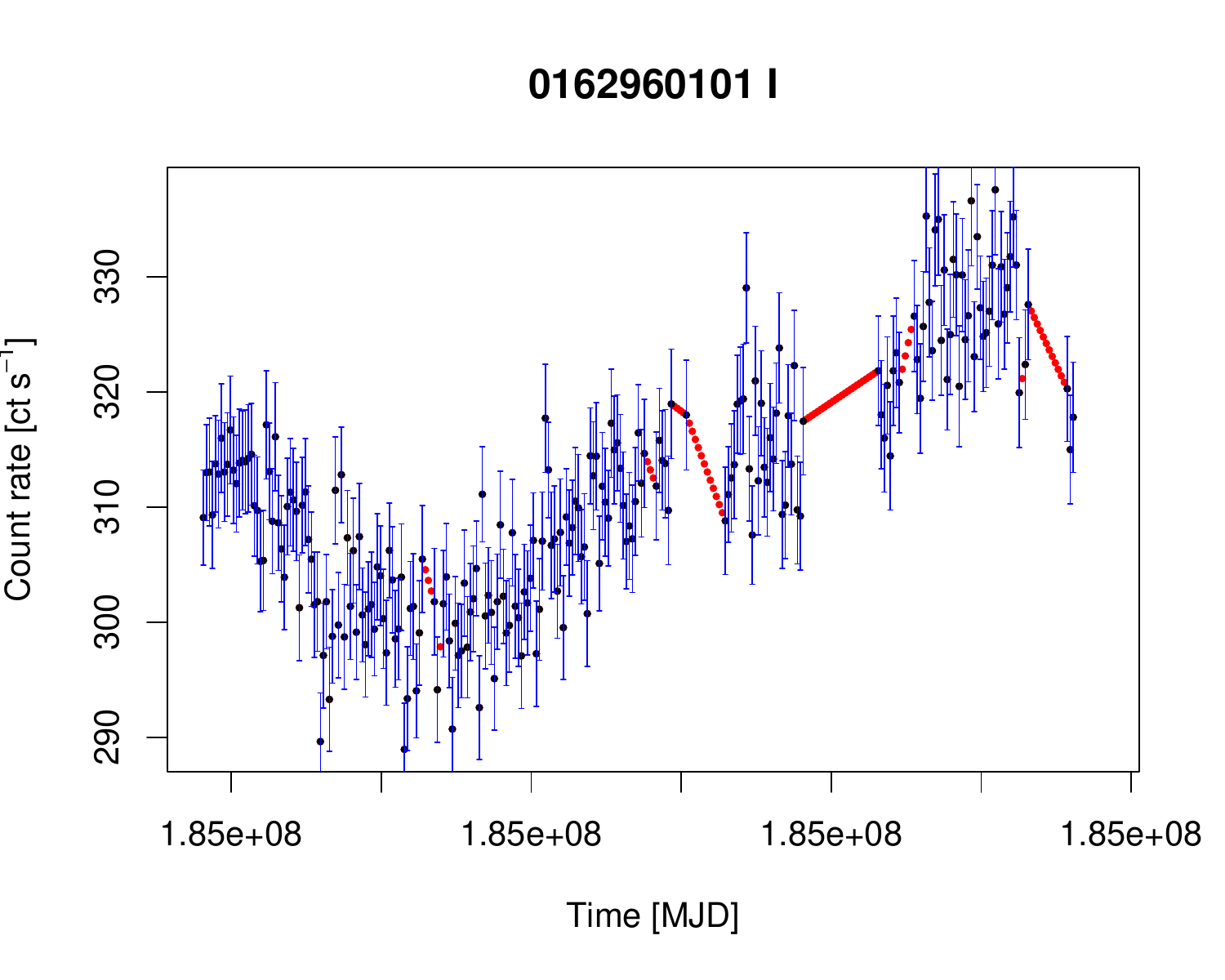}
	\end{minipage}
	\begin{minipage}{.38\textwidth} 
		\centering 
		\includegraphics[width=0.891\linewidth]{ 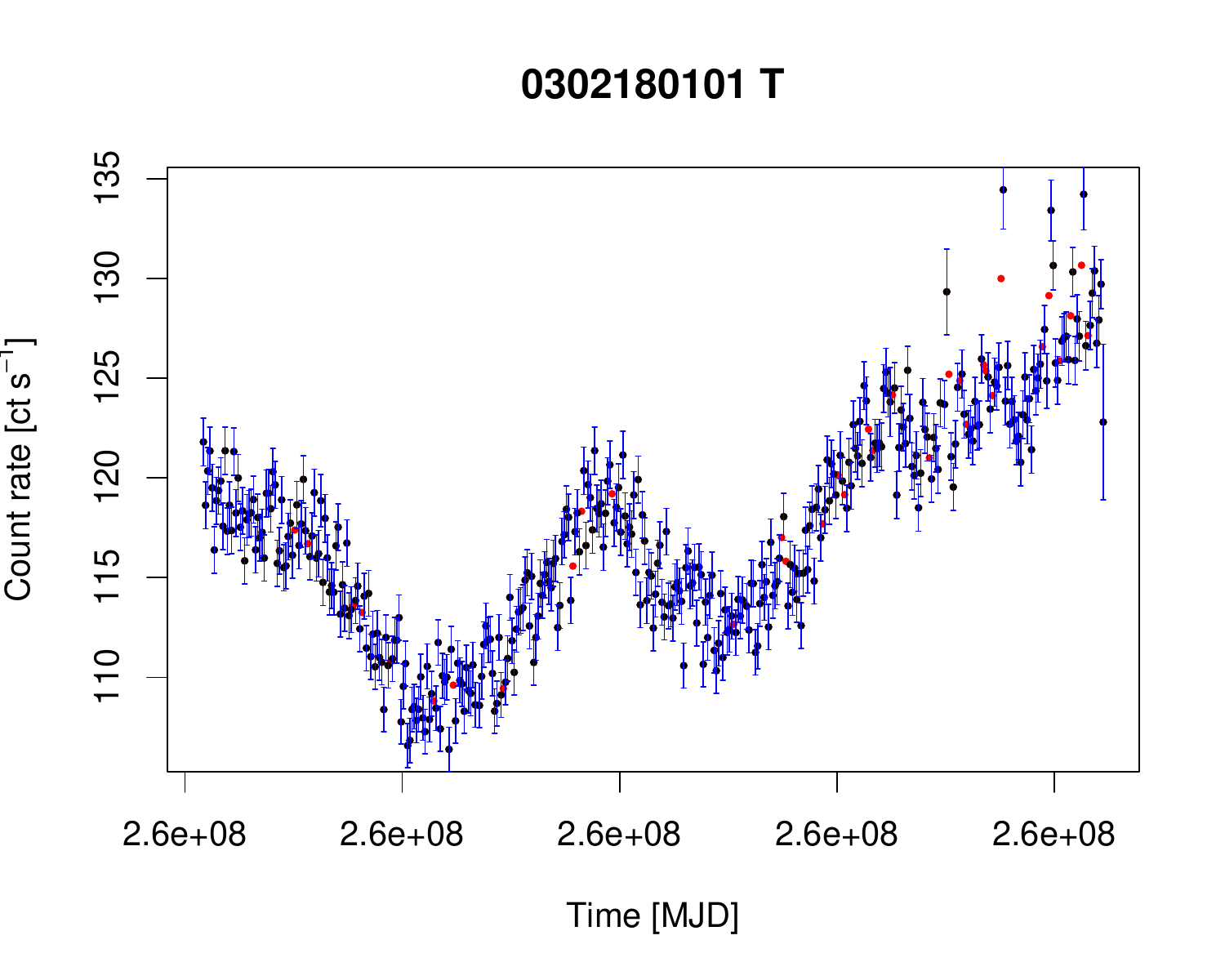}
	\end{minipage}
	\begin{minipage}{.38\textwidth} 
		\centering 
		\includegraphics[width=0.891\linewidth]{ 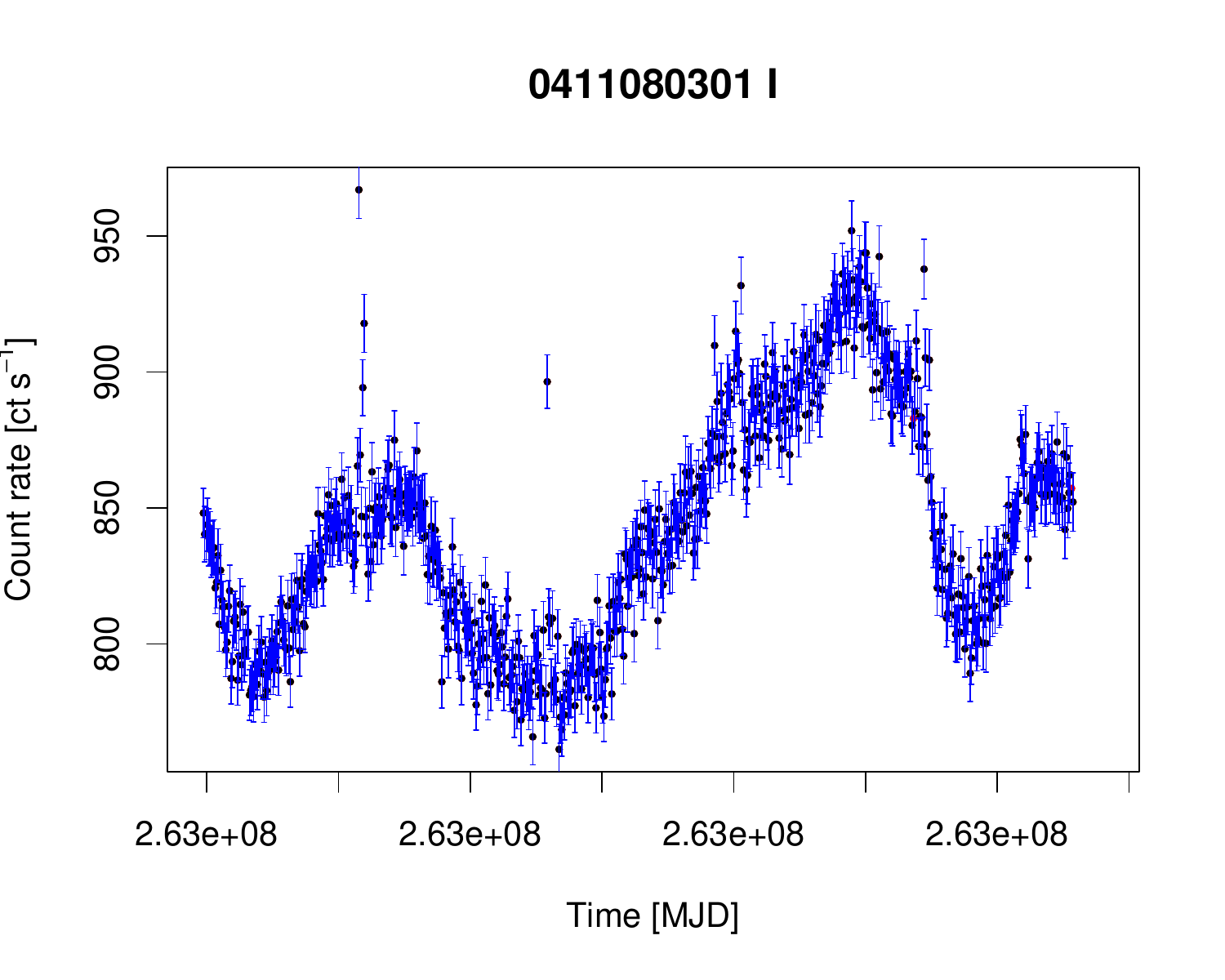}
	\end{minipage}
		\begin{minipage}{.38\textwidth} 
		\centering 
		\includegraphics[width=0.891\linewidth]{ 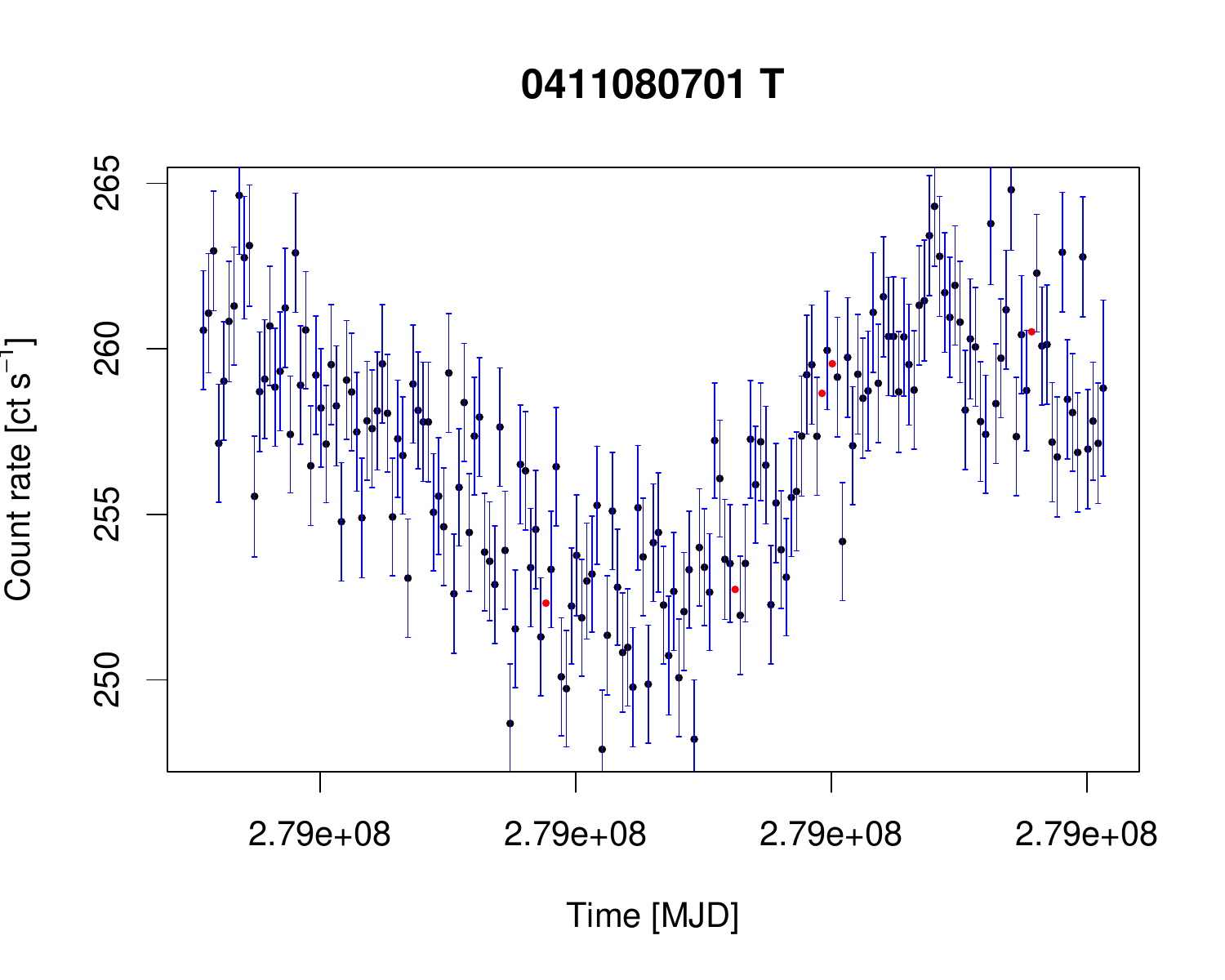}
	\end{minipage}
		\begin{minipage}{.38\textwidth} 
		\centering 
		\includegraphics[width=0.891\linewidth]{ 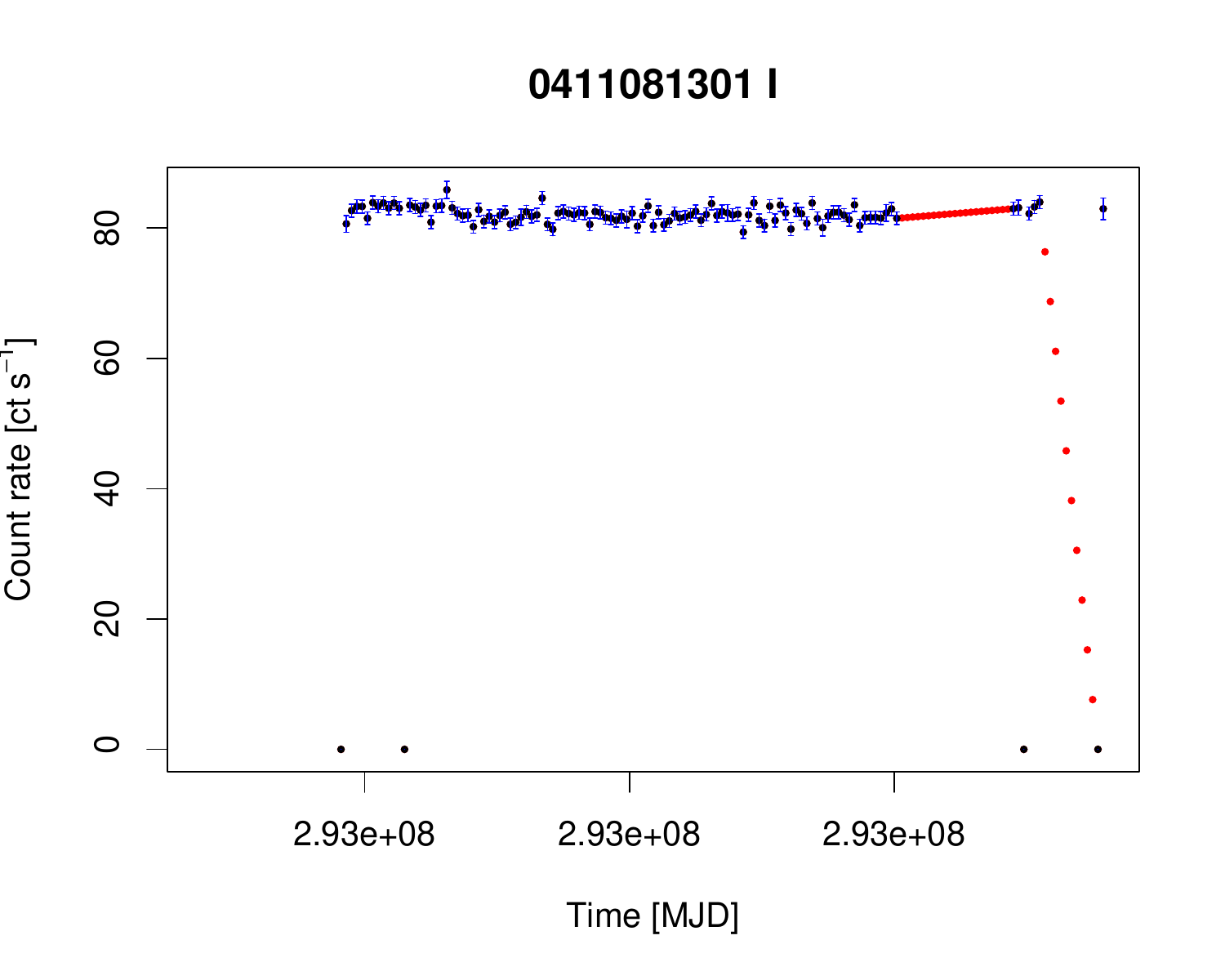}
	\end{minipage}
	\begin{minipage}{.38\textwidth} 
		\centering 
		\includegraphics[width=0.891\linewidth]{ 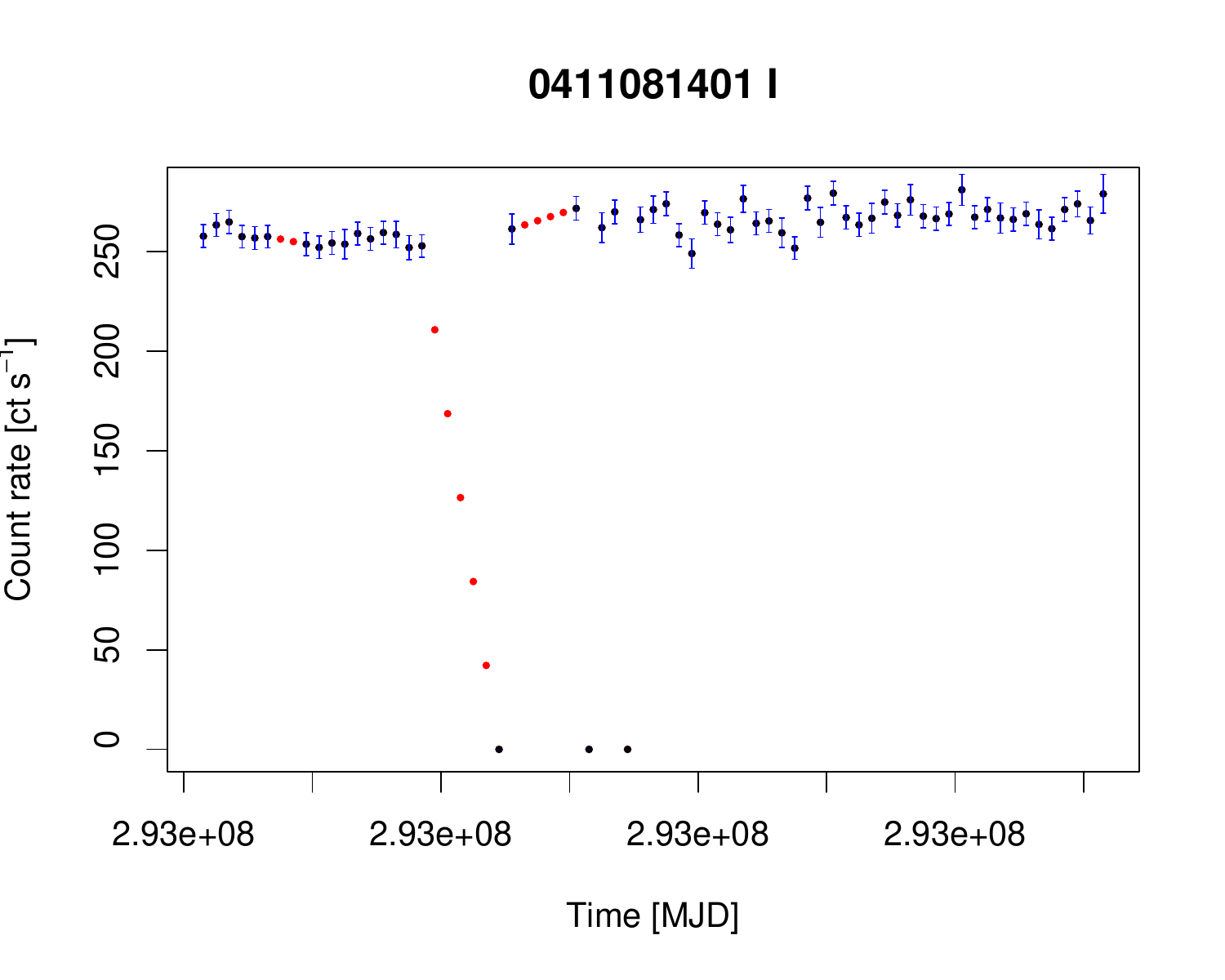}
	\end{minipage}
	\begin{minipage}{.38\textwidth} 
		\centering 
		\includegraphics[width=0.891\linewidth]{ 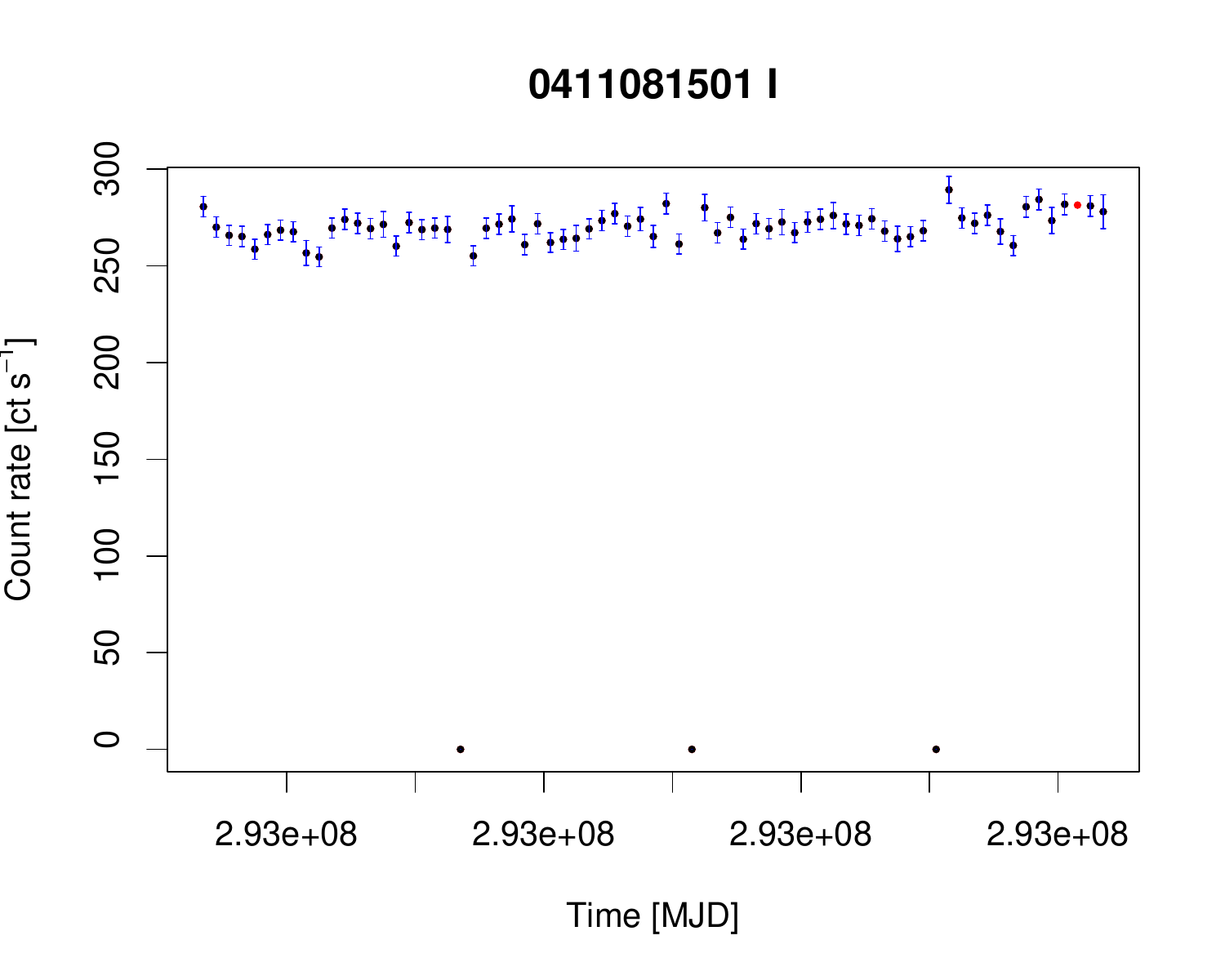}
	\end{minipage}
	\begin{minipage}{.38\textwidth} 
		\centering 
		\includegraphics[width=0.891\linewidth]{ 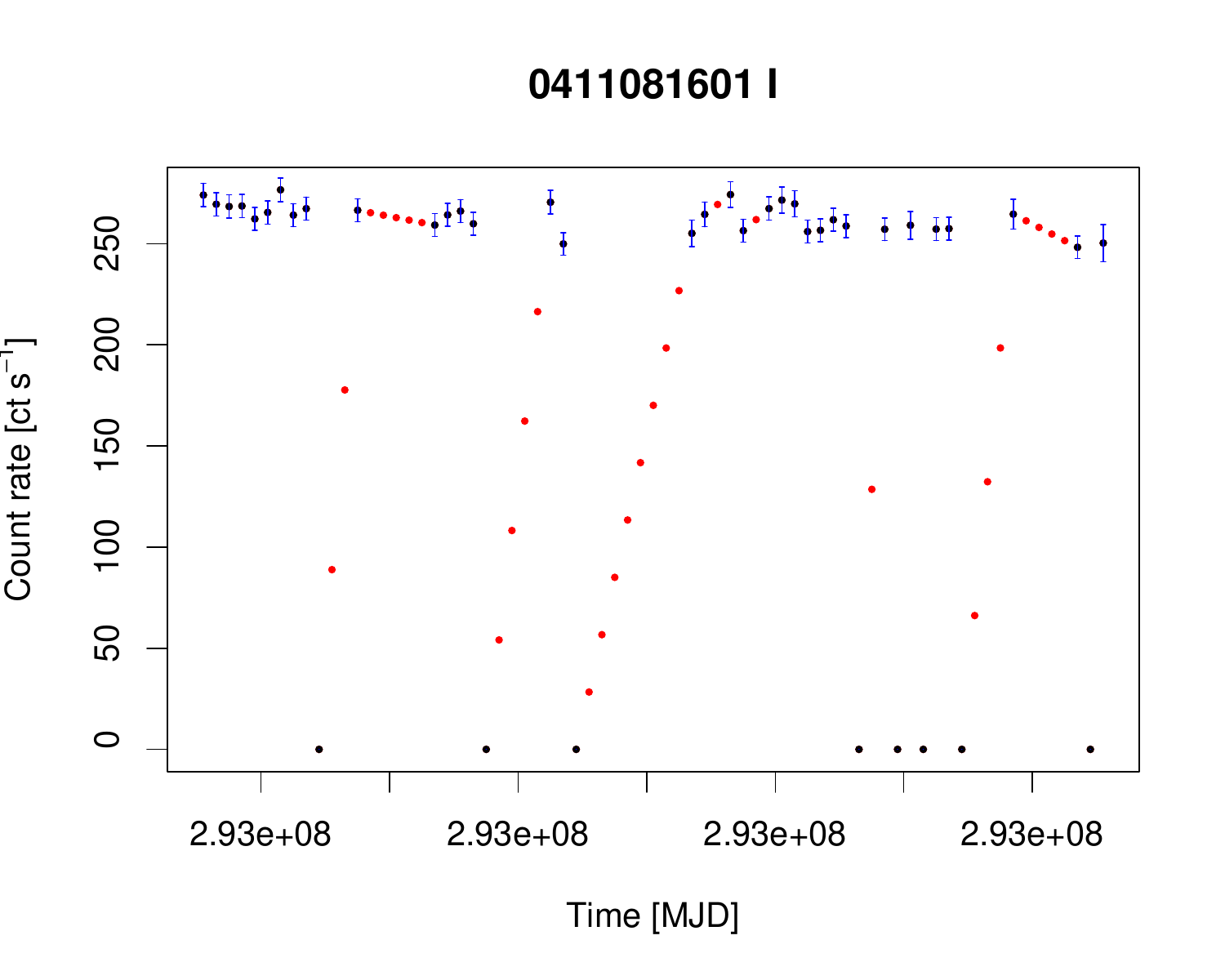}
	\end{minipage}
	\begin{minipage}{.38\textwidth} 
		\centering 
		\includegraphics[width=0.891\linewidth]{ 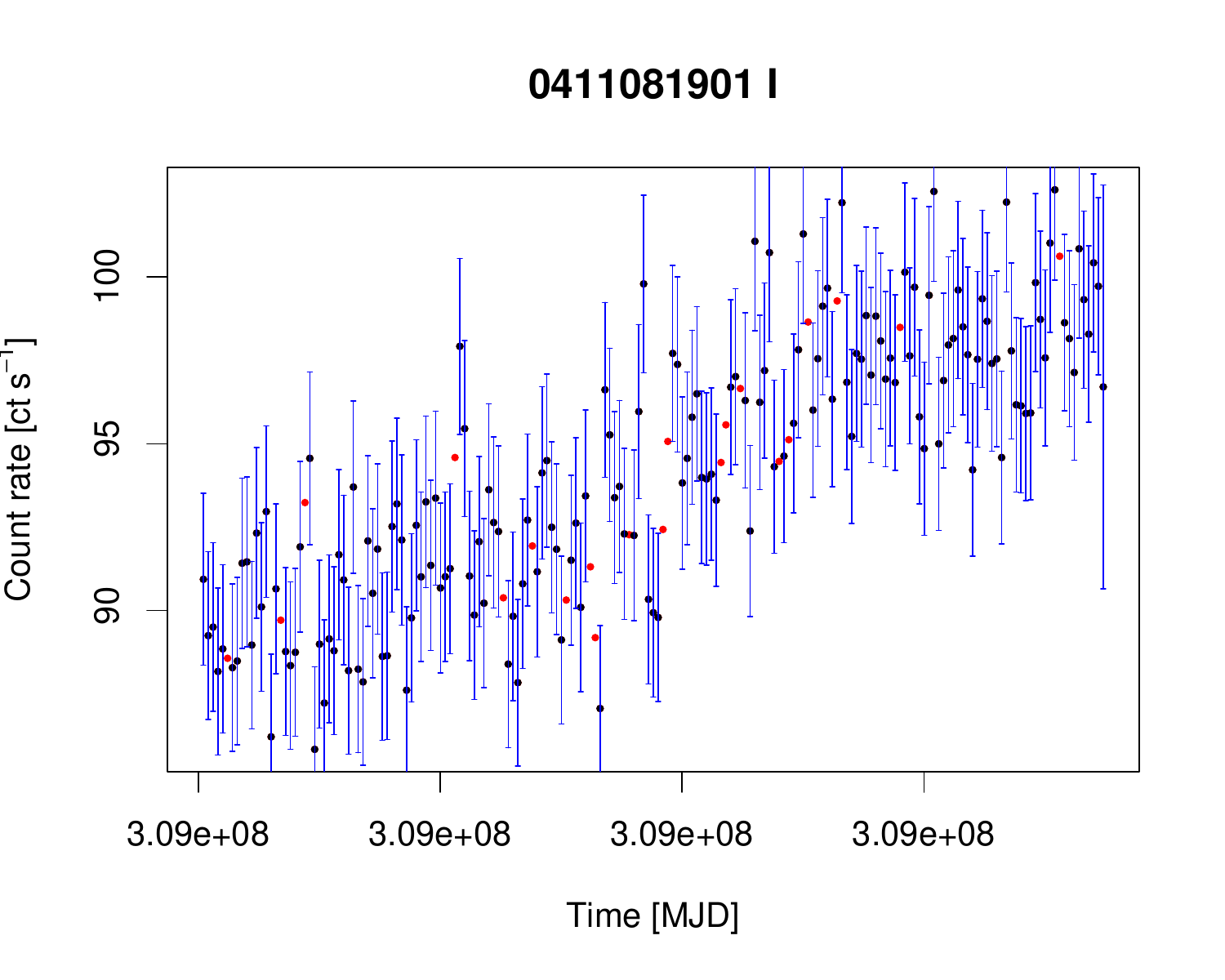}
	\end{minipage}
  \caption{ Analyzed light curves 19-28 of Mrk~421, the red color denotes linearly interpolated data.   \label{LCS2}}
\end{figure*}

\begin{figure*}[!htb] 
		\centering 
	\begin{minipage}{.38\textwidth} 
		\centering 
		\includegraphics[width=0.891\linewidth]{ 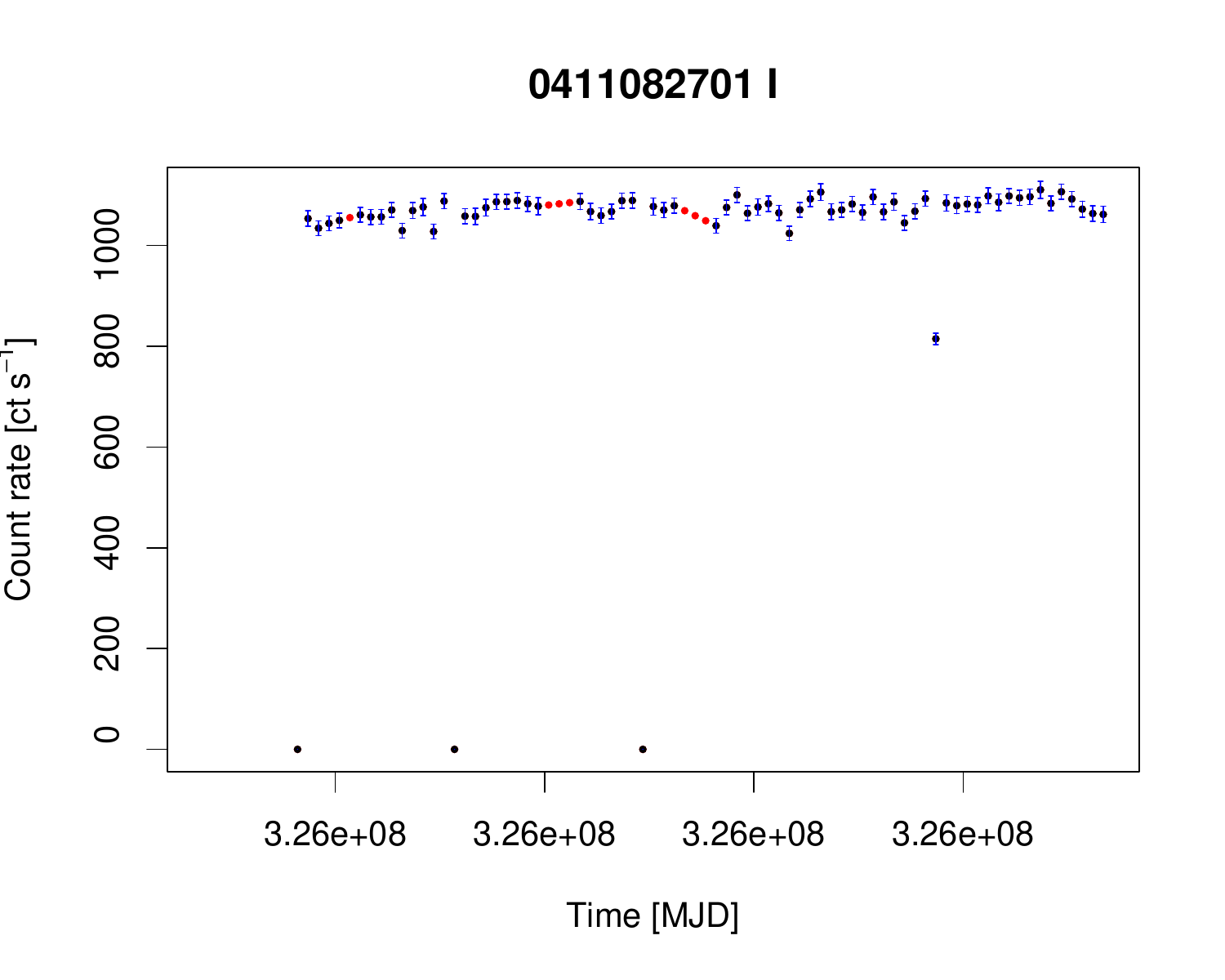}
	\end{minipage}
	\begin{minipage}{.38\textwidth} 
		\centering 
		\includegraphics[width=0.891\linewidth]{ 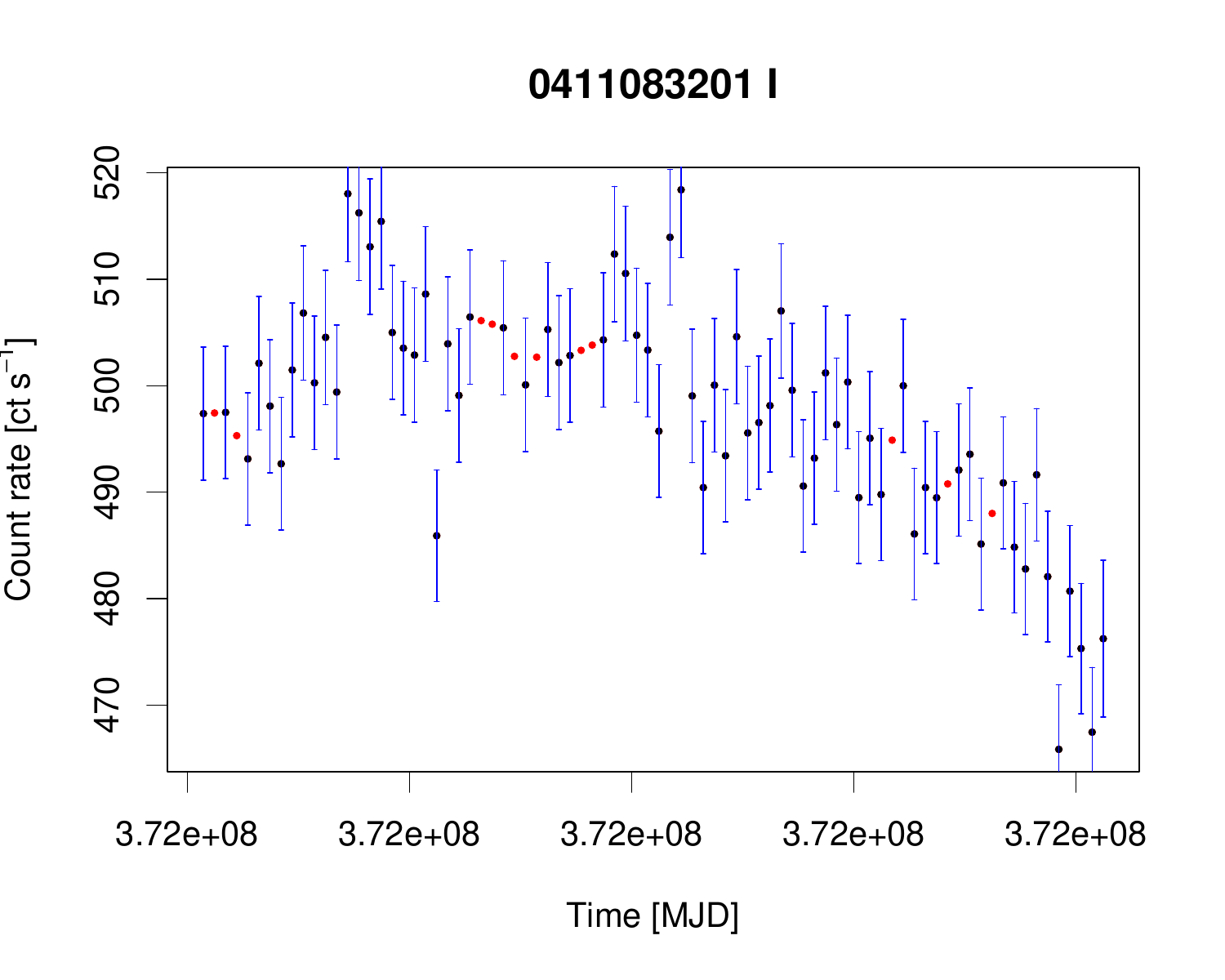}
	\end{minipage}
	\begin{minipage}{.38\textwidth} 
		\centering 
		\includegraphics[width=0.891\linewidth]{ 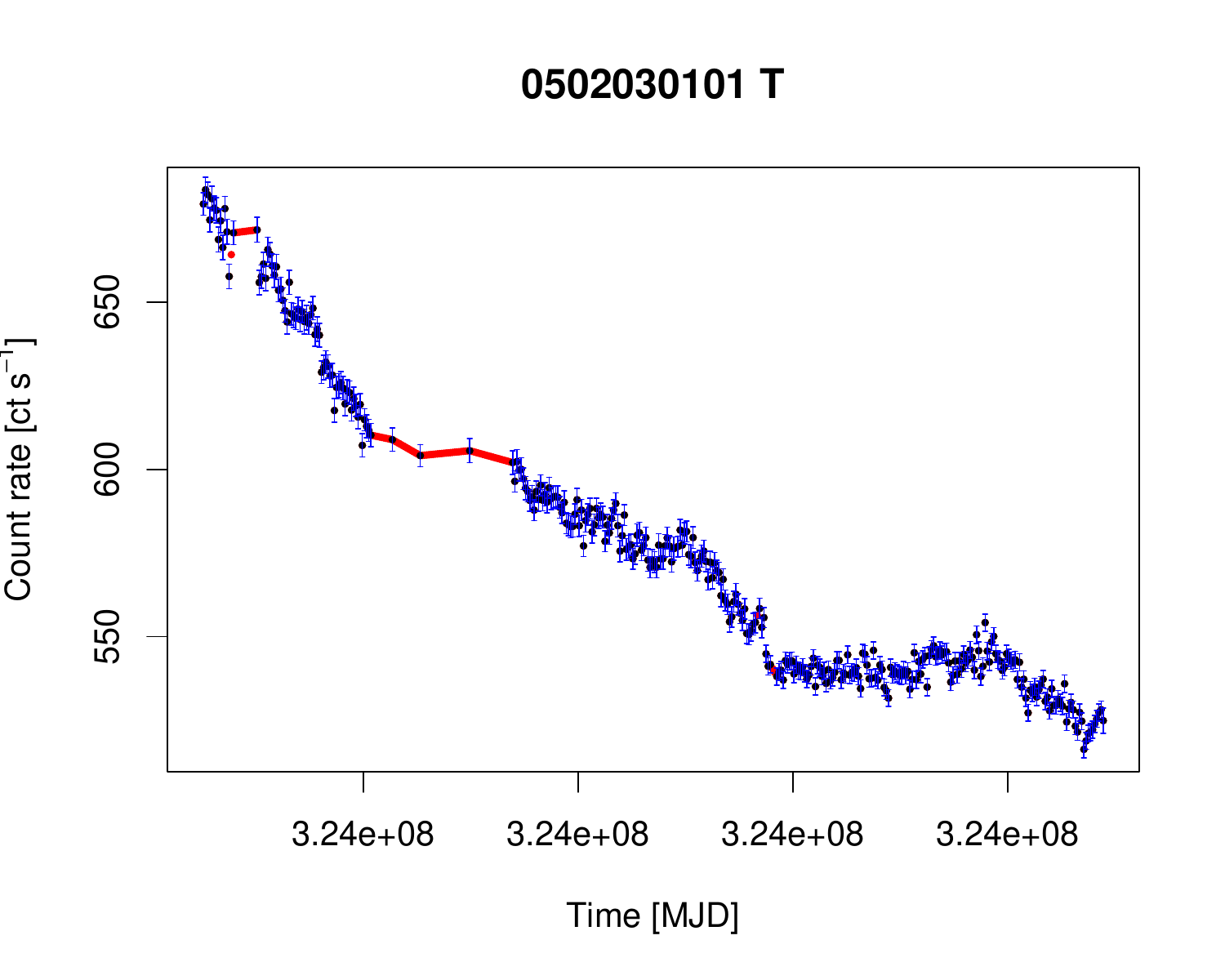}
	\end{minipage}
	\begin{minipage}{.38\textwidth} 
		\centering 
		\includegraphics[width=0.891\linewidth]{ 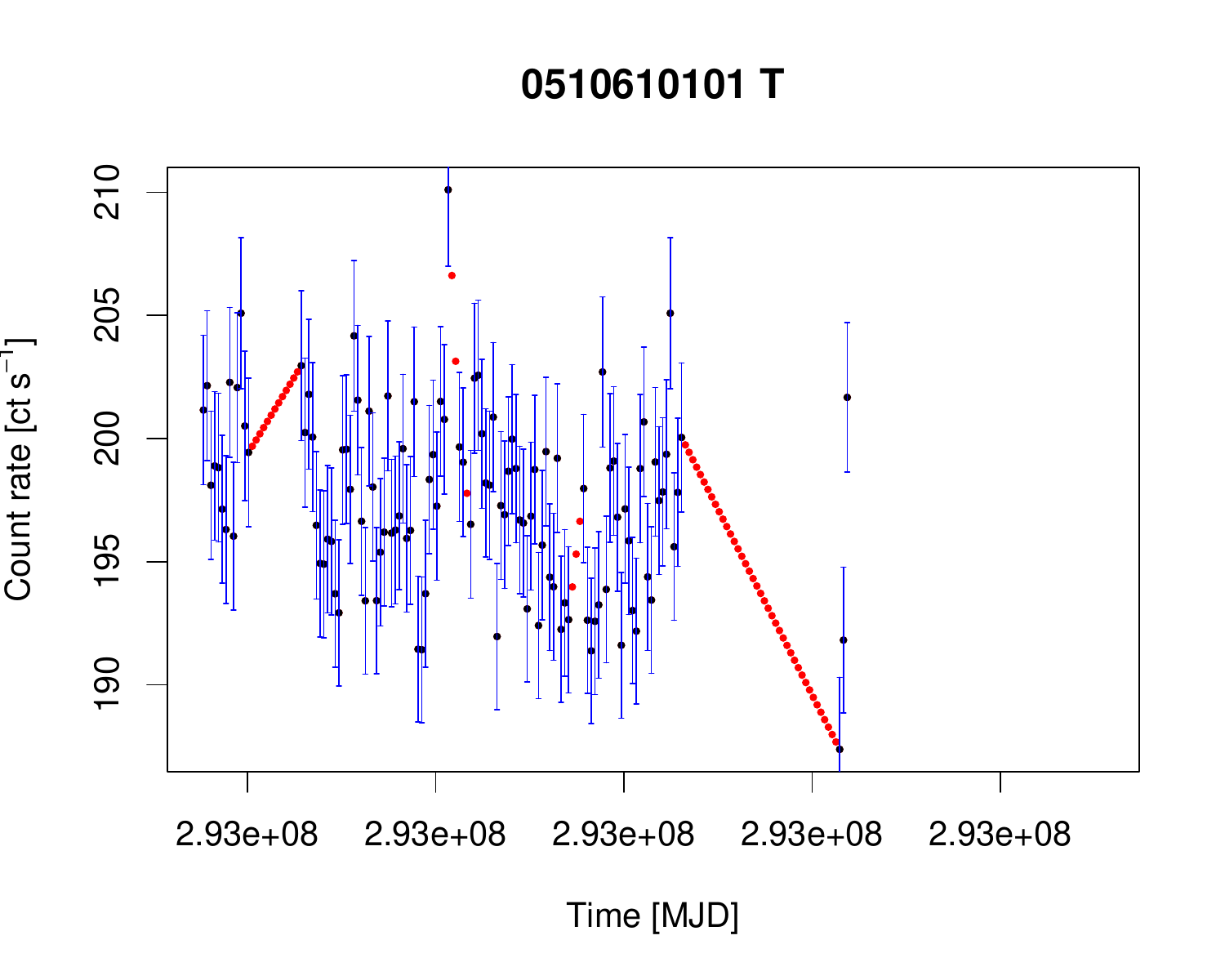}
	\end{minipage}
		\begin{minipage}{.38\textwidth} 
		\centering 
		\includegraphics[width=0.891\linewidth]{ 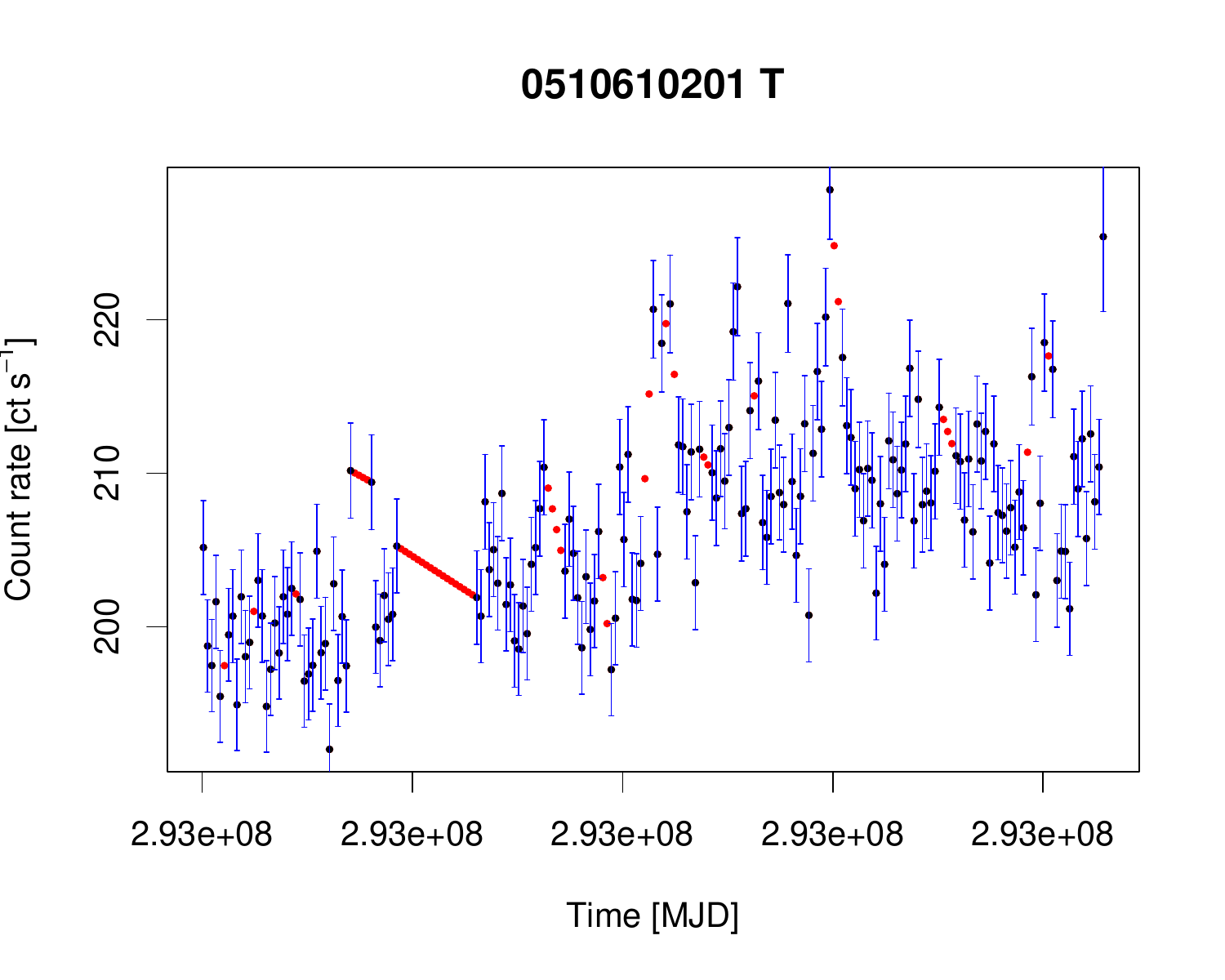}
	\end{minipage}
		\begin{minipage}{.38\textwidth} 
		\centering 
		\includegraphics[width=0.891\linewidth]{ 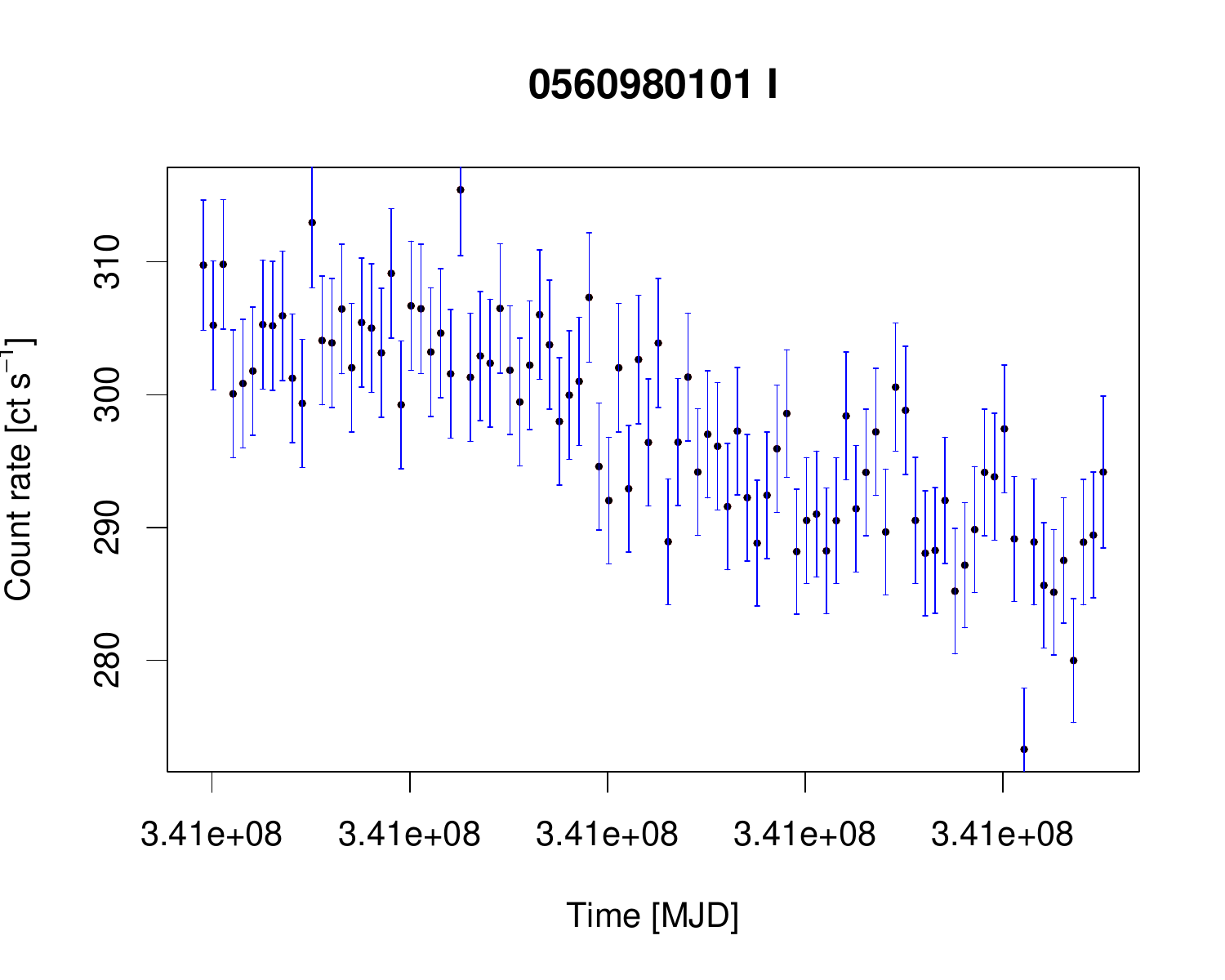}
	\end{minipage}
	\begin{minipage}{.38\textwidth} 
		\centering 
		\includegraphics[width=0.891\linewidth]{ 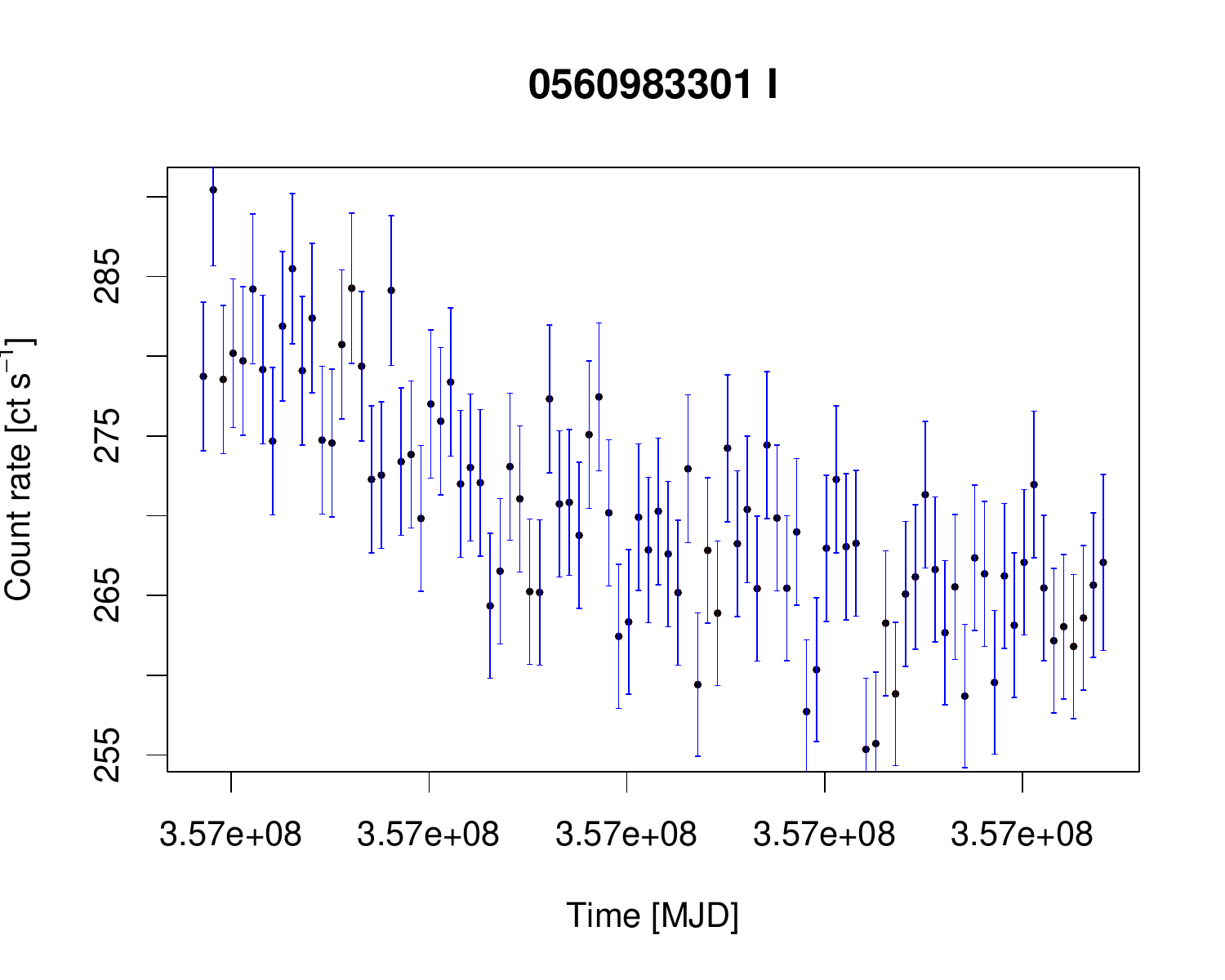}
	\end{minipage}
	\begin{minipage}{.38\textwidth} 
		\centering 
		\includegraphics[width=0.891\linewidth]{ 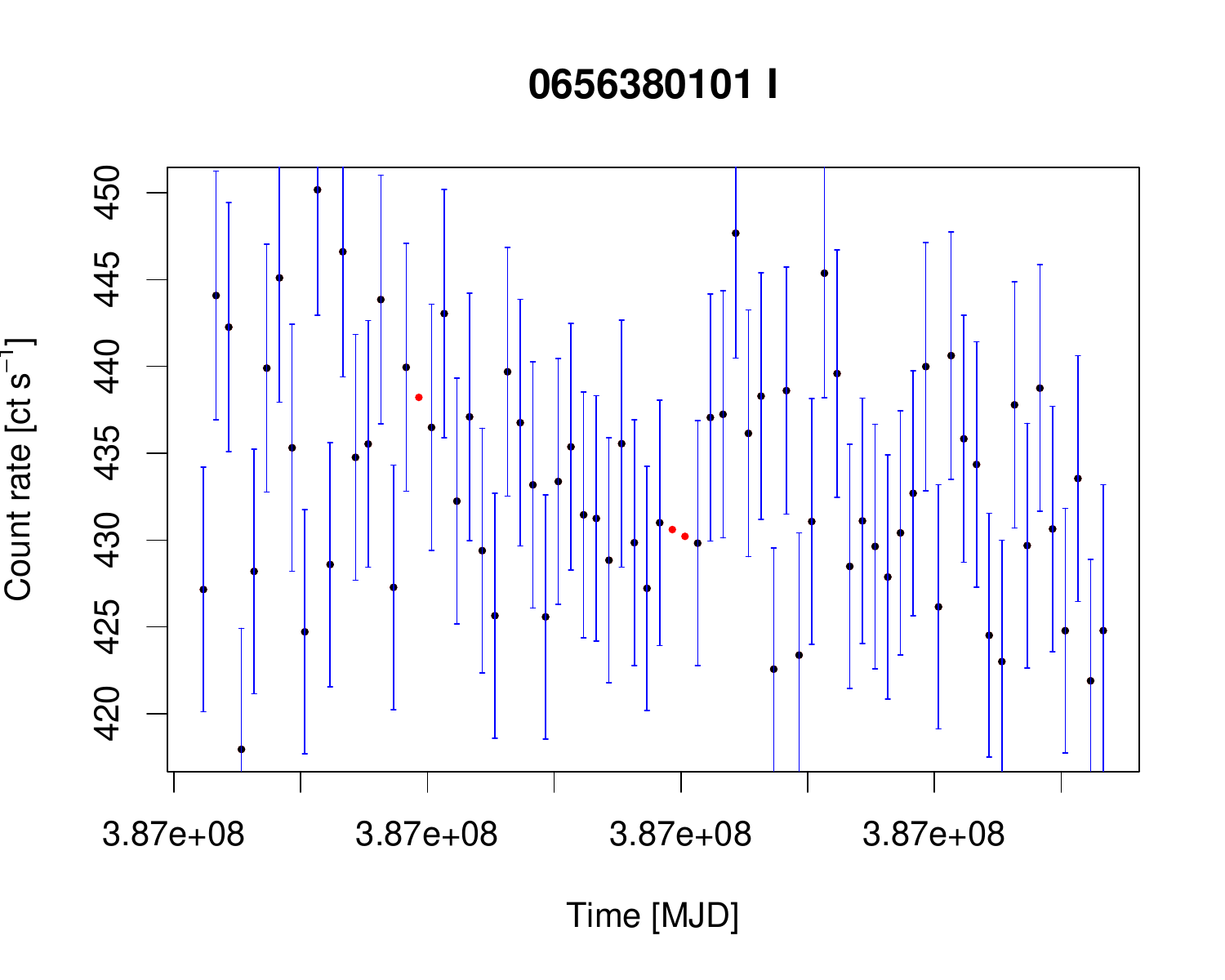}
	\end{minipage}
	\begin{minipage}{.38\textwidth} 
		\centering 
		\includegraphics[width=0.891\linewidth]{ 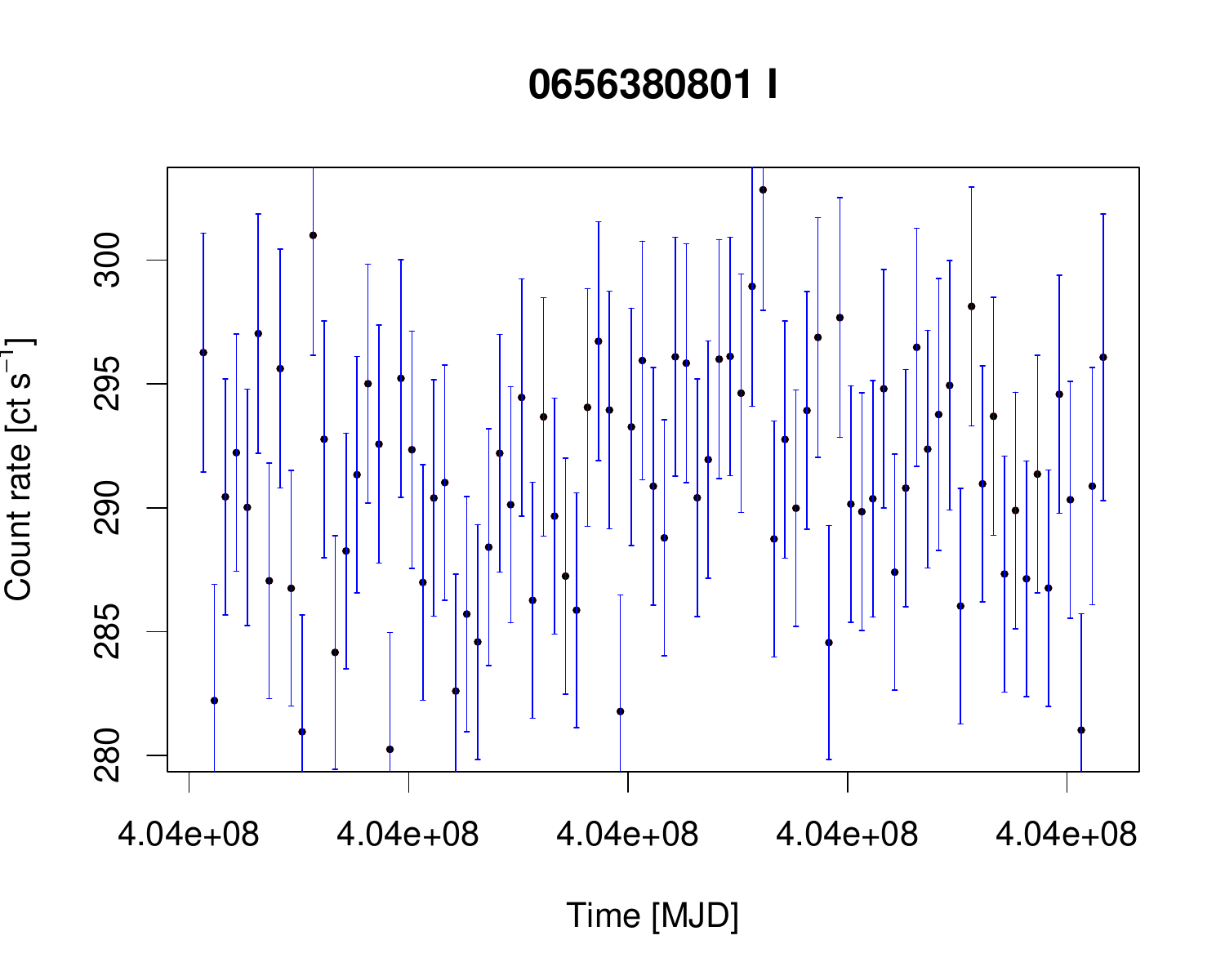}
	\end{minipage}
	\begin{minipage}{.38\textwidth} 
		\centering 
		\includegraphics[width=0.891\linewidth]{ 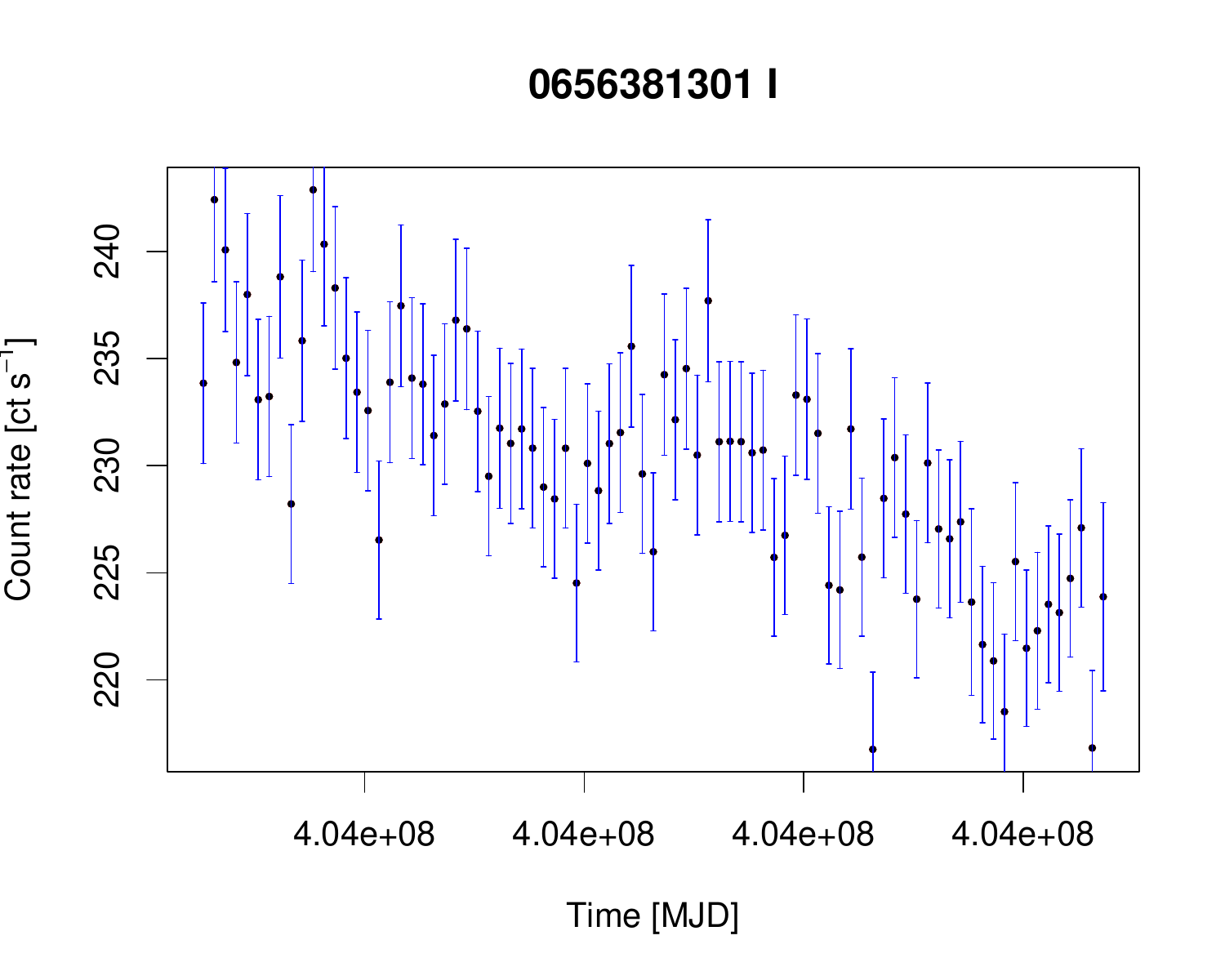}
	\end{minipage}
  \caption{ Analysed light curves 29-38 of Mrk~421, the red color denotes linearly interpolated data.   \label{LCS3}}
\end{figure*}

\begin{figure*}[!htb] 
		\centering 
	\begin{minipage}{.38\textwidth} 
		\centering 
		\includegraphics[width=0.891\linewidth]{ 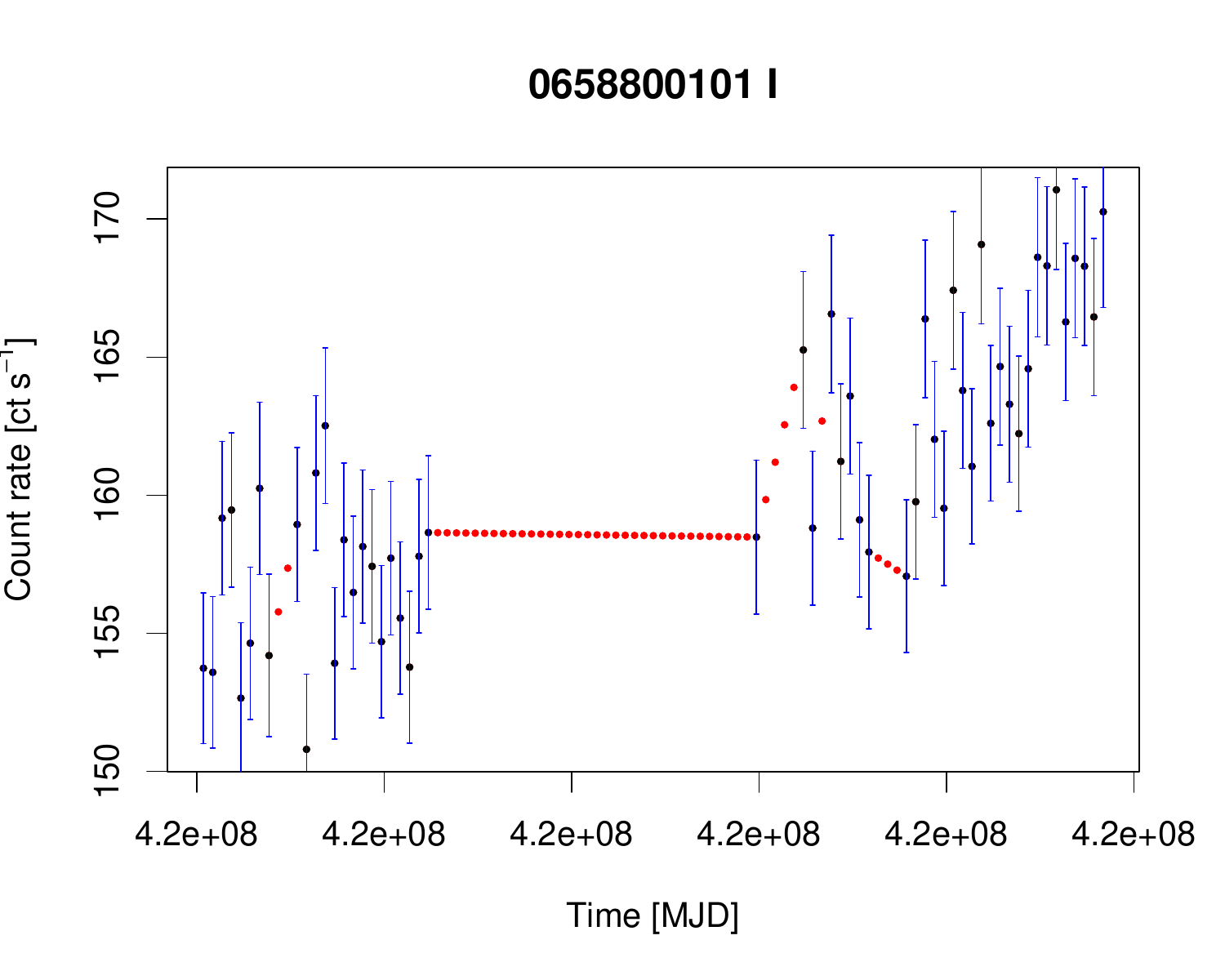}
	\end{minipage}
	\begin{minipage}{.38\textwidth} 
		\centering 
		\includegraphics[width=0.891\linewidth]{ 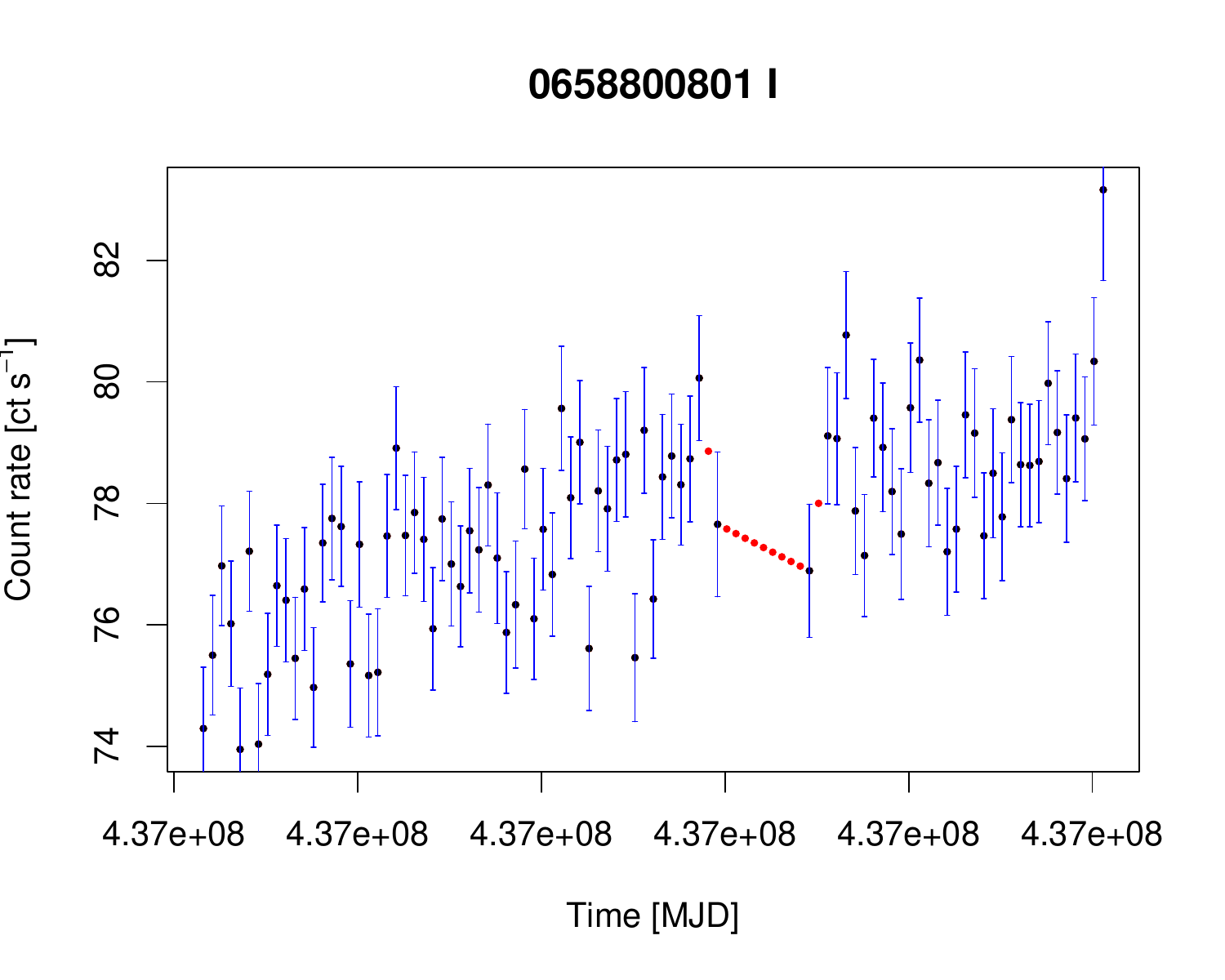}
	\end{minipage}
	\begin{minipage}{.38\textwidth} 
		\centering 
		\includegraphics[width=0.891\linewidth]{ 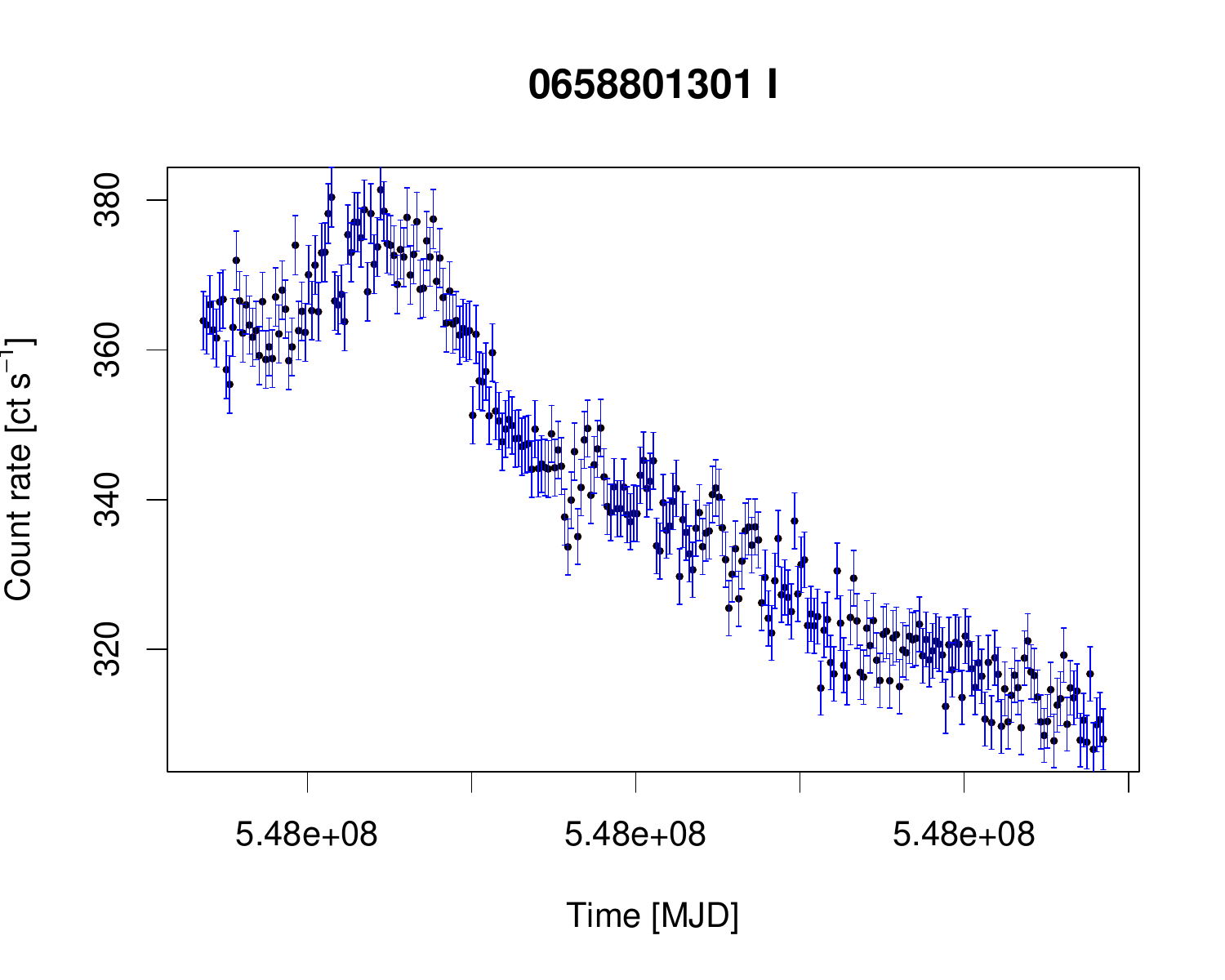}
	\end{minipage}
	\begin{minipage}{.38\textwidth} 
		\centering 
		\includegraphics[width=0.891\linewidth]{ 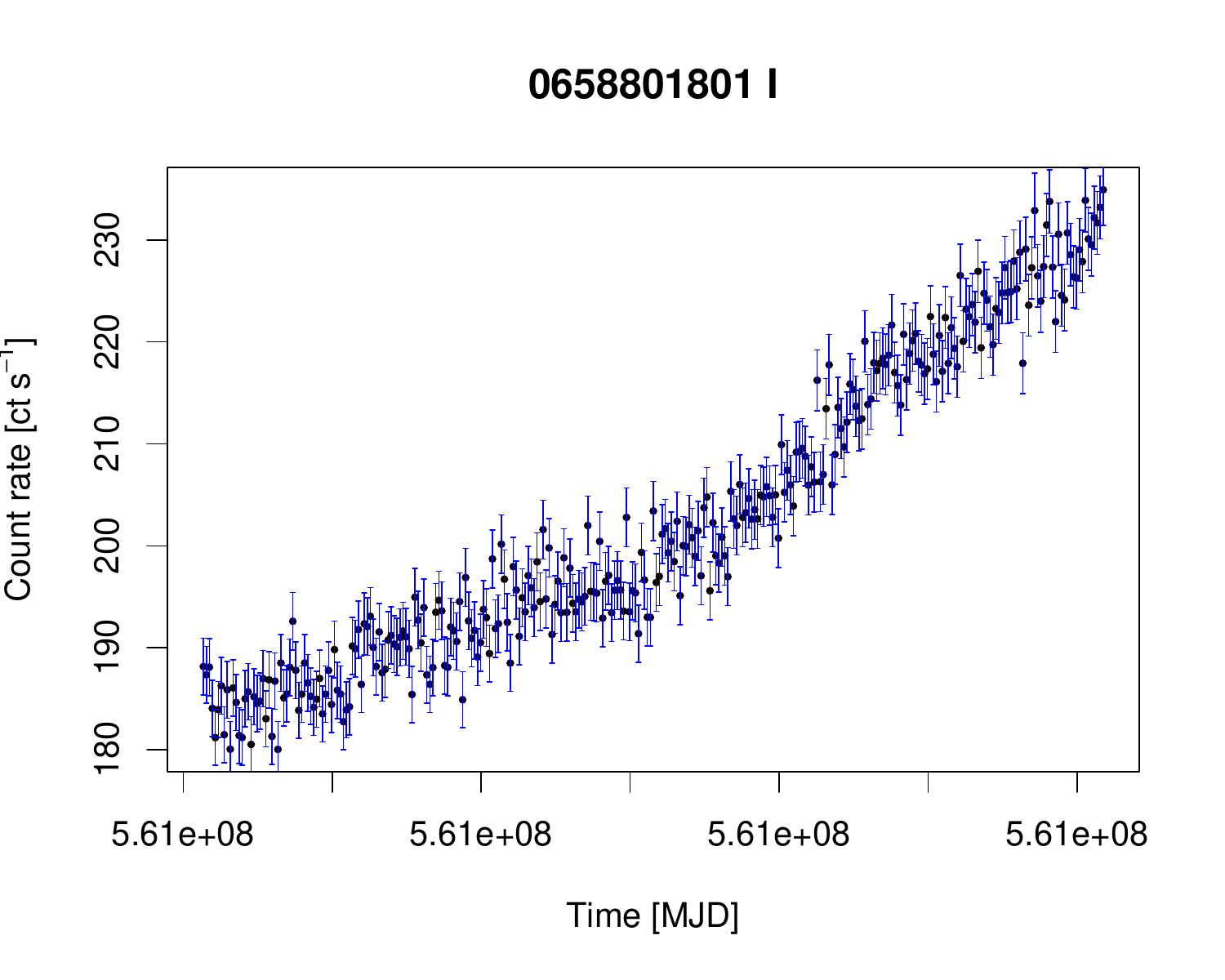}
	\end{minipage}
		\begin{minipage}{.38\textwidth} 
		\centering 
		\includegraphics[width=0.891\linewidth]{ 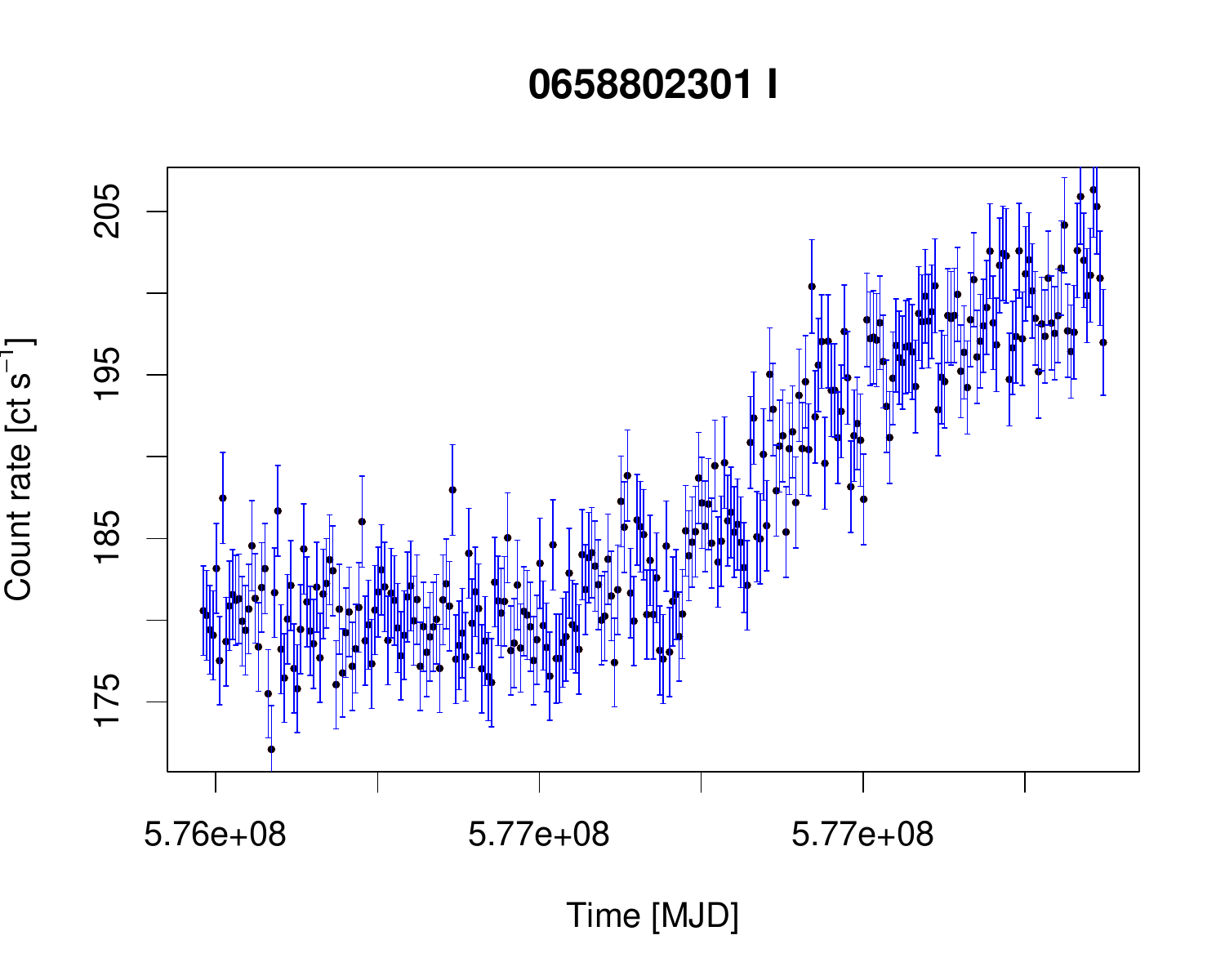}
	\end{minipage}
		\begin{minipage}{.38\textwidth} 
		\centering 
		\includegraphics[width=0.891\linewidth]{ 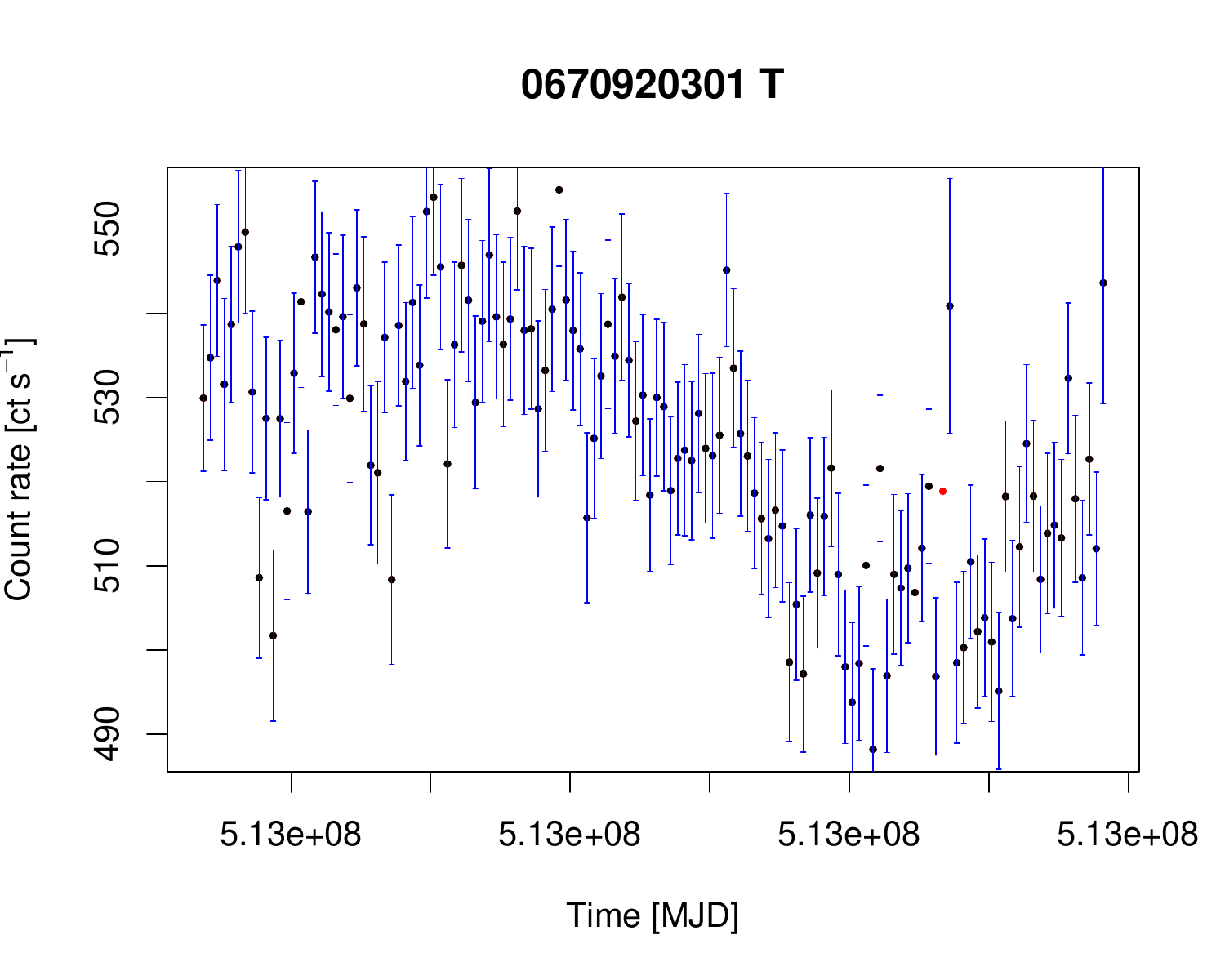}
	\end{minipage}
		\begin{minipage}{.38\textwidth} 
		\centering 
		\includegraphics[width=0.891\linewidth]{ 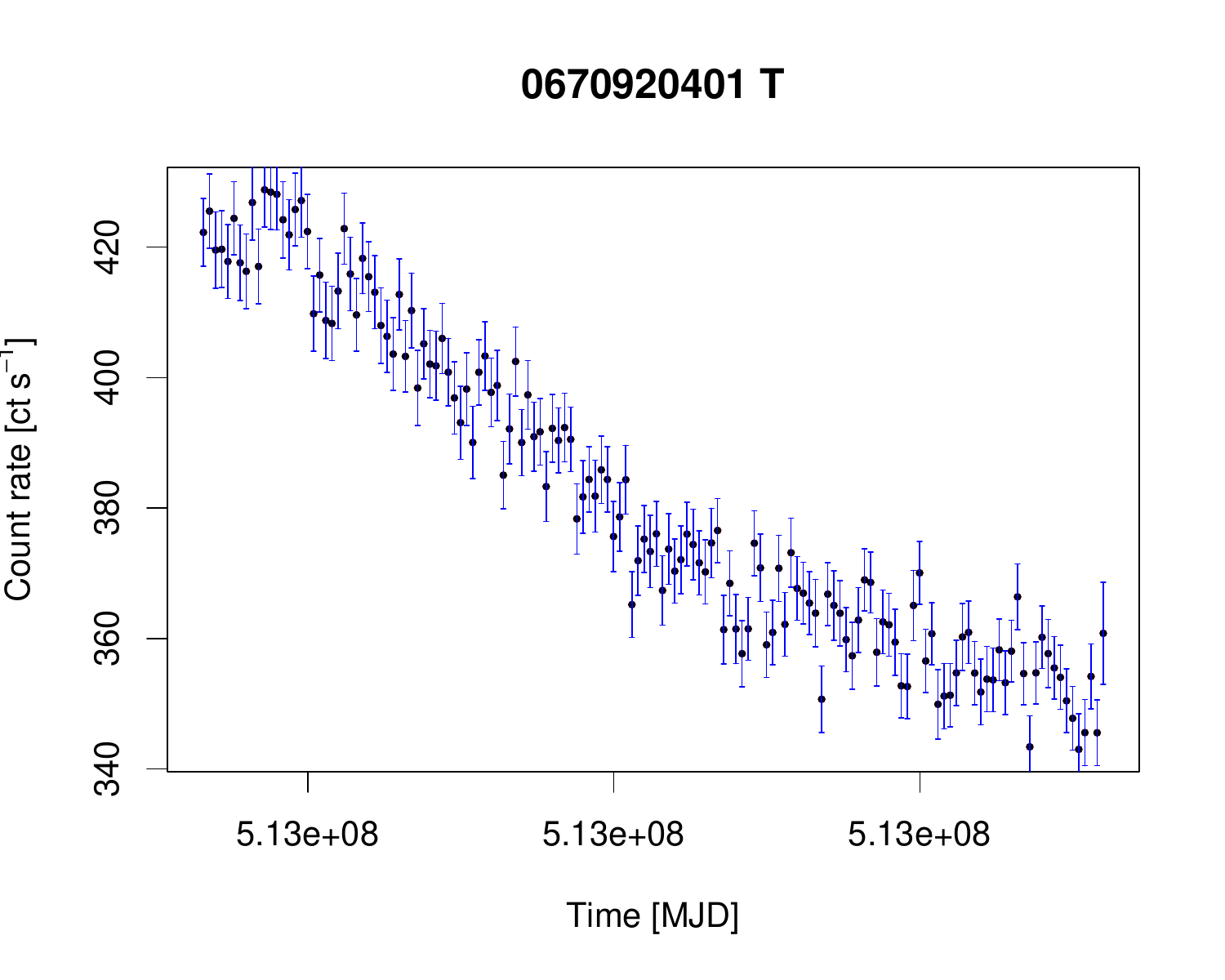}
	\end{minipage}
		\begin{minipage}{.38\textwidth} 
		\centering 
		\includegraphics[width=0.891\linewidth]{ 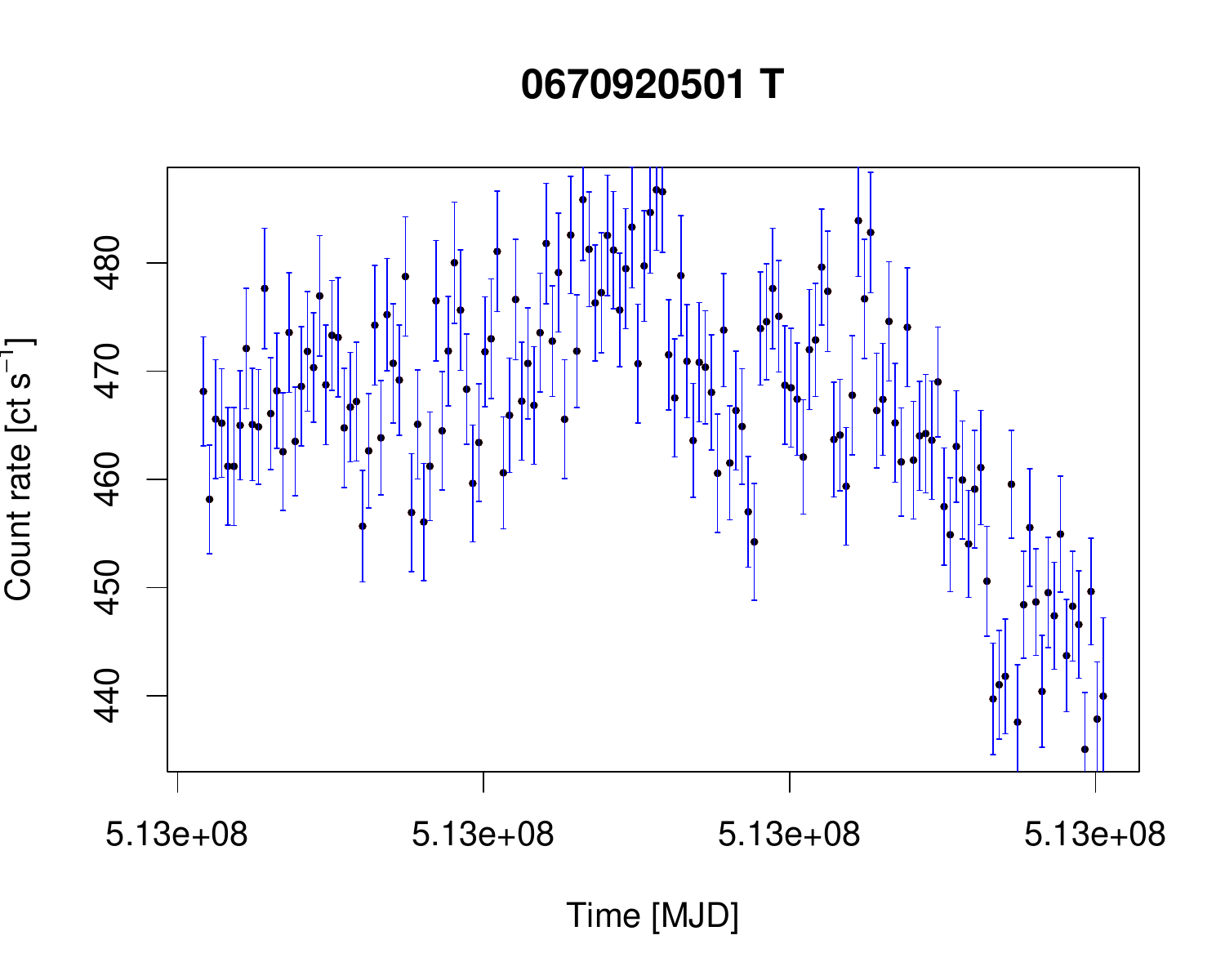}
	\end{minipage}
		\begin{minipage}{.38\textwidth} 
		\centering 
		\includegraphics[width=0.891\linewidth]{ 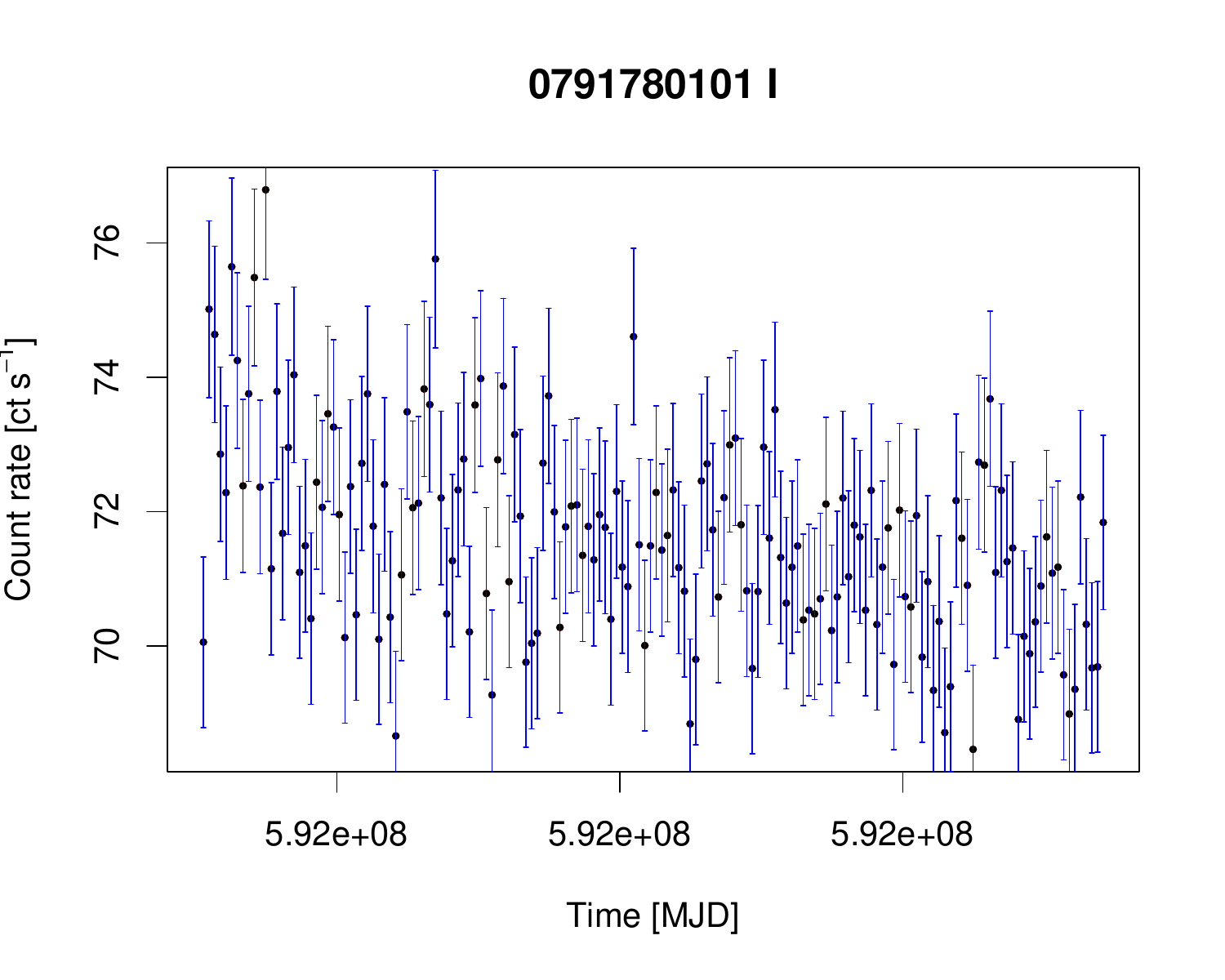}
	\end{minipage}
		\begin{minipage}{.38\textwidth} 
		\centering 
		\includegraphics[width=0.891\linewidth]{ 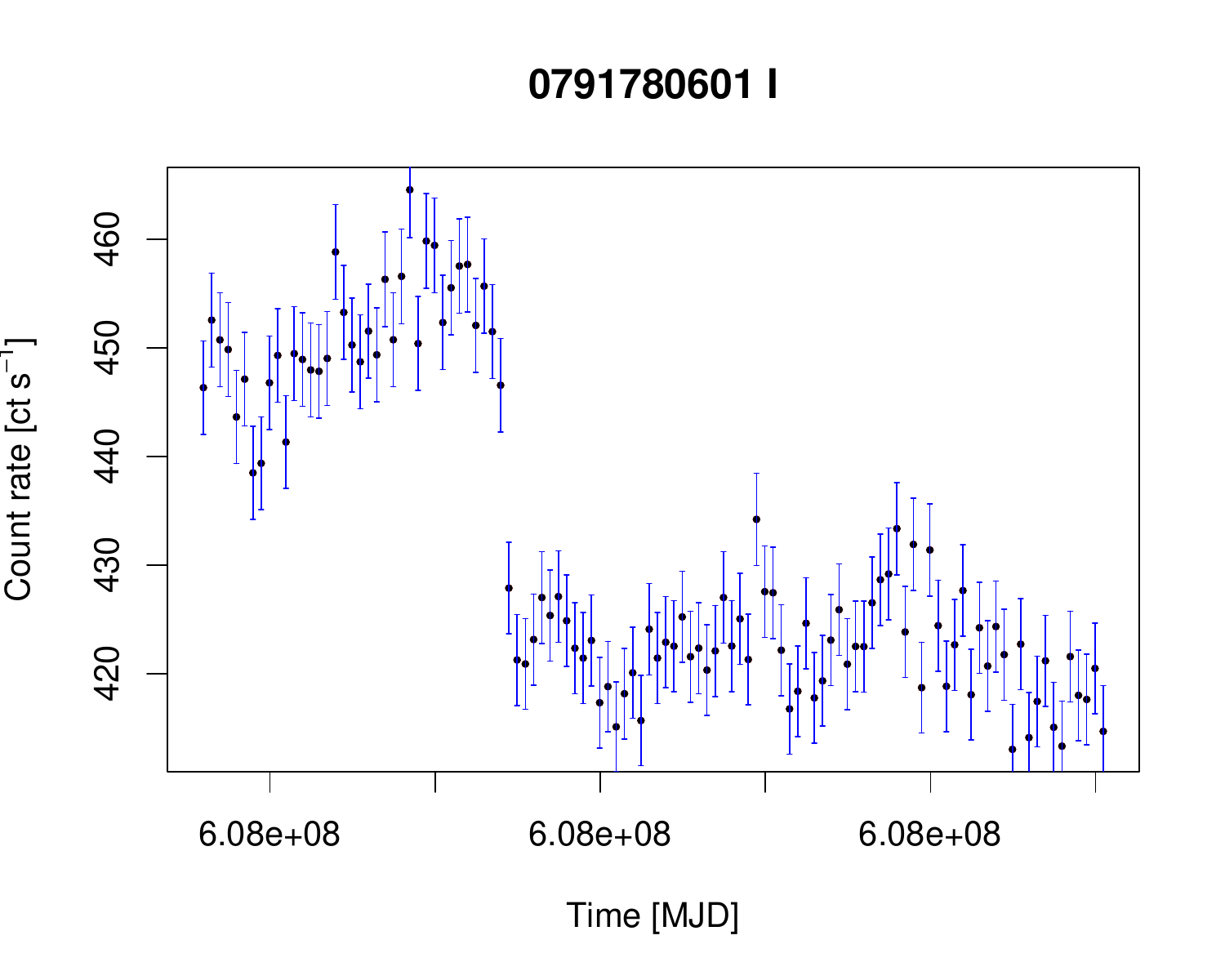}
	\end{minipage}
\caption{Analyzed light curves 39–48 of Mrk 421. The red color denotes linearly interpolated data.}
\label{LCS4}
\end{figure*}

\begin{figure*}[!htb] 
		\centering 
	\begin{minipage}{.38\textwidth} 
		\centering 
		\includegraphics[width=0.891\linewidth]{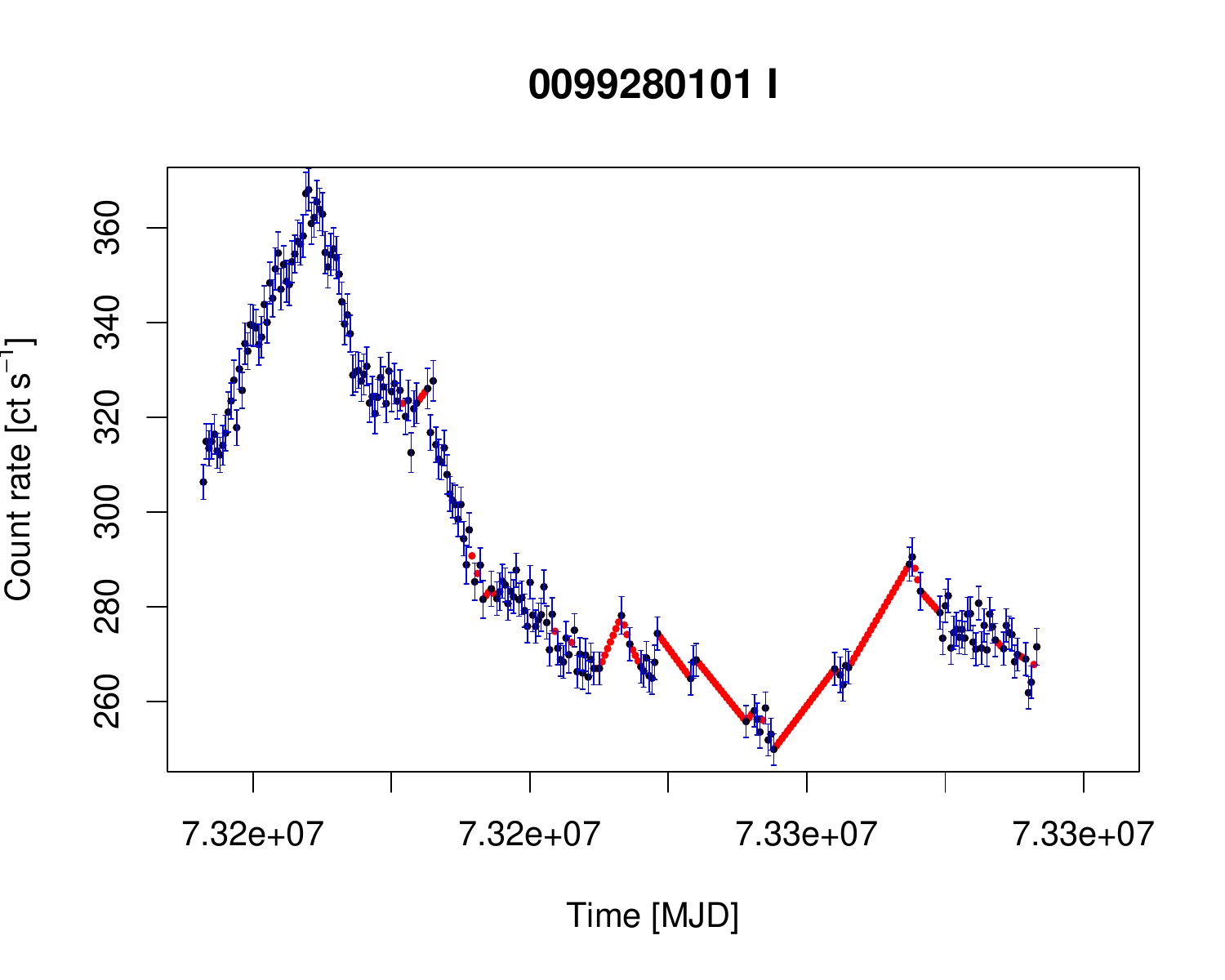}
	\end{minipage}
	\begin{minipage}{.38\textwidth} 
		\centering 
		\includegraphics[width=0.891\linewidth]{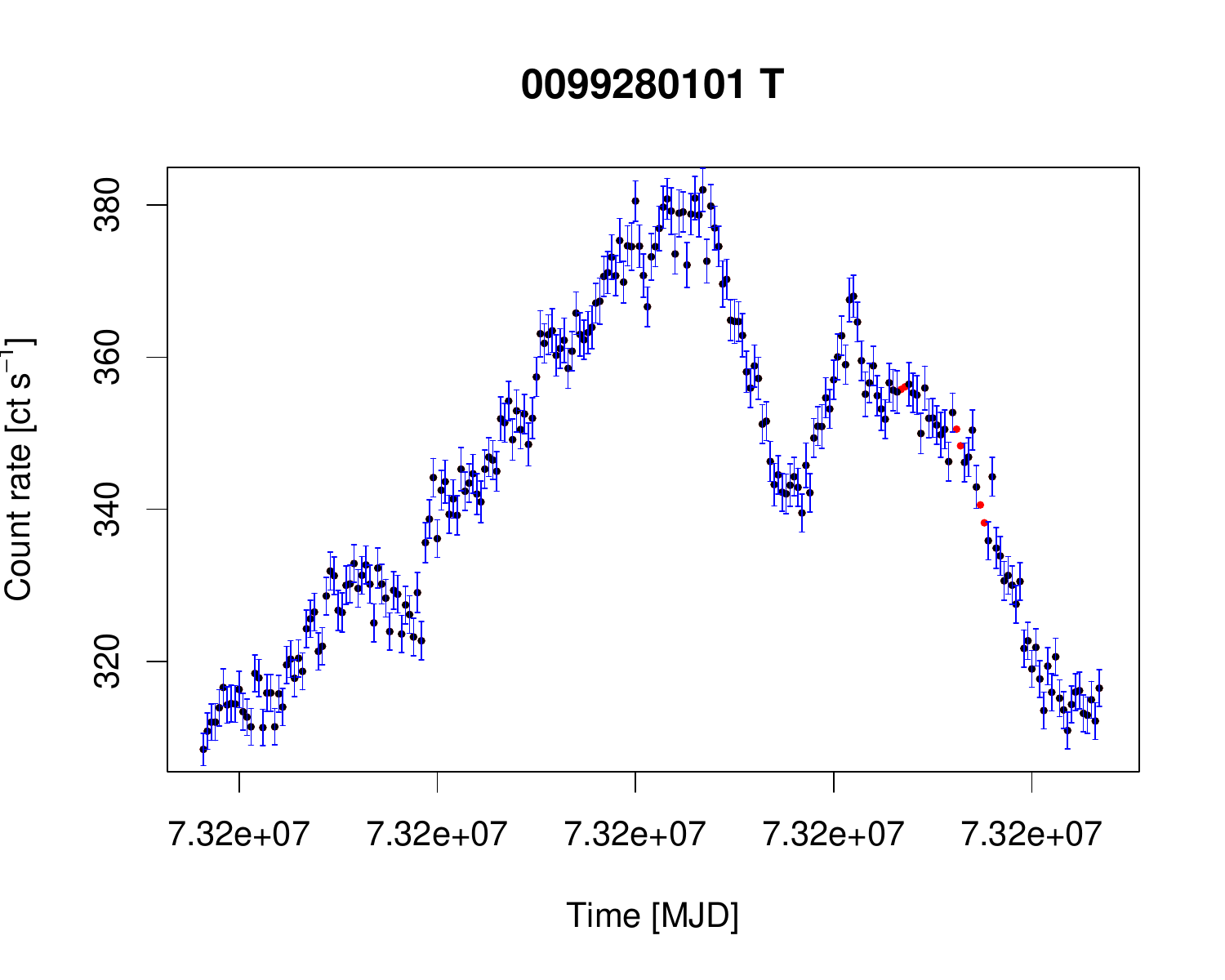}
	\end{minipage}
  \caption{Analyzed light curves 49-50 of Mrk~421, the red color denotes linearly interpolated data.  \label{LCS5}}
\end{figure*}

\end{appendix}




\begin{thebibliography}{}
\expandafter\ifx\csname natexlab\endcsname\relax\def\natexlab#1{#1}\fi
\providecommand{\url}[1]{\href{#1}{#1}}
\providecommand{\dodoi}[1]{doi:~\href{http://doi.org/#1}{\nolinkurl{#1}}}
\providecommand{\doeprint}[1]{\href{http://ascl.net/#1}{\nolinkurl{http://ascl.net/#1}}}
\providecommand{\doarXiv}[1]{\href{https://arxiv.org/abs/#1}{\nolinkurl{https://arxiv.org/abs/#1}}}

\bibitem[{{Abdo} {et~al.}(2011){Abdo}, {Ackermann}, {Ajello}, {Baldini},
  {Ballet}, {Barbiellini}, {Bastieri}, {Bechtol}, {Bellazzini}, {Berenji},
  {Blandford}, {Bloom}, {Bonamente}, {Borgland}, {Bouvier}, {Bregeon}, {Brez},
  {Brigida}, {Bruel}, {Buehler}, {Buson}, {Caliandro}, {Cameron}, {Cannon},
  {Caraveo}, {Carrigan}, {Casandjian}, {Cavazzuti}, {Cecchi}, {{\c{C}}elik},
  {Charles}, {Chekhtman}, {Chiang}, {Ciprini}, {Claus}, {Cohen-Tanugi},
  {Conrad}, {Cutini}, {de Angelis}, {de Palma}, {Dermer}, {Silva}, {Drell},
  {Dubois}, {Dumora}, {Escande}, {Favuzzi}, {Fegan}, {Finke}, {Focke},
  {Fortin}, {Frailis}, {Fuhrmann}, {Fukazawa}, {Fukuyama}, {Funk}, {Fusco},
  {Gargano}, {Gasparrini}, {Gehrels}, {Georganopoulos}, {Germani}, {Giebels},
  {Giglietto}, {Giommi}, {Giordano}, {Giroletti}, {Glanzman}, {Godfrey},
  {Grenier}, {Guiriec}, {Hadasch}, {Hayashida}, {Hays}, {Horan}, {Hughes},
  {J{\'o}hannesson}, {Johnson}, {Johnson}, {Kadler}, {Kamae}, {Katagiri},
  {Kataoka}, {Kn{\"o}dlseder}, {Kuss}, {Lande}, {Latronico}, {Lee}, {Longo},
  {Loparco}, {Lott}, {Lovellette}, {Lubrano}, {Madejski}, {Makeev},
  {Max-Moerbeck}, {Mazziotta}, {McEnery}, {Mehault}, {Michelson},
  {Mitthumsiri}, {Mizuno}, {Monte}, {Monzani}, {Morselli}, {Moskalenko},
  {Murgia}, {Nakamori}, {Naumann-Godo}, {Nishino}, {Nolan}, {Norris}, {Nuss},
  {Ohsugi}, {Okumura}, {Omodei}, {Orlando}, {Ormes}, {Ozaki}, {Paneque},
  {Panetta}, {Parent}, {Pavlidou}, {Pearson}, {Pelassa}, {Pepe},
  {Pesce-Rollins}, {Pierbattista}, {Piron}, {Porter}, {Rain{\`o}}, {Rando},
  {Razzano}, {Readhead}, {Reimer}, {Reimer}, {Reyes}, {Richards}, {Ritz},
  {Roth}, {Sadrozinski}, {Sanchez}, {Sander}, {Sgr{\`o}}, {Siskind}, {Smith},
  {Spandre}, {Spinelli}, {Stawarz}, {Stevenson}, {Strickman}, {Suson},
  {Takahashi}, {Takahashi}, {Tanaka}, {Thayer}, {Thayer}, {Thompson},
  {Tibaldo}, {Torres}, {Tosti}, {Tramacere}, {Troja}, {Usher}, {Vandenbroucke},
  {Vasileiou}, {Vianello}, {Vilchez}, {Vitale}, {Waite}, {Wang}, {Wehrle},
  {Winer}, {Wood}, {Yang}, {Yatsu}, {Ylinen}, {Zensus}, {Ziegler}, {Fermi LAT
  Collaboration}, {Aleksi{\'c}}, {Antonelli}, {Antoranz}, {Backes}, {Barrio},
  {Becerra Gonz{\'a}lez}, {Bednarek}, {Berdyugin}, {Berger}, {Bernardini},
  {Biland}, {Blanch}, {Bock}, {Boller}, {Bonnoli}, {Bordas}, {Borla Tridon},
  {Bosch-Ramon}, {Bose}, {Braun}, {Bretz}, {Camara}, {Carmona}, {Carosi},
  {Colin}, {Colombo}, {Contreras}, {Cortina}, {Covino}, {Dazzi}, {de Angelis},
  {De Cea del Pozo}, {Delgado Mendez}, {De Lotto}, {De Maria}, {De Sabata},
  {Diago Ortega}, {Doert}, {Dom{\'\i}nguez}, {Dominis Prester}, {Dorner},
  {Doro}, {Elsaesser}, {Ferenc}, {Fonseca}, {Font}, {Garc{\'\i}a L{\'o}pez},
  {Garczarczyk}, {Gaug}, {Giavitto}, {Godinovi}, {Hadasch}, {Herrero},
  {Hildebrand}, {H{\"o}hne-M{\"o}nch}, {Hose}, {Hrupec}, {Jogler}, {Klepser},
  {Kr{\"a}henb{\"u}hl}, {Kranich}, {Krause}, {La Barbera}, {Leonardo},
  {Lindfors}, {Lombardi}, {L{\'o}pez}, {Lorenz}, {Majumdar}, {Makariev},
  {Maneva}, {Mankuzhiyil}, {Mannheim}, {Maraschi}, {Mariotti}, {Mart{\'\i}nez},
  {Mazin}, {Meucci}, {Miranda}, {Mirzoyan}, {Miyamoto}, {Mold{\'o}n},
  {Moralejo}, {Nieto}, {Nilsson}, {Orito}, {Oya}, {Paoletti}, {Paredes},
  {Partini}, {Pasanen}, {Pauss}, {Pegna}, {Perez-Torres}, {Persic}, {Peruzzo},
  {Pochon}, {Prada}, {Prada Moroni}, {Prandini}, {Puchades}, {Puljak},
  {Reichardt}, {Rhode}, {Rib{\'o}}, {Rico}, {Rissi}, {R{\"u}gamer}, {Saggion},
  {Saito}, {Saito}, {Salvati}, {S{\'a}nchez-Conde}, {Satalecka}, {Scalzotto},
  {Scapin}, {Schultz}, {Schweizer}, {Shayduk}, {Shore}, {Sierpowska-Bartosik},
  {Sillanp{\"a}{\"a}}, {Sitarek}, {Sobczynska}, {Spanier}, {Spiro}, {Stamerra},
  {Steinke}, {Storz}, {Strah}, {Struebig}, {Suric}, {Takalo}, {Tavecchio},
  {Temnikov}, {Terzi{\'c}}, {Tescaro}, {Teshima}, {Vankov}, {Wagner},
  {Weitzel}, {Zabalza}, {Zandanel}, {Zanin}, {MAGIC Collaboration}, {Villata},
  {Raiteri}, {Aller}, {Aller}, {Chen}, {Jordan}, {Koptelova}, {Kurtanidze},
  {L{\"a}hteenm{\"a}ki}, {McBreen}, {Larionov}, {Lin}, {Nikolashvili},
  {Reinthal}, {Angelakis}, {Capalbi}, {Carrami{\~n}ana}, {Carrasco}, {Cassaro},
  {Cesarini}, {Falcone}, {Gurwell}, {Hovatta}, {Kovalev}, {Kovalev},
  {Krichbaum}, {Krimm}, {Lister}, {Moody}, {Maccaferri}, {Mori}, {Nestoras},
  {Orlati}, {Pace}, {Pagani}, {Pearson}, {Perri}, {Piner}, {Ros}, {Sadun},
  {Sakamoto}, {Tammi}, \& {Zook}}]{2011ApJ...736..131A}
{Abdo}, A.~A., {Ackermann}, M., {Ajello}, M., {et~al.} 2011, \apj, 736, 131,
  \dodoi{10.1088/0004-637X/736/2/131}

\bibitem[{{Abeysekara} {et~al.}(2020){Abeysekara}, {Benbow}, {Bird}, {Brill},
  {Brose}, {Buchovecky}, {Buckley}, {Christiansen}, {Chromey}, {Daniel},
  {Dumm}, {Falcone}, {Feng}, {Finley}, {Fortson}, {Furniss}, {Galante}, {Gent},
  {Gillanders}, {Giuri}, {Gueta}, {Hassan}, {Hervet}, {Holder}, {Hughes},
  {Humensky}, {Johnson}, {Kaaret}, {Kar}, {Kelley-Hoskins}, {Kertzman},
  {Kieda}, {Krause}, {Krennrich}, {Kumar}, {Lang}, {Moriarty}, {Mukherjee},
  {Nelson}, {Nieto}, {Nievas-Rosillo}, {O'Brien}, {Ong}, {Otte}, {Park},
  {Petrashyk}, {Pichel}, {Pohl}, {Prado}, {Pueschel}, {Quinn}, {Ragan},
  {Reynolds}, {Richards}, {Roache}, {Rovero}, {Rulten}, {Sadeh}, {Santander},
  {Sembroski}, {Shahinyan}, {Stevenson}, {Sushch}, {Tyler}, {Vassiliev},
  {Wakely}, {Weinstein}, {Wells}, {Wilcox}, {Wilhelm}, {Williams}, {Zitzer},
  {Acciari}, {Ansoldi}, {Antonelli}, {Arbet Engels}, {Baack}, {Babi{\'c}},
  {Banerjee}, {Barres de Almeida}, {Barrio}, {Becerra Gonz{\'a}lez},
  {Bednarek}, {Bellizzi}, {Bernardini}, {Berti}, {Besenrieder},
  {Bhattacharyya}, {Bigongiari}, {Biland}, {Blanch}, {Bonnoli}, {Busetto},
  {Carosi}, {Ceribella}, {Chai}, {Cikota}, {Colak}, {Colin}, {Colombo},
  {Contreras}, {Cortina}, {Covino}, {D'Elia}, {Da Vela}, {Dazzi}, {De Angelis},
  {De Lotto}, {Delfino}, {Delgado}, {Di Pierro}, {Do Souto Espi{\~n}era},
  {Dominis Prester}, {Dorner}, {Doro}, {Einecke}, {Elsaesser}, {Fallah
  Ramazani}, {Fattorini}, {Fern{\'a}ndez-Barral}, {Ferrara}, {Fidalgo},
  {Foffano}, {Fonseca}, {Font}, {Fruck}, {Galindo}, {Gallozzi}, {Garc{\'\i}a
  L{\'o}pez}, {Garczarczyk}, {Gasparyan}, {Gaug}, {Godinovi{\'c}}, {Green},
  {Guberman}, {Hadasch}, {Hahn}, {Herrera}, {Hoang}, {Hrupec}, {Inoue},
  {Ishio}, {Iwamura}, {Kubo}, {Kushida}, {Lamastra}, {Lelas}, {Leone},
  {Lindfors}, {Lombardi}, {Longo}, {L{\'o}pez}, {L{\'o}pez-Coto},
  {L{\'o}pez-Oramas}, {Machado de Oliveira Fraga}, {Maggio}, {Majumdar},
  {Makariev}, {Mallamaci}, {Maneva}, {Manganaro}, {Mannheim}, {Maraschi},
  {Mariotti}, {Mart{\'\i}nez}, {Masuda}, {Mazin}, {Miceli}, {Minev}, {Miranda},
  {Mirzoyan}, {Molina}, {Moralejo}, {Morcuende}, {Moreno}, {Moretti},
  {Munar-Adrover}, {Neustroev}, {Niedzwiecki}, {Nievas Rosillo}, {Nigro},
  {Nilsson}, {Ninci}, {Nishijima}, {Noda}, {Nogu{\'e}s}, {N{\"o}the}, {Paiano},
  {Palacio}, {Palatiello}, {Paneque}, {Paoletti}, {Paredes}, {Pe{\~n}il},
  {Peresano}, {Persic}, {Prada Moroni}, {Prandini}, {Puljak}, {Rhode},
  {Rib{\'o}}, {Rico}, {Righi}, {Rugliancich}, {Saha}, {Sahakyan}, {Saito},
  {Satalecka}, {Schweizer}, {Sitarek}, {{\v{S}}nidari{\'c}}, {Sobczynska},
  {Somero}, {Stamerra}, {Strom}, {Strzys}, {Sun}, {Suri{\'c}}, {Tavecchio},
  {Temnikov}, {Terzi{\'c}}, {Teshima}, {Torres-Alb{\`a}}, {Tsujimoto}, {van
  Scherpenberg}, {Vanzo}, {Vazquez Acosta}, {Vovk}, {Will}, {Zari{\'c}},
  {Aller}, {Aller}, {Carini}, {Horan}, {Jordan}, {Jorstad}, {Kurtanidze},
  {Kurtanidze}, {L{\"a}hteenm{\"a}ki}, {Larionov}, {Larionova}, {Madejski},
  {Marscher}, {Max-Moerbeck}, {Moody}, {Morozova}, {Nikolashvili}, {Raiteri},
  {Readhead}, {Richards}, {Sadun}, {Sakamoto}, {Sigua}, {Smith}, {Talvikki},
  {Tammi}, {Tornikoski}, {Troitsky}, \& {Villata}}]{2020ApJ...890...97A}
{Abeysekara}, A.~U., {Benbow}, W., {Bird}, R., {et~al.} 2020, \apj, 890, 97,
  \dodoi{10.3847/1538-4357/ab6612}

\bibitem[{{Ackermann} {et~al.}(2015){Ackermann}, {Ajello}, {Atwood}, {Baldini},
  {Ballet}, {Barbiellini}, {Bastieri}, {Becerra Gonzalez}, {Bellazzini},
  {Bissaldi}, {Blandford}, {Bloom}, {Bonino}, {Bottacini}, {Brandt}, {Bregeon},
  {Britto}, {Bruel}, {Buehler}, {Buson}, {Caliandro}, {Cameron}, {Caragiulo},
  {Caraveo}, {Carpenter}, {Casandjian}, {Cavazzuti}, {Cecchi}, {Charles},
  {Chekhtman}, {Cheung}, {Chiang}, {Chiaro}, {Ciprini}, {Claus},
  {Cohen-Tanugi}, {Cominsky}, {Conrad}, {Cutini}, {D'Abrusco}, {D'Ammando}, {de
  Angelis}, {Desiante}, {Digel}, {Di Venere}, {Drell}, {Favuzzi}, {Fegan},
  {Ferrara}, {Finke}, {Focke}, {Franckowiak}, {Fuhrmann}, {Fukazawa},
  {Furniss}, {Fusco}, {Gargano}, {Gasparrini}, {Giglietto}, {Giommi},
  {Giordano}, {Giroletti}, {Glanzman}, {Godfrey}, {Grenier}, {Grove},
  {Guiriec}, {Hewitt}, {Hill}, {Horan}, {Itoh}, {J{\'o}hannesson}, {Johnson},
  {Johnson}, {Kataoka}, {Kawano}, {Krauss}, {Kuss}, {La Mura}, {Larsson},
  {Latronico}, {Leto}, {Li}, {Li}, {Longo}, {Loparco}, {Lott}, {Lovellette},
  {Lubrano}, {Madejski}, {Mayer}, {Mazziotta}, {McEnery}, {Michelson},
  {Mizuno}, {Moiseev}, {Monzani}, {Morselli}, {Moskalenko}, {Murgia}, {Nuss},
  {Ohno}, {Ohsugi}, {Ojha}, {Omodei}, {Orienti}, {Orlando}, {Paggi}, {Paneque},
  {Perkins}, {Pesce-Rollins}, {Piron}, {Pivato}, {Porter}, {Rain{\`o}},
  {Rando}, {Razzano}, {Razzaque}, {Reimer}, {Reimer}, {Romani}, {Salvetti},
  {Schaal}, {Schinzel}, {Schulz}, {Sgr{\`o}}, {Siskind}, {Sokolovsky}, {Spada},
  {Spandre}, {Spinelli}, {Stawarz}, {Suson}, {Takahashi}, {Takahashi},
  {Tanaka}, {Thayer}, {Thayer}, {Tibaldo}, {Torres}, {Torresi}, {Tosti},
  {Troja}, {Uchiyama}, {Vianello}, {Winer}, {Wood}, \&
  {Zimmer}}]{2015ApJ...810...14A}
{Ackermann}, M., {Ajello}, M., {Atwood}, W.~B., {et~al.} 2015, \apj, 810, 14,
  \dodoi{10.1088/0004-637X/810/1/14}

\bibitem[{{Akaike}(1974)}]{1974ITAC...19..716A}
{Akaike}, H. 1974, IEEE Transactions on Automatic Control, 19, 716

\bibitem[{{Aleksi{\'c}} {et~al.}(2015){Aleksi{\'c}}, {Ansoldi}, {Antonelli},
  {Antoranz}, {Babic}, {Bangale}, {Barres de Almeida}, {Barrio}, {Becerra
  Gonz{\'a}lez}, {Bednarek}, {Berger}, {Bernardini}, {Biland}, {Blanch},
  {Bock}, {Bonnefoy}, {Bonnoli}, {Borracci}, {Bretz}, {Carmona}, {Carosi},
  {Carreto Fidalgo}, {Colin}, {Colombo}, {Contreras}, {Cortina}, {Covino}, {Da
  Vela}, {Dazzi}, {De Angelis}, {De Caneva}, {De Lotto}, {Delgado Mendez},
  {Doert}, {Dom{\'\i}nguez}, {Dominis Prester}, {Dorner}, {Doro}, {Einecke},
  {Eisenacher}, {Elsaesser}, {Farina}, {Ferenc}, {Fonseca}, {Font}, {Frantzen},
  {Fruck}, {Garc{\'\i}a L{\'o}pez}, {Garczarczyk}, {Garrido Terrats}, {Gaug},
  {Giavitto}, {Godinovi{\'c}}, {Gonz{\'a}lez Mu{\~n}oz}, {Gozzini}, {Hadamek},
  {Hadasch}, {Herrero}, {Hildebrand}, {Hose}, {Hrupec}, {Idec}, {Kadenius},
  {Kellermann}, {Knoetig}, {Krause}, {Kushida}, {La Barbera}, {Lelas},
  {Lewandowska}, {Lindfors}, {Longo}, {Lombardi}, {L{\'o}pez},
  {L{\'o}pez-Coto}, {L{\'o}pez-Oramas}, {Lorenz}, {Lozano}, {Makariev},
  {Mallot}, {Maneva}, {Mankuzhiyil}, {Mannheim}, {Maraschi}, {Marcote},
  {Mariotti}, {Mart{\'\i}nez}, {Mazin}, {Menzel}, {Meucci}, {Miranda},
  {Mirzoyan}, {Moralejo}, {Munar-Adrover}, {Nakajima}, {Niedzwiecki},
  {Nilsson}, {Nowak}, {Orito}, {Overkemping}, {Paiano}, {Palatiello},
  {Paneque}, {Paoletti}, {Paredes}, {Paredes-Fortuny}, {Partini}, {Persic},
  {Prada}, {Prada Moroni}, {Prandini}, {Preziuso}, {Puljak}, {Reinthal},
  {Rhode}, {Rib{\'o}}, {Rico}, {RodriguezGarcia}, {R{\"u}gamer}, {Saggion},
  {Saito}, {Salvati}, {Satalecka}, {Scalzotto}, {Scapin}, {Schultz},
  {Schweizer}, {Shore}, {Sillanp{\"a}{\"a}}, {Sitarek}, {Snidaric},
  {Sobczynska}, {Spanier}, {Stamatescu}, {Stamerra}, {Steinbring}, {Storz},
  {Sun}, {Suri{\'c}}, {Takalo}, {Tavecchio}, {Temnikov}, {Terzi{\'c}},
  {Tescaro}, {Teshima}, {Thaele}, {Tibolla}, {Torres}, {Toyama}, {Treves},
  {Uellenbeck}, {Vogler}, {Wagner}, {Zandanel}, {Zanin}, {MAGIC Collaboration},
  {Archambault}, {Behera}, {Beilicke}, {Benbow}, {Bird}, {Buckley}, {Bugaev},
  {Cerruti}, {Chen}, {Ciupik}, {Collins-Hughes}, {Cui}, {Dumm}, {Eisch},
  {Falcone}, {Federici}, {Feng}, {Finley}, {Fleischhack}, {Fortin}, {Fortson},
  {Furniss}, {Griffin}, {Griffiths}, {Grube}, {Gyuk}, {Hanna}, {Holder},
  {Hughes}, {Humensky}, {Johnson}, {Kaaret}, {Kertzman}, {Khassen}, {Kieda},
  {Krawczynski}, {Krennrich}, {Kumar}, {Lang}, {Maier}, {McArthur}, {Meagher},
  {Moriarty}, {Mukherjee}, {Ong}, {Otte}, {Park}, {Pichel}, {Pohl}, {Popkow},
  {Prokoph}, {Quinn}, {Ragan}, {Rajotte}, {Reynolds}, {Richards}, {Roache},
  {Rovero}, {Sembroski}, {Shahinyan}, {Staszak}, {Telezhinsky}, {Theiling},
  {Tucci}, {Tyler}, {Varlotta}, {Wakely}, {Weekes}, {Weinstein}, {Welsing},
  {Wilhelm}, {Williams}, {Zitzer}, {VERITAS Collaboration}, {Villata},
  {Raiteri}, {Aller}, {Aller}, {Chen}, {Jordan}, {Koptelova}, {Kurtanidze},
  {L{\"a}hteenm{\"a}ki}, {McBreen}, {Larionov}, {Lin}, {Nikolashvili},
  {Angelakis}, {Capalbi}, {Carrami{\~n}ana}, {Carrasco}, {Cassaro}, {Cesarini},
  {Fuhrmann}, {Giroletti}, {Hovatta}, {Krichbaum}, {Krimm}, {Max-Moerbeck},
  {Moody}, {Maccaferri}, {Mori}, {Nestoras}, {Orlati}, {Pace}, {Pearson},
  {Perri}, {Readhead}, {Richards}, {Sadun}, {Sakamoto}, {Tammi}, {Tornikoski},
  {Yatsu}, \& {Zook}}]{2015A&A...576A.126A}
{Aleksi{\'c}}, J., {Ansoldi}, S., {Antonelli}, L.~A., {et~al.} 2015, \aap, 576,
  A126, \dodoi{10.1051/0004-6361/201424216}

\bibitem[{{Arbet-Engels} {et~al.}(2021){Arbet-Engels}, {Baack}, {Balbo},
  {Biland}, {Blank}, {Bretz}, {Bruegge}, {Bulinski}, {Buss}, {Doerr}, {Dorner},
  {Elsaesser}, {Hildebrand}, {Mannheim}, {Mueller}, {Neise}, {Noethe},
  {Paravac}, {Rhode}, {Schleicher}, {Sedlaczek}, {Shukla}, {Sliusar}, {Walter},
  \& {von Willert}}]{2021A&A...647A..88A}
{Arbet-Engels}, A., {Baack}, D., {Balbo}, M., {et~al.} 2021, \aap, 647, A88,
  \dodoi{10.1051/0004-6361/201935557}

\bibitem[{Babaei {et~al.}(2014)Babaei, Zarghami, Sedighikamal,
  Sotudeh-Gharebagh, \& Mostoufi}]{Babaei2014Selection}
Babaei, B., Zarghami, R., Sedighikamal, H., Sotudeh-Gharebagh, R., \& Mostoufi,
  N. 2014, Physica A-statistical Mechanics and Its Applications, 395, 112,
  \dodoi{10.1016/J.PHYSA.2013.10.016}

\bibitem[{{Bednarek} \& {Protheroe}(1997)}]{1997MNRAS.290..139B}
{Bednarek}, W., \& {Protheroe}, R.~J. 1997, \mnras, 290, 139,
  \dodoi{10.1093/mnras/290.1.139}

\bibitem[{Bhatta(2021)}]{Bhatta_2021}
Bhatta, G. 2021, The Astrophysical Journal, 923, 7,
  \dodoi{10.3847/1538-4357/ac2819}

\bibitem[{Bhatta \& Dhital(2020)}]{bhatta2020}
Bhatta, G., \& Dhital, N. 2020, The Astrophysical Journal, 891, 120

\bibitem[{{Bhatta} {et~al.}(2020){Bhatta}, {P{\'a}nis}, \&
  {Stuchl{\'\i}k}}]{2020ApJ...905..160B}
{Bhatta}, G., {P{\'a}nis}, R., \& {Stuchl{\'\i}k}, Z. 2020, \apj, 905, 160,
  \dodoi{10.3847/1538-4357/abc625}

\bibitem[{{Bhatta} \& {Webb}(2018)}]{2018Galax...6....2B}
{Bhatta}, G., \& {Webb}, J. 2018, Galaxies, 6, 2,
  \dodoi{10.3390/galaxies6010002}

\bibitem[{{Bhattacharyya} {et~al.}(2020){Bhattacharyya}, {Ghosh}, {Chatterjee},
  \& {Das}}]{2020ApJ...897...25B}
{Bhattacharyya}, S., {Ghosh}, R., {Chatterjee}, R., \& {Das}, N. 2020, \apj,
  897, 25, \dodoi{10.3847/1538-4357/ab91a8}

\bibitem[{Bhattacharyya {et~al.}(2020)Bhattacharyya, Ghosh, Chatterjee, \&
  Das}]{bhattacharyya2020blazar}
Bhattacharyya, S., Ghosh, R., Chatterjee, R., \& Das, N. 2020, The
  Astrophysical Journal, 897, 25

\bibitem[{{Box, G. E. P. and Jenkins, G. M. and Reinsel, G.
  C.}(1976)}]{1976tsaf.conf.....B}
{Box, G. E. P. and Jenkins, G. M. and Reinsel, G. C.}, ed. 1976, {Time Series
  Analysis: Forecasting and Control} ({San Francisco, CA, USA}: {Holden-Day})

\bibitem[{{Bradley} \& {Kantz}(2015)}]{2015Chaos..25i7610B}
{Bradley}, E., \& {Kantz}, H. 2015, Chaos, 25, 097610,
  \dodoi{10.1063/1.4917289}

\bibitem[{{Britzen} {et~al.}(2017){Britzen}, {Fendt}, {Eckart}, \&
  {Karas}}]{2017A&A...601A..52B}
{Britzen}, S., {Fendt}, C., {Eckart}, A., \& {Karas}, V. 2017, \aap, 601, A52,
  \dodoi{10.1051/0004-6361/201629469}

\bibitem[{{Britzen} {et~al.}(2023){Britzen}, {Zaja{\v{c}}ek}, {Gopal-Krishna},
  {Fendt}, {Kun}, {Jaron}, {Sillanp{\"a}{\"a}}, \&
  {Eckart}}]{2023ApJ...951..106B}
{Britzen}, S., {Zaja{\v{c}}ek}, M., {Gopal-Krishna}, {et~al.} 2023, \apj, 951,
  106, \dodoi{10.3847/1538-4357/accbbc}

\bibitem[{{Britzen} {et~al.}(2021){Britzen}, {Zaja{\v{c}}ek}, {Popovi{\'c}},
  {Fendt}, {Tramacere}, {Pashchenko}, {Jaron}, {P{\'a}nis}, {Petrov}, {Aller},
  \& {Aller}}]{2021MNRAS.503.3145B}
{Britzen}, S., {Zaja{\v{c}}ek}, M., {Popovi{\'c}}, L.~{\v{C}}., {et~al.} 2021,
  \mnras, 503, 3145, \dodoi{10.1093/mnras/stab589}

\bibitem[{{Cao}(1997)}]{1997PhyD..110...43C}
{Cao}, L. 1997, Physica D Nonlinear Phenomena, 110, 43,
  \dodoi{10.1016/S0167-2789(97)00118-8}

\bibitem[{Coffman {et~al.}(2000)Coffman, Kundu, \&
  Wootters}]{PhysRevA.61.052306}
Coffman, V., Kundu, J., \& Wootters, W.~K. 2000, Phys. Rev. A, 61, 052306,
  \dodoi{10.1103/PhysRevA.61.052306}

\bibitem[{Dickey \& Fuller(1979)}]{Dickey1979DistributionOT}
Dickey, D.~A., \& Fuller, W.~A. 1979, Journal of the American Statistical
  Association, 74, 427.
\newblock \url{https://api.semanticscholar.org/CorpusID:56458593}

\bibitem[{{Dinesh} {et~al.}(2023){Dinesh}, {Bhatta}, {Adhikari}, {Mohorian},
  {Dhital}, {Chaudhary}, {P{\'a}nis}, \& {G{\'o}ra}}]{2023ApJ...955..121D}
{Dinesh}, A., {Bhatta}, G., {Adhikari}, T.~P., {et~al.} 2023, \apj, 955, 121,
  \dodoi{10.3847/1538-4357/acf316}

\bibitem[{{Eckmann} {et~al.}(1987){Eckmann}, {Oliffson Kamphorst}, \&
  {Ruelle}}]{1987EL......4..973E}
{Eckmann}, J.~P., {Oliffson Kamphorst}, S., \& {Ruelle}, D. 1987, EPL
  (Europhysics Letters), 4, 973, \dodoi{10.1209/0295-5075/4/9/004}

\bibitem[{{Emmanoulopoulos}(2007)}]{2007PhDT........70E}
{Emmanoulopoulos}, D. 2007, PhD thesis, University of Heidelberg, Astronomical
  Observatory

\bibitem[{{Feigelson} {et~al.}(2018){Feigelson}, {Babu}, \&
  {Caceres}}]{2018FrP.....6...80F}
{Feigelson}, E.~D., {Babu}, G.~J., \& {Caceres}, G.~A. 2018, Frontiers in
  Physics, 6, 80, \dodoi{10.3389/fphy.2018.00080}

\bibitem[{Garcia \& Sawitzki(2020)}]{garcia2nonlineartseries}
Garcia, C.~A., \& Sawitzki, G. 2020, nonlinearTseries: Nonlinear Time Series
  Analysis.
\newblock \url{https://CRAN.R-project.org/package=nonlinearTseries}

\bibitem[{{Ghisellini}(1999)}]{1999ASPC..159..311G}
{Ghisellini}, G. 1999, in Astronomical Society of the Pacific Conference
  Series, Vol. 159, BL Lac Phenomenon, ed. L.~O. {Takalo} \&
  A.~{Sillanp{\"a}{\"a}}, 311, \dodoi{10.48550/arXiv.astro-ph/9810230}

\bibitem[{{Giommi} \& {Padovani}(2015)}]{2015mgm..conf.1015G}
{Giommi}, P., \& {Padovani}, P. 2015, in Thirteenth Marcel Grossmann Meeting:
  On Recent Developments in Theoretical and Experimental General Relativity,
  Astrophysics and Relativistic Field Theories, 1015--1018,
  \dodoi{10.1142/9789814623995_0069}

\bibitem[{{Goyal} {et~al.}(2018){Goyal}, {Stawarz}, {Zola}, {Marchenko},
  {Soida}, {Nilsson}, {Ciprini}, {Baran}, {Ostrowski}, {Wiita},
  {Gopal-Krishna}, {Siemiginowska}, {Sobolewska}, {Jorstad}, {Marscher},
  {Aller}, {Aller}, {Hovatta}, {Caton}, {Reichart}, {Matsumoto}, {Sadakane},
  {Gazeas}, {Kidger}, {Piirola}, {Jermak}, {Alicavus}, {Baliyan}, {Baransky},
  {Berdyugin}, {Blay}, {Boumis}, {Boyd}, {Bufan}, {Campas Torrent}, {Campos},
  {Carrillo G{\'o}mez}, {Dalessio}, {Debski}, {Dimitrov}, {Drozdz}, {Er},
  {Erdem}, {Escartin P{\'e}rez}, {Fallah Ramazani}, {Filippenko}, {Gafton},
  {Garcia}, {Godunova}, {G{\'o}mez Pinilla}, {Gopinathan}, {Haislip}, {Haque},
  {Harmanen}, {Hudec}, {Hurst}, {Ivarsen}, {Joshi}, {Kagitani}, {Karaman},
  {Karjalainen}, {Kaur}, {Kozie{\l}-Wierzbowska}, {Kuligowska}, {Kundera},
  {Kurowski}, {Kvammen}, {LaCluyze}, {Lee}, {Liakos}, {Lozano de Haro},
  {Moore}, {Mugrauer}, {Naves Nogues}, {Neely}, {Ogloza}, {Okano}, {Pajdosz},
  {Pandey}, {Perri}, {Poyner}, {Provencal}, {Pursimo}, {Raj}, {Rajkumar},
  {Reinthal}, {Reynolds}, {Saario}, {Sadegi}, {Sakanoi}, {Salto Gonz{\'a}lez},
  {Sameer}, {Simon}, {Siwak}, {Schweyer}, {Sold{\'a}n Alfaro}, {Sonbas},
  {Strobl}, {Takalo}, {Tremosa Espasa}, {Valdes}, {Vasylenko}, {Verrecchia},
  {Webb}, {Yoneda}, {Zejmo}, {Zheng}, {Zielinski}, {Janik}, {Chavushyan},
  {Mohammed}, {Cheung}, \& {Giroletti}}]{2018ApJ...863..175G}
{Goyal}, A., {Stawarz}, {\L}., {Zola}, S., {et~al.} 2018, \apj, 863, 175,
  \dodoi{10.3847/1538-4357/aad2de}

\bibitem[{{Henriksen} \& {Irwin}(2019)}]{2019arXiv190208704H}
{Henriksen}, R.~N., \& {Irwin}, J.~A. 2019, arXiv e-prints, arXiv:1902.08704,
  \dodoi{10.48550/arXiv.1902.08704}

\bibitem[{{Hota} {et~al.}(2021){Hota}, {Shah}, {Khatoon}, {Misra}, {Pradhan},
  \& {Gogoi}}]{2021MNRAS.508.5921H}
{Hota}, J., {Shah}, Z., {Khatoon}, R., {et~al.} 2021, \mnras, 508, 5921,
  \dodoi{10.1093/mnras/stab2903}

\bibitem[{Iwanski \& Bradley(1998)}]{Iwanski1998RecurrencePO}
Iwanski, J.~S., \& Bradley, E. 1998, Chaos, 8 4, 861.
\newblock \url{https://api.semanticscholar.org/CorpusID:15272545}

\bibitem[{{Joshi} {et~al.}(2002){Joshi}, {Baliyan}, \&
  {Ganesh}}]{2002BASI...30..301J}
{Joshi}, U.~C., {Baliyan}, K.~S., \& {Ganesh}, S. 2002, Bulletin of the
  Astronomical Society of India, 30, 301

\bibitem[{{Kadowaki} {et~al.}(2021){Kadowaki}, {de Gouveia Dal Pino},
  {Medina-Torrej{\'o}n}, {Mizuno}, \& {Kushwaha}}]{2021ApJ...912..109K}
{Kadowaki}, L. H.~S., {de Gouveia Dal Pino}, E.~M., {Medina-Torrej{\'o}n},
  T.~E., {Mizuno}, Y., \& {Kushwaha}, P. 2021, \apj, 912, 109,
  \dodoi{10.3847/1538-4357/abee7a}

\bibitem[{{Kapanadze} {et~al.}(2018){Kapanadze}, {Vercellone}, {Romano},
  {Hughes}, {Aller}, {Aller}, {Kharshiladze}, {Kapanadze}, \&
  {Tabagari}}]{2018ApJ...854...66K}
{Kapanadze}, B., {Vercellone}, S., {Romano}, P., {et~al.} 2018, \apj, 854, 66,
  \dodoi{10.3847/1538-4357/aaa75d}

\bibitem[{KEENAN(1985)}]{10.1093/biomet/72.1.39}
KEENAN, D.~M. 1985, Biometrika, 72, 39, \dodoi{10.1093/biomet/72.1.39}

\bibitem[{{Kirk} \& {Mochol}(2011)}]{2011ApJ...729..104K}
{Kirk}, J.~G., \& {Mochol}, I. 2011, \apj, 729, 104,
  \dodoi{10.1088/0004-637X/729/2/104}

\bibitem[{{Kravchenko} {et~al.}(2020){Kravchenko}, {G{\'o}mez}, {Kovalev}, \&
  {Voitsik}}]{2020AdSpR..65..720K}
{Kravchenko}, E.~V., {G{\'o}mez}, J.~L., {Kovalev}, Y.~Y., \& {Voitsik}, P.~A.
  2020, Advances in Space Research, 65, 720, \dodoi{10.1016/j.asr.2019.01.042}

\bibitem[{{Lyutikov}(2003)}]{2003NewAR..47..513L}
{Lyutikov}, M. 2003, \nar, 47, 513, \dodoi{10.1016/S1387-6473(03)00083-6}

\bibitem[{{MAGIC Collaboration} {et~al.}(2021){MAGIC Collaboration}, {Acciari},
  {Ansoldi}, {Antonelli}, {Arbet Engels}, {Artero}, {Asano}, {Babi{\'c}},
  {Baquero}, {Barres de Almeida}, {Barrio}, {Batkovi{\'c}}, {Becerra
  Gonz{\'a}lez}, {Bednarek}, {Bellizzi}, {Bernardini}, {Bernardos}, {Berti},
  {Besenrieder}, {Bhattacharyya}, {Bigongiari}, {Blanch}, {Bo{\v{s}}njak},
  {Busetto}, {Carosi}, {Ceribella}, {Cerruti}, {Chai}, {Chilingarian},
  {Cikota}, {Colak}, {Colombo}, {Contreras}, {Cortina}, {Covino}, {D'Amico},
  {D'Elia}, {da Vela}, {Dazzi}, {de Angelis}, {de Lotto}, {Delfino}, {Delgado},
  {Delgado Mendez}, {Depaoli}, {di Pierro}, {di Venere}, {Do Souto
  Espi{\~n}eira}, {Dominis Prester}, {Donini}, {Doro}, {Fallah Ramazani},
  {Fattorini}, {Ferrara}, {Fonseca}, {Font}, {Fruck}, {Fukami}, {Garc{\'\i}a
  L{\'o}pez}, {Garczarczyk}, {Gasparyan}, {Gaug}, {Giglietto}, {Giordano},
  {Gliwny}, {Godinovi{\'c}}, {Green}, {Green}, {Hadasch}, {Hahn}, {Heckmann},
  {Herrera}, {Hoang}, {Hrupec}, {H{\"u}tten}, {Inada}, {Inoue}, {Ishio},
  {Iwamura}, {Jim{\'e}nez}, {Jormanainen}, {Jouvin}, {Kajiwara}, {Karjalainen},
  {Kerszberg}, {Kobayashi}, {Kubo}, {Kushida}, {Lamastra}, {Lelas}, {Leone},
  {Lindfors}, {Lombardi}, {Longo}, {L{\'o}pez-Coto}, {L{\'o}pez-Moya},
  {L{\'o}pez-Oramas}, {Loporchio}, {Machado de Oliveira Fraga}, {Maggio},
  {Majumdar}, {Makariev}, {Mallamaci}, {Maneva}, {Manganaro}, {Maraschi},
  {Mariotti}, {Mart{\'\i}nez}, {Mazin}, {Menchiari}, {Mender},
  {Mi{\'c}anovi{\'c}}, {Miceli}, {Miener}, {Minev}, {Miranda}, {Mirzoyan},
  {Molina}, {Moralejo}, {Morcuende}, {Moreno}, {Moretti}, {Neustroev}, {Nigro},
  {Nilsson}, {Nishijima}, {Noda}, {Nozaki}, {Ohtani}, {Oka}, {Otero-Santos},
  {Paiano}, {Palatiello}, {Paneque}, {Paoletti}, {Paredes}, {Pavleti{\'c}},
  {Pe{\~n}il}, {Perennes}, {Persic}, {Prada Moroni}, {Prandini}, {Priyadarshi},
  {Puljak}, {Rib{\'o}}, {Rico}, {Righi}, {Rugliancich}, {Saha}, {Sahakyan},
  {Saito}, {Sakurai}, {Satalecka}, {Saturni}, {Schmidt}, {Schweizer},
  {Sitarek}, {{\v{S}}nidari{\'c}}, {Sobczynska}, {Spolon}, {Stamerra}, {Strom},
  {Strzys}, {Suda}, {Suri{\'c}}, {Takahashi}, {Tavecchio}, {Temnikov},
  {Terzi{\'c}}, {Teshima}, {Tosti}, {Truzzi}, {Tutone}, {Ubach}, {van
  Scherpenberg}, {Vanzo}, {Vazquez Acosta}, {Ventura}, {Verguilov}, {Vigorito},
  {Vitale}, {Vovk}, {Will}, {Wunderlich}, {Zari{\'c}}, {FACT Collaboration},
  {Baack}, {Balbo}, {Biederbeck}, {Biland}, {Bretz}, {Buss}, {Dorner},
  {Eisenberger}, {Elsaesser}, {Hildebrand}, {Iotov}, {Mannheim}, {Neise},
  {Noethe}, {Paravac}, {Rhode}, {Schleicher}, {Sliusar}, {Walter}, {D'Ammando},
  {Horan}, {Lien}, {Balokovi{\'c}}, {Madejski}, {Perri}, {Verrecchia}, {Leto},
  {L{\"a}hteenm{\"a}ki}, {Tornikoski}, {Ramakrishnan}, {J{\"a}rvel{\"a}},
  {Vera}, {Chamani}, {Villata}, {Raiteri}, {Gupta}, {Pandey}, {Fuentes},
  {Agudo}, {Casadio}, {Semkov}, {Ibryamov}, {Marchini}, {Bachev}, {Strigachev},
  {Ovcharov}, {Bozhilov}, {Valcheva}, {Zaharieva}, {Damljanovic}, {Vince},
  {Larionov}, {Borman}, {Grishina}, {Hagen-Thorn}, {Kopatskaya}, {Larionova},
  {Larionova}, {Morozova}, {Nikiforova}, {Savchenko}, {Troitskiy},
  {Troitskaya}, {Vasilyev}, {Merkulova}, {Chen}, {Samal}, {Lin}, {Moody},
  {Sadun}, {Jorstad}, {Marscher}, {Weaver}, {Feige}, {Kania}, {Kopp}, {Kunkel},
  {Reinhart}, {Scherbantin}, {Schneider}, {Lorey}, {Acosta-Pulido},
  {Carnerero}, {Carosati}, {Kurtanidze}, {Kurtanidze}, {Nikolashvili},
  {Chigladze}, {Ivanidze}, {Kimeridze}, {Sigua}, {Joner}, {Spencer},
  {Giroletti}, {Marchili}, {Righini}, {Rizzi}, \&
  {Bonnoli}}]{2021A&A...655A..89M}
{MAGIC Collaboration}, {Acciari}, V.~A., {Ansoldi}, S., {et~al.} 2021, \aap,
  655, A89, \dodoi{10.1051/0004-6361/202141004}

\bibitem[{{Mannattil} {et~al.}(2016){Mannattil}, {Gupta}, \&
  {Chakraborty}}]{2016ApJ...833..208M}
{Mannattil}, M., {Gupta}, H., \& {Chakraborty}, S. 2016, \apj, 833, 208,
  \dodoi{10.3847/1538-4357/833/2/208}

\bibitem[{{Mannheim}(1999)}]{1999APh....11...49M}
{Mannheim}, K. 1999, Astroparticle Physics, 11, 49,
  \dodoi{10.1016/S0927-6505(99)00024-9}

\bibitem[{{Manoharan} \& {Kokkotas}(2023)}]{2023arXiv230713063M}
{Manoharan}, P., \& {Kokkotas}, K.~D. 2023, arXiv e-prints, arXiv:2307.13063,
  \dodoi{10.48550/arXiv.2307.13063}

\bibitem[{{Maraschi}(1999)}]{1999NuPhS..69..389M}
{Maraschi}, L. 1999, Nuclear Physics B Proceedings Supplements, 69, 389,
  \dodoi{10.1016/S0920-5632(98)00247-3}

\bibitem[{{Maraschi} \& {Tavecchio}(2001)}]{2001ASPC..234..437M}
{Maraschi}, L., \& {Tavecchio}, F. 2001, in Astronomical Society of the Pacific
  Conference Series, Vol. 234, X-ray Astronomy 2000, ed. R.~{Giacconi},
  S.~{Serio}, \& L.~{Stella}, 437, \dodoi{10.48550/arXiv.astro-ph/0107566}

\bibitem[{{Markowitz} {et~al.}(2022){Markowitz}, {Nalewajko}, {Bhatta},
  {Dewangan}, {Chandra}, {Dorner}, {Schleicher}, {Pajdosz-{\'S}mierciak},
  {Stawarz}, {Zola}, {Ostrowski}, {Carosati}, {Krishnan}, {Bachev},
  {Ben{\'\i}tez}, {Gazeas}, {Hiriart}, {Hu}, {Larionov}, {Marchini},
  {Matsumoto}, {Nikiforova}, {Pursimo}, {Raiteri}, {Reichart}, {Rodriguez},
  {Semkov}, {Strigachev}, {Sugiura}, {Villata}, {Webb}, {Arbet-Engels},
  {Baack}, {Balbo}, {Biland}, {Bretz}, {Buss}, {Eisenberger}, {Elsaesser},
  {Hildebrand}, {Iotov}, {Kalenski}, {Mannheim}, {Mitchell}, {Neise}, {Noethe},
  {Paravac}, {Rhode}, {Sliusar}, \& {Walter}}]{2022MNRAS.513.1662M}
{Markowitz}, A.~G., {Nalewajko}, K., {Bhatta}, G., {et~al.} 2022, \mnras, 513,
  1662, \dodoi{10.1093/mnras/stac917}

\bibitem[{{Marscher}(2014)}]{2014ApJ...780...87M}
{Marscher}, A.~P. 2014, \apj, 780, 87, \dodoi{10.1088/0004-637X/780/1/87}

\bibitem[{{Marwan}(2008)}]{2008EPJST.164....3M}
{Marwan}, N. 2008, European Physical Journal Special Topics, 164, 3,
  \dodoi{10.1140/epjst/e2008-00829-1}

\bibitem[{{Marwan} {et~al.}(2007){Marwan}, {Carmen Romano}, {Thiel}, \&
  {Kurths}}]{2007PhR...438..237M}
{Marwan}, N., {Carmen Romano}, M., {Thiel}, M., \& {Kurths}, J. 2007, \physrep,
  438, 237, \dodoi{10.1016/j.physrep.2006.11.001}

\bibitem[{{Mohorian} {et~al.}(2022){Mohorian}, {Bhatta}, {Adhikari}, {Dhital},
  {P{\'a}nis}, {Dinesh}, {Chaudhary}, {Bachchan}, \&
  {Stuchl{\'\i}k}}]{2022MNRAS.510.5280M}
{Mohorian}, M., {Bhatta}, G., {Adhikari}, T.~P., {et~al.} 2022, \mnras, 510,
  5280, \dodoi{10.1093/mnras/stab3738}

\bibitem[{{Moreno} {et~al.}(2019){Moreno}, {Vogeley}, {Richards}, \&
  {Yu}}]{2019PASP..131f3001M}
{Moreno}, J., {Vogeley}, M.~S., {Richards}, G.~T., \& {Yu}, W. 2019, \pasp,
  131, 063001, \dodoi{10.1088/1538-3873/ab1597}

\bibitem[{{Noel} {et~al.}(2022){Noel}, {Gaur}, {Gupta}, {Wierzcholska},
  {Ostrowski}, {Dhiman}, \& {Bhatta}}]{2022ApJS..262....4N}
{Noel}, A.~P., {Gaur}, H., {Gupta}, A.~C., {et~al.} 2022, \apjs, 262, 4,
  \dodoi{10.3847/1538-4365/ac7799}

\bibitem[{{O' Riordan} {et~al.}(2017){O' Riordan}, {Pe'er}, \&
  {McKinney}}]{2017ApJ...843...81O}
{O' Riordan}, M., {Pe'er}, A., \& {McKinney}, J.~C. 2017, \apj, 843, 81,
  \dodoi{10.3847/1538-4357/aa7339}

\bibitem[{{P{\'a}nis} {et~al.}(2023){P{\'a}nis}, {Ad{\'a}mek}, \&
  {Marwan}}]{2023EPJST.232...47P}
{P{\'a}nis}, R., {Ad{\'a}mek}, K., \& {Marwan}, N. 2023, European Physical
  Journal Special Topics, 232, 47, \dodoi{10.1140/epjs/s11734-022-00686-4}

\bibitem[{{P{\'a}nis} {et~al.}(2019){P{\'a}nis}, {Kolo{\v{s}}}, \&
  {Stuchl{\'\i}k}}]{2019EPJC...79..479P}
{P{\'a}nis}, R., {Kolo{\v{s}}}, M., \& {Stuchl{\'\i}k}, Z. 2019, European
  Physical Journal C, 79, 479, \dodoi{10.1140/epjc/s10052-019-6961-7}

\bibitem[{{Papadakis} {et~al.}(2002){Papadakis}, {Brinkmann}, {Negoro}, \&
  {Gliozzi}}]{2002A&A...382L...1P}
{Papadakis}, I.~E., {Brinkmann}, W., {Negoro}, H., \& {Gliozzi}, M. 2002, \aap,
  382, L1, \dodoi{10.1051/0004-6361:20011763}

\bibitem[{{Petropoulou} {et~al.}(2016){Petropoulou}, {Coenders}, \&
  {Dimitrakoudis}}]{2016APh....80..115P}
{Petropoulou}, M., {Coenders}, S., \& {Dimitrakoudis}, S. 2016, Astroparticle
  Physics, 80, 115, \dodoi{10.1016/j.astropartphys.2016.04.001}

\bibitem[{{Phillipson} {et~al.}(2023){Phillipson}, {Vogeley}, \&
  {Boyd}}]{2023MNRAS.518.4372P}
{Phillipson}, R.~A., {Vogeley}, M.~S., \& {Boyd}, P.~T. 2023, \mnras, 518,
  4372, \dodoi{10.1093/mnras/stac3419}

\bibitem[{{Plaschke} {et~al.}(2018){Plaschke}, {Hietala}, {Archer},
  {Blanco-Cano}, {Kajdi{\v{c}}}, {Karlsson}, {Lee}, {Omidi}, {Palmroth},
  {Roytershteyn}, {Schmid}, {Sergeev}, \& {Sibeck}}]{2018SSRv..214...81P}
{Plaschke}, F., {Hietala}, H., {Archer}, M., {et~al.} 2018, \ssr, 214, 81,
  \dodoi{10.1007/s11214-018-0516-3}

\bibitem[{Psaradakis \& Spagnolo(2002)}]{Psaradakis2002Power}
Psaradakis, Z., \& Spagnolo, N. 2002, Studies in Nonlinear Dynamics \&
  Econometrics, 6, \dodoi{10.2202/1558-3708.1091}

\bibitem[{{Ripperda} {et~al.}(2020){Ripperda}, {Bacchini}, \&
  {Philippov}}]{2020ApJ...900..100R}
{Ripperda}, B., {Bacchini}, F., \& {Philippov}, A.~A. 2020, \apj, 900, 100,
  \dodoi{10.3847/1538-4357/ababab}

\bibitem[{{Roy} {et~al.}(2019){Roy}, {Chatterjee}, {Joshi}, \&
  {Ghosh}}]{2019MNRAS.482..743R}
{Roy}, N., {Chatterjee}, R., {Joshi}, M., \& {Ghosh}, A. 2019, \mnras, 482,
  743, \dodoi{10.1093/mnras/sty2748}

\bibitem[{Shumway \& Stoffer(2005)}]{10.5555/1088844}
Shumway, R.~H., \& Stoffer, D.~S. 2005, Time Series Analysis and Its
  Applications (Springer Texts in Statistics) (Berlin, Heidelberg:
  Springer-Verlag)

\bibitem[{{Smithsonian Astrophysical Observatory}(2000)}]{2000ascl.soft03002S}
{Smithsonian Astrophysical Observatory}. 2000, {SAOImage DS9: A utility for
  displaying astronomical images in the X11 window environment}, Astrophysics
  Source Code Library, record ascl:0003.002.
\newblock \doeprint{0003.002}

\bibitem[{{Sobolewska} {et~al.}(2014){Sobolewska}, {Siemiginowska}, {Kelly}, \&
  {Nalewajko}}]{2014ApJ...786..143S}
{Sobolewska}, M.~A., {Siemiginowska}, A., {Kelly}, B.~C., \& {Nalewajko}, K.
  2014, \apj, 786, 143, \dodoi{10.1088/0004-637X/786/2/143}

\bibitem[{{Stuchl{\'\i}k} \& {Kolo{\v{s}}}(2016)}]{2016EPJC...76...32S}
{Stuchl{\'\i}k}, Z., \& {Kolo{\v{s}}}, M. 2016, European Physical Journal C,
  76, 32, \dodoi{10.1140/epjc/s10052-015-3862-2}

\bibitem[{{Tavecchio}(2017)}]{2017AIPC.1792b0007T}
{Tavecchio}, F. 2017, in American Institute of Physics Conference Series, Vol.
  1792, 6th International Symposium on High Energy Gamma-Ray Astronomy, 020007,
  \dodoi{10.1063/1.4968892}

\bibitem[{{Tavecchio} {et~al.}(2007){Tavecchio}, {Maraschi}, {Wolter},
  {Cheung}, {Sambruna}, \& {Urry}}]{2007ApJ...662..900T}
{Tavecchio}, F., {Maraschi}, L., {Wolter}, A., {et~al.} 2007, \apj, 662, 900,
  \dodoi{10.1086/518085}

\bibitem[{TSAY(1986)}]{10.1093/biomet/73.2.461}
TSAY, R.~S. 1986, Biometrika, 73, 461, \dodoi{10.1093/biomet/73.2.461}

\bibitem[{{Urry} \& {Padovani}(1995)}]{1995PASP..107..803U}
{Urry}, C.~M., \& {Padovani}, P. 1995, \pasp, 107, 803, \dodoi{10.1086/133630}

\bibitem[{{Uttley} {et~al.}(2005){Uttley}, {McHardy}, \&
  {Vaughan}}]{2005MNRAS.359..345U}
{Uttley}, P., {McHardy}, I.~M., \& {Vaughan}, S. 2005, \mnras, 359, 345,
  \dodoi{10.1111/j.1365-2966.2005.08886.x}

\bibitem[{{Vio} {et~al.}(2005){Vio}, {Kristensen}, {Madsen}, \&
  {Wamsteker}}]{2005A&A...435..773V}
{Vio}, R., {Kristensen}, N.~R., {Madsen}, H., \& {Wamsteker}, W. 2005, \aap,
  435, 773, \dodoi{10.1051/0004-6361:20042154}

\bibitem[{{Vio} {et~al.}(2020){Vio}, {Nagler}, \&
  {Andreani}}]{2020A&A...642A.156V}
{Vio}, R., {Nagler}, T.~W., \& {Andreani}, P. 2020, \aap, 642, A156,
  \dodoi{10.1051/0004-6361/202038585}

\bibitem[{Webber \& Marwan(2014)}]{webber2014recurrence}
Webber, C.~L., \& Marwan, N. 2014, Recurrence Quantification Analysis: Theory
  and Best Practices, Understanding Complex Systems (Cham, Switzerland:
  Springer), \dodoi{10.1007/978-3-319-07155-8}

\bibitem[{{Yan} {et~al.}(2018){Yan}, {Yang}, {Zhang}, {Dai}, {Wang}, \&
  {Zhang}}]{2018ApJ...864..164Y}
{Yan}, D., {Yang}, S., {Zhang}, P., {et~al.} 2018, \apj, 864, 164,
  \dodoi{10.3847/1538-4357/aadd01}

\bibitem[{{Zbilut} {et~al.}(1998){Zbilut}, {Giuliani}, \&
  {Webber}}]{1998PhLA..237..131Z}
{Zbilut}, J.~P., {Giuliani}, A., \& {Webber}, C.~L. 1998, Physics Letters A,
  237, 131, \dodoi{10.1016/S0375-9601(97)00843-8}

\bibitem[{{Zbilut} \& {Webber}(1992)}]{1992PhLA..171..199Z}
{Zbilut}, J.~P., \& {Webber}, C.~L. 1992, Physics Letters A, 171, 199,
  \dodoi{10.1016/0375-9601(92)90426-M}

\end{thebibliography}
\end{document}